\documentclass[aip,reprint]{revtex4-1}  

\usepackage{graphicx}

\usepackage{dcolumn}    

\usepackage[T1]{fontenc}
\usepackage{mathptmx}
\usepackage{mathtools}  
\usepackage{siunitx}

\DeclarePairedDelimiter\norm{\lvert}{\rvert}  
\DeclarePairedDelimiter\ket{\lvert}{\rangle}  
\DeclarePairedDelimiter\expect{\langle}{\rangle}  
\DeclareMathOperator{\erf}{\mathrm{erf}}  

\DeclareSIUnit{\dBm}{dBm}
\DeclareSIUnit{\charge}{\ensuremath{\mathit{q}}}
\DeclareSIUnit{\elementarycharge}{\ensuremath{\mathit{e}}}
\DeclareSIUnit{\atomicmassunit}{u}
\DeclareSIUnit{\linepair}{line pair}

\begin{document}

\title{Cold highly charged ions in a radio-frequency trap with superconducting magnetic shielding}

\author{Elwin~A. Dijck}
\email{elwin.dijck@mpi-hd.mpg.de}
\affiliation{Max Planck Institute for Nuclear Physics, Saupfercheckweg~1, 69117~Heidelberg, Germany}

\author{Christian Warnecke}
\affiliation{Max Planck Institute for Nuclear Physics, Saupfercheckweg~1, 69117~Heidelberg, Germany}
\affiliation{Heidelberg Graduate School for Physics, Ruprecht Karl University, Im Neuenheimer Feld~226, 69120~Heidelberg, Germany}
\affiliation{Physikalisch-Technische Bundesanstalt, Bundesallee~100, 38116~Braunschweig, Germany}

\author{Malte Wehrheim}
\affiliation{Max Planck Institute for Nuclear Physics, Saupfercheckweg~1, 69117~Heidelberg, Germany}
\affiliation{Physikalisch-Technische Bundesanstalt, Bundesallee~100, 38116~Braunschweig, Germany}

\author{Ruben~B. Henninger}
\affiliation{Max Planck Institute for Nuclear Physics, Saupfercheckweg~1, 69117~Heidelberg, Germany}

\author{Julia Eff}
\affiliation{Max Planck Institute for Nuclear Physics, Saupfercheckweg~1, 69117~Heidelberg, Germany}

\author{Kostas Georgiou}
\affiliation{Max Planck Institute for Nuclear Physics, Saupfercheckweg~1, 69117~Heidelberg, Germany}
\affiliation{School of Physics and Astronomy, University of Birmingham, Edgbaston, Birmingham~B15~2TT, United Kingdom}

\author{Andrea Graf}
\affiliation{Max Planck Institute for Nuclear Physics, Saupfercheckweg~1, 69117~Heidelberg, Germany}

\author{Stepan Kokh}
\affiliation{Max Planck Institute for Nuclear Physics, Saupfercheckweg~1, 69117~Heidelberg, Germany}

\author{Lakshmi~P. {Kozhiparambil~Sajith}}
\affiliation{Max Planck Institute for Nuclear Physics, Saupfercheckweg~1, 69117~Heidelberg, Germany}
\affiliation{Department of Physics, Humboldt University of Berlin, Newtonstra\ss{}e~15, 12489~Berlin, Germany}
\affiliation{Deutsches Elektronen-Synchrotron DESY, Platanenallee~6, 15738~Zeuthen, Germany}

\author{Christopher Mayo}
\affiliation{Max Planck Institute for Nuclear Physics, Saupfercheckweg~1, 69117~Heidelberg, Germany}
\affiliation{School of Physics and Astronomy, University of Birmingham, Edgbaston, Birmingham~B15~2TT, United Kingdom}

\author{Vera~M. Sch\"{a}fer}
\affiliation{Max Planck Institute for Nuclear Physics, Saupfercheckweg~1, 69117~Heidelberg, Germany}

\author{Claudia Volk}
\affiliation{Max Planck Institute for Nuclear Physics, Saupfercheckweg~1, 69117~Heidelberg, Germany}

\author{Piet~O. Schmidt}
\affiliation{Physikalisch-Technische Bundesanstalt, Bundesallee~100, 38116~Braunschweig, Germany}
\affiliation{Institute of Quantum Optics, Leibniz University, Welfengarten~1, 30167~Hannover, Germany}

\author{Thomas Pfeifer}
\affiliation{Max Planck Institute for Nuclear Physics, Saupfercheckweg~1, 69117~Heidelberg, Germany}

\author{Jos\'{e}~R. {Crespo~L\'{o}pez-Urrutia}}
\affiliation{Max Planck Institute for Nuclear Physics, Saupfercheckweg~1, 69117~Heidelberg, Germany}

\date{\today}

\begin{abstract}
We implement sympathetic cooling of highly charged ions (HCI) by fully enclosing a linear Paul trap within a superconducting radio-frequency resonator. A quantization magnetic field applied while cooling down into the superconducting state remains present in the trap for centuries and external electromagnetic fluctuations are greatly suppressed. A magnetic field decay rate at the~$10^{-10}$\;s$^{-1}$ level is found using trapped Doppler-cooled Be$^+$ ions as hyperfine-structure (hfs) qubits. Ramsey interferometry and spin-echo measurements on magnetically-sensitive hfs transitions yield coherence times of \SI{>400}{\milli\second}, showing excellent passive shielding at frequencies down to DC. For sympathetic cooling of HCI, we extract them from an electron beam ion trap (EBIT) and co-crystallize one together with Doppler-cooled Be$^+$ ions. By subsequently ejecting all but one Be$^+$ ions, we prepare single HCI for quantum logic spectroscopy towards frequency metrology and qubit operations with a great variety of HCI species.
\end{abstract}


\maketitle

\section{Introduction}
\label{sec:introduction}

Testing fundamental physics in the low-energy, high-precision regime requires sensitive atomic systems in stable and well-characterized environments.
Highly charged ions (HCI) feature optical transitions with enhanced sensitivity to potential variation of fundamental constants \cite{Kozlov2018} by virtue of intrinsically large relativistic effects of their fine-structure transitions \cite{Rehbehn2021,Liang2021}, crossings of orbital levels \cite{Berengut2010,Berengut2012,Bekker2019,Porsev2020}, and extreme hyperfine effects \cite{Shabaev1997,Klaft1994,Crespo1998,Beiersdorfer2001,Schiller2007,Oreshkina2017,Skripnikov2018,Noertershaeuser2019} induced by the overlap of the active electron wave function with the nucleus. 
In addition, systematic frequency shifts caused by AC Stark shifts are strongly suppressed \cite{Kozlov2018} by up to an order of $Z^{4}$ within an isoelectronic sequence, where $Z$ is the atomic number.
This bears promise for optical atomic clocks reaching relative frequency uncertainties~$\Delta\nu/\nu$ at or below the $10^{-20}$~level \cite{Yudin2014}.
Theory has identified many HCI candidates for testing physics beyond the Standard Model \cite{Kozlov2018} among different isoelectronic sequences, multiplying the number of species having forbidden ground-level transitions within the optical range necessary for an optical clock.

A very general method applied for optical frequency metrology is quantum logic spectroscopy (QLS) \cite{Schmidt2005}.
Combined with sympathetic cooling of HCI \cite{Schmoeger2015,Schmoeger2015a} within a Coulomb crystal, the first coherent laser spectroscopy of any HCI has recently been demonstrated \cite{Micke2020}. 
Time dilation frequency shifts were eliminated through algorithmic cooling of weakly-coupled motional modes \cite{King2021}, culminating in a HCI-based optical clock \cite{King2022} reaching a systematic frequency uncertainty of $2.2 \times 10^{-17}$.
The application of QLS is rather universal, and could be extended to frequency metrology in the extreme ultraviolet (XUV) range \cite{Crespo2016}.
The abundance of highly forbidden, ultra-narrow XUV transitions in HCI, and the inception of XUV frequency combs \cite{Jones2005,Gohle2005,Pupeza2021} has triggered the development of such instruments for HCI frequency metrology \cite{Nauta2021,Lyu2020}. 

Radio-frequency (RF) ion traps, also known as Paul traps for their Nobel-laureate inventor \cite{Paul1990}, enabled the development of quantum optics \cite{Dehmelt1983,Diedrich1987}, and more recently, of quantum computing \cite{Cirac1995,Monroe1995,Georgescu2020} and frequency metrology \cite{Ludlow2015}. 
Characteristics of the trapping environment often set the limit of clock performance \cite{King2022}.
Various electromagnetic perturbations affect the resonance frequency of the reference atomic species for time keeping.
To reduce these and minimize systematic frequency shifts, we developed a novel type of RF ion trap, CryPTEx-SC (Cryogenic Paul Trap Experiment -- Superconducting) \cite{Stark2021}.
It comprises a quasi-monolithic superconducting RF resonator with built-in linear quadrupole trap, which filters the RF drive, shields magnetic field fluctuations, `freezes' the static field present at the onset of superconductivity, and enables coherent operations without the need for external fields and their stabilization.


Here we present our current CryPTEx-SC setup, including its cryogenic imaging optics (Section~\ref{sec:design}), and show selected measurements characterizing the ion trap and verifying the cooling of  re-trapped HCI to millikelvin temperature (Section~\ref{sec:characterization}).
Finally, we use microwave spectroscopy on Be$^+$ ions to quantify the effectiveness of the superconducting magnetic shielding and the resulting qubit coherence time (Section~\ref{sec:superconducting_shielding}).

\section{Experimental design}
\label{sec:design}

\begin{figure*}
\includegraphics[width=\textwidth]{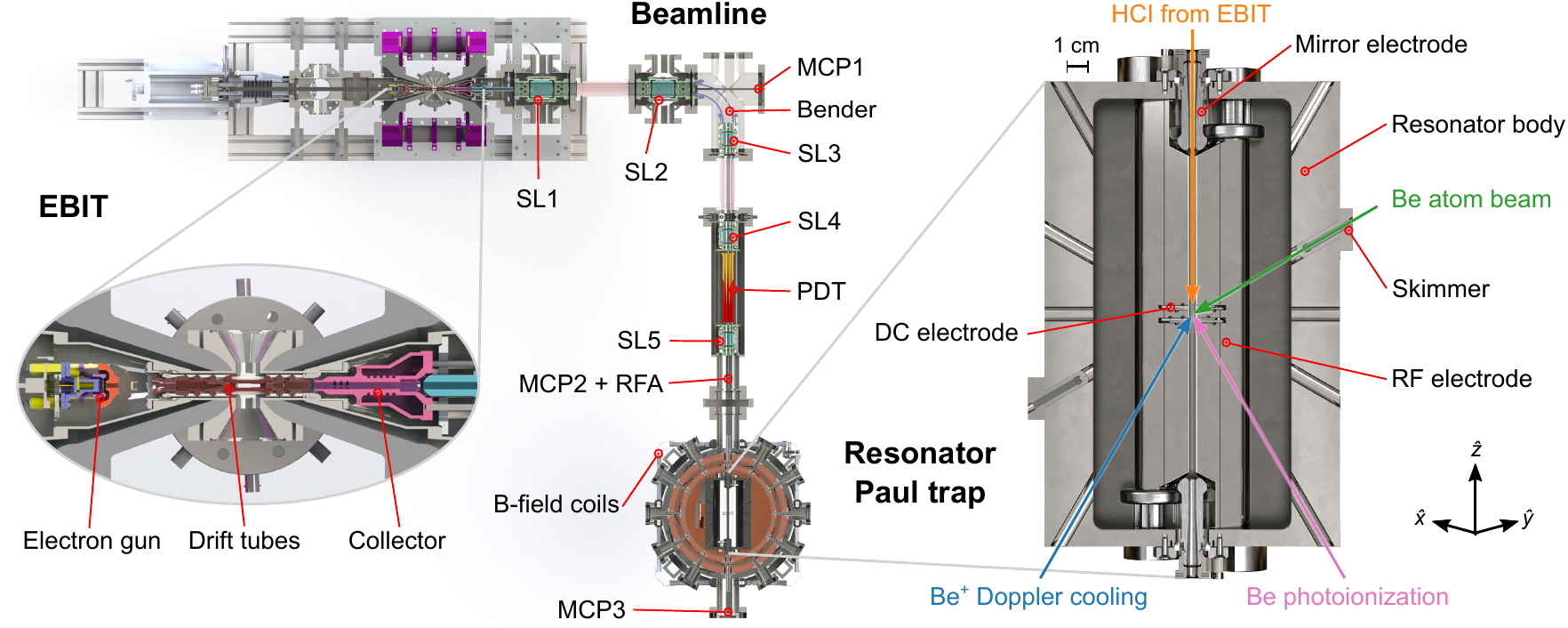}
\caption{\label{fig:schematic_overview}%
Overview of the experiment.
An electron beam ion trap (EBIT) generates HCI by electron-impact ionization.
The ions are transferred to the CryPTEx-SC Paul trap through a beamline equipped with five Sikler lenses (SL), an electrostatic bender, and a pulsed drift tube (PDT) for ion deceleration and bunching.
Three micro-channel plate (MCP) detectors are used for diagnostics; the second one is equipped with a retarding field analyzer (RFA) for kinetic energy determination.
After injection into the Paul trap, HCI are confined by mirror electrodes at both ends of the four RF blade electrodes.
Eight DC electrodes shape the central axial potential where HCI are sympathetically cooled by Be$^+$ ions.
We define a coordinate system aligned with the principal trap axes as indicated; $\hat{x}$ and $\hat{y}$ are rotated 45\textdegree{} out of the plane.
}
\end{figure*}

We produce HCI in a compact electron beam ion trap \cite{Micke2018}. For laser spectroscopy, the HCI, which are generated with megakelvin temperatures in the EBIT, have to be cooled down to the millikelvin range and below.
For this, they have to be re-trapped and sympathetically cooled by a Coulomb crystal of Be$^+$ ions prepared in the cryogenic Paul trap \cite{Schmoeger2015,Schmoeger2015a}.

\subsection{Electron beam ion trap}

Within the compact EBIT \cite{Micke2018}, permanent magnets generate a magnetic field of ~\SI{0.86}{\tesla} at the trap center, which compresses an electron beam of several mA for ionizing injected atoms to high charge states.
These ions are then radially confined by the negative space charge of the electron beam. Voltages applied to a set of six drift tubes (see Fig.~\ref{fig:schematic_overview}) confine the HCI axially. The highest charge state is the one with an ionization energy higher than the electron beam energy (up to $< \SI{6}{\kilo\electronvolt}$ in our device).
The ionization time, typically a fraction of a second, is chosen depending on the desired charge state. By pulsing the central drift tube, the HCI inventory is dumped from the trap into the beamline with an initial kinetic energy of about $E_\text{ion} \approx \SI{700}{\volt\times\charge}$, where $q$ is their electric charge. 

\subsection{HCI transfer beamline}

A beamline guides the HCI to the Paul trap, comprising several Sikler deflector lenses \cite{Mandal2011}, an electrostatic bender, and a gradient pulsed drift tube (PDT) for deceleration and compression of the ion bunch, as in CryPTEx-I \cite{Schmoeger2015,Schmoeger2015a}.
The pulsed extraction allows for time-of-flight selection of the HCI charge state and isotope among the different species produced in the EBIT.
Three micro-channel plate (MCP) detectors along the beamline (see Fig.~\ref{fig:schematic_overview}) are used for diagnostics.
The first one is set behind an aperture in the electrostatic bender in line with the first straight section of the beamline.
The other two are mounted on manipulators before and after the spectroscopy trap and can be inserted into the beam path as needed.
The second MCP has also a retarding field analyzer (RFA) consisting of two wire-mesh grids mounted in front of it for measuring the HCI bunch kinetic energy.
For this, the first grid is grounded while the second, close to the detector surface is set to an adjustable positive potential~$\Phi_\text{grid}$, which repels ions with kinetic energy $E_\text{ion} < \Phi_\text{grid} \times q$, while letting the others pass through.
Observing the ion signal while sweeping $\Phi_\text{grid}$ delivers the kinetic-energy distribution of the ion beam.

\subsection{Superconducting resonator Paul trap} \label{sec:setup_trap}

CryPTEx-SC combines a quasi-monolithic superconducting RF resonator with a linear Paul trap and is described in detail elsewhere\cite{Stark2021}. With a loaded quality factor of $Q \approx \num{3e4}$, the niobium RF resonator works as a band-pass filter around the trap drive frequency $\Omega_\text{RF} = 2\pi \times \SI{34.3}{\mega\hertz}$. This suppresses RF noise at the sidebands $\Omega_\text{RF} \pm \omega_i$ due to the secular frequencies $\omega_i$ of ions in the trap, which are induced by the RF drive and lead to motional heating of the ions \cite{Brownnutt2015,Paasche2003}.  

Four electrodes produce the quadrupole RF field for radial confinement (see Fig.~\ref{fig:schematic_overview}).
Each one encloses a co-axial cylindrical inner electrode of the opposite RF phase separated by a narrow \SI{300}{\micro\meter} gap. This increases the capacitance of the resonator and brings its resonance frequency to a value suitable for ion trapping while much reducing the required size.
Static voltages applied to eight DC electrodes mounted within the quadrupole electrodes confine the ions along the trap axis.

An antenna inserted into the niobium resonator \cite{Stark2021} sends microwaves for driving the Be$^+$ $^2\text{S}_{1/2}$ $(F=1)$ $\leftrightarrow$ $(F=2)$ at \SI{1.25}{\giga\hertz} (see Fig.~\ref{fig:be_qubit}). It is made of niobium wire with a length of \SI{57}{\milli\meter}, forming a $\lambda/4$ antenna mounted under a $\approx 20$\textdegree{} angle to the horizontal plane in order to couple to all polarization modes.
Its efficiency is affected by higher-order resonances of the resonator trap; the return loss over the operational frequency range is a few~dB.
Microwaves are fed from a function generator (Sinara Urukul) referenced to a GPS-disciplined quartz oscillator (TimeTech RefGen) and amplified by a \SI{50}{\deci\bel} amplifier (Mini-Circuits HPA-25W-272+). 

Magnetic fields for trap quantization are applied by three orthogonal pairs of coils approximating a Helmholtz configuration, and can reach up to about \SI{200}{\micro\tesla}.

A frequency-quadrupled solid state laser system (Toptica TA~FHG~pro) delivers several mW of \SI{235}{\nano\meter} light to produce Be$^+$ ions inside the trap by resonant two-step photoionization. The Be atoms come from a resistively-heated oven mounted about \SI{93}{\centi\meter} away from the trap center. It emits an atomic beam collimated by several apertures before reaching the trap region through a \SI{800}{\micro\meter} diameter skimmer (see Fig.~\ref{fig:schematic_overview}), preventing Be contamination of the superconducting surfaces.

\subsection{Doppler-cooling laser} \label{sec:lasers}

A second frequency-quadrupled UV laser system (Toptica TA~FHG~pro) is used for Doppler cooling on the $\text{Be}^+$ $\text{1s}^2\;^\text{2}\text{S}_\text{1/2}$ -- $\text{1s}^2\;^\text{2}\text{P}_\text{3/2}$ transition at \SI{313}{\nano\meter}. An acousto-optic modulator (AOM) stabilizes its power and allows fast switching of the light.
When the quantization axis defined by the magnetic field is well aligned with the laser propagation axis and the laser is circularly polarized, a closed transition between the states $^\text{2}\text{S}_\text{1/2}$ ($F=2$, $m_F=\pm2$) -- $^\text{2}\text{P}_\text{3/2}$ ($F=3$, $m_F=\pm3$) can be driven.
In general, some repumping from the $F=1$ upper ground state (see Fig.~\ref{fig:be_qubit}) is necessary.
To this end, we split off some \SI{313}{\nano\meter} laser light and shift its frequency by \SI{1.25}{\giga\hertz} with a combination of AOMs.
Both laser beams are stably pointed to the trap by piezo-driven mirrors (MRC Systems) controlled by a pair of position-sensitive detectors.
The beams are overlapped, circularly polarized and enter the trapping region under an angle of \SI{30}{\degree} to the trap axis (see Fig.~\ref{fig:schematic_overview}).
The handedness of polarization is set by a $\lambda/4$ waveplate mounted in a motorized rotation stage.

\subsection{Cryogenic imaging optics}
\label{sec:optics}

\begin{figure*}
\includegraphics[width=\textwidth]{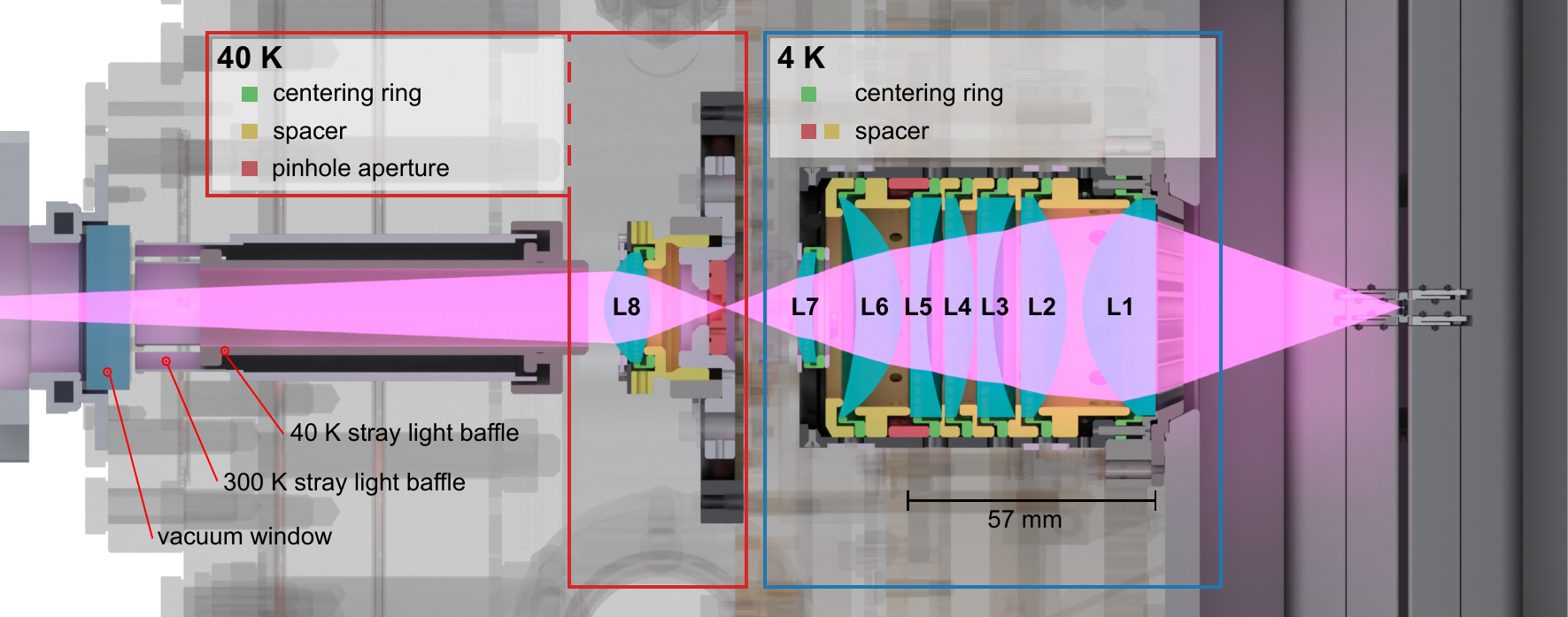}
\caption{\label{fig:optics_render}
Cross section through the cryogenic objective (rotated 90\textdegree{} for clarity). Starting from the right, lenses L1--L7 are held by centering rings (green) and separated by spacers (yellow, red), projecting an intermediate image of the laser-cooled Be$^+$ ions in the trap through a \SI{2}{\milli\meter} pinhole (red, 40\;K stage) at~$\SI{157}{\milli\meter}$ from the trap center. This aperture reduces heating of the 4\;K stage by room-temperature black-body radiation.
Custom bi-aspheric lens L8 is mounted on the 40\;K stage for relaying the intermediate image to the EMCCD camera focal plane and the PMTs for fluorescence detection. The holder for L8 (light gray) is radially positioned by four set screws and the focal distance is adjustable by a thread (yellow). Its $\SI{5}{\milli\meter}$~axial travel allows a magnification on the final image in a range of $8\times$ to $20\times$. 
Stray light from the lasers is blocked by two coaxial, thermally decoupled cylindrical shields attached to the 40\;K stage and vacuum window
flange, respectively.
}
\end{figure*}

Non-destructive detection of ions inside a Paul trap is usually accomplished by imaging with lens systems \cite{Alt2002,Noek2013,Pyka2013,WongCampos2016,Li2020,Nordmann2023} in combination with cameras and photo-multiplier tubes (PMTs).  
For cryogenic environments, single, bi-aspheric lenses \cite{Leopold2019} and reflecting Schwarzschild objectives \cite{Dubielzig2021} have been developed.
These systems are usually separated only a few millimeters from the ions to ease manufacturing requirements and reduce their size.
Here, we have designed an optical system (see Fig.~\ref{fig:optics_render}) to accommodate the relatively long distance to the trapped ions of about 60\,mm.
Fluorescence emitted by the Be$^+$ ions at \SI{313}{\nano\meter} is refocused by a $\mathrm{NA}=0.36$ eight-lens objective.
For this wavelength, we use UV fused silica (UVFS) and calcium fluoride (CaF$_2$) lenses with a maximum diameter of~\SI{50.8}{mm}, most of them stock models (see Table~\ref{tab:lenses}).

\begin{table}
\caption{\label{tab:lenses}%
Table of lenses. The last column indicates the axial distance from the preceding element at the lens center. Magnification is set by the distance to lens~L8; the quoted distance yields about $10\times$.
}
\begin{ruledtabular}
\begin{tabular}{ l c c r }
 Lens & Vendor & Product ID & Distance (\si{\milli\meter}) \\
\hline
 L1 & Edmund Optics & 67268 & \num{57.0} \\
 L2 & Newport &  SPX043 & \num{3.5} \\
 L3 & Knight Optical &  67112 & \num{6.6} \\  
 L4 & Lambda Research & PCX-50.8U-150 & \num{0.5} \\
 L5 & Eksma Optics & 112-5519E & \num{4.3} \\
 L6 & Thorlabs & LE4412 & \num{1.8} \\  
 L7 & Thorlabs & LE5414 & \num{10.3} \\ 
 L8 & Asphericon & 150111-000-03C & \num{33.0} \\ 
\end{tabular}
\end{ruledtabular}
\end{table}

To avoid a reduction of the quality factor of the resonator trap by dielectric materials inside it, the 4-K objective is mounted on its top with a superconducting chevron with 90\% transmission for light but  blocking RF. The resulting working distance is~\SI{57}{\milli\meter}. The whole stack can be adjusted radially by~\SI{1}{\milli\meter} to compensate misalignments.
Each optical element is centered by a partially segmented ring with 40 contact blades manufactured from anodized aluminum (\SI{0.25}{\milli\meter}~thick) to absorb differential thermal contraction of the Al, CaF$_2$ and UVFS materials during the cool-down.
In axial direction, spacer rings made from anodized aluminum in direct contact with the lenses set their separation and guarantee guided alignment. Their thicknesses are adjusted to measured lens dimensions to compensate manufacturing tolerances.
A non-anodized aluminum cylindrical cage holds rings, spacers, and lenses, and shields anodized parts from thermal radiation.
Apertures on its surface ensure efficient pumping between the lenses.
A~\SI{3}{\milli\meter} longitudinal cut reduces radial compression of the inner elements during cool-down.

\begin{figure}
\includegraphics[width=\columnwidth]{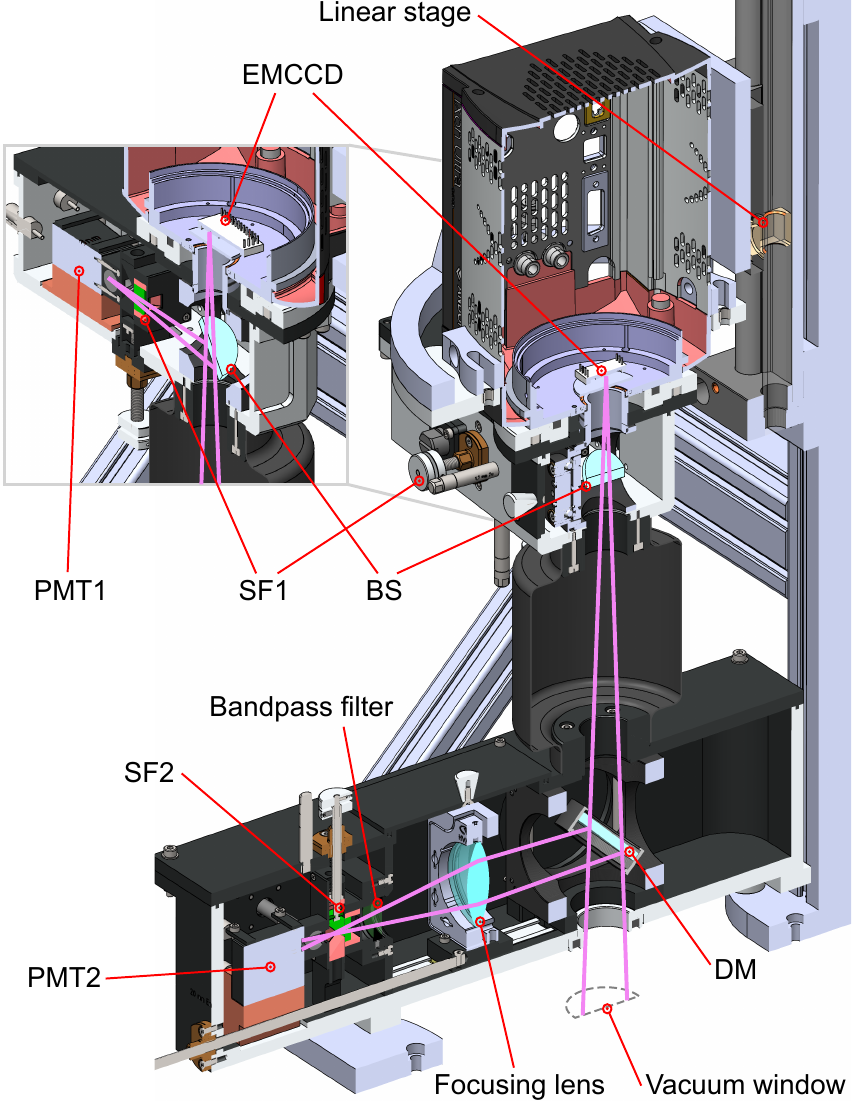}
\caption{\label{fig:EMCCD-Setup}%
Air-side imaging and fluorescence-detection setup. EMCCD: electron-multiplication charge-coupled device; PMT1: photo-multiplier tube for cooling-ion fluorescence detection; PMT2: same, for spectroscopy ion fluorescence; SF1, SF2: adjustable spatial filters; DM: dichroic mirror; BS: 50/50 beam splitter.
}
\end{figure}

On the air side, a detection unit consisting of two PMTs and a camera is mounted on top of the vacuum chamber (see Fig.~\ref{fig:EMCCD-Setup}).
The EMCCD (electron-multiplying CCD) camera (Andor iXon Ultra 888) has a pixel pitch of~\SI{13}{\micro\meter}, capable of resolving the ion positions in the trap. A rubber bellows and a linear stage are used for bringing it into focus. 

A 50/50 beam splitter moving along with the camera sends \SI{313}{\nano\meter} light to the first PMT. Its housing contains a beam block for protection against over-exposure as well as a spatial filter located in the focal plane of L8 for rejecting background light. It consists of two knife edges in vertical and two in horizontal direction that can be moved independently using fine threads. A second PMT setup (e.g., to detect light from $\text{Be}^\text{+}$ photoionization or for HCI spectroscopy) sits directly on top of the vacuum chamber, and includes a lens for refocusing light of other wavelengths.

\begin{figure}
\includegraphics[width=\columnwidth]{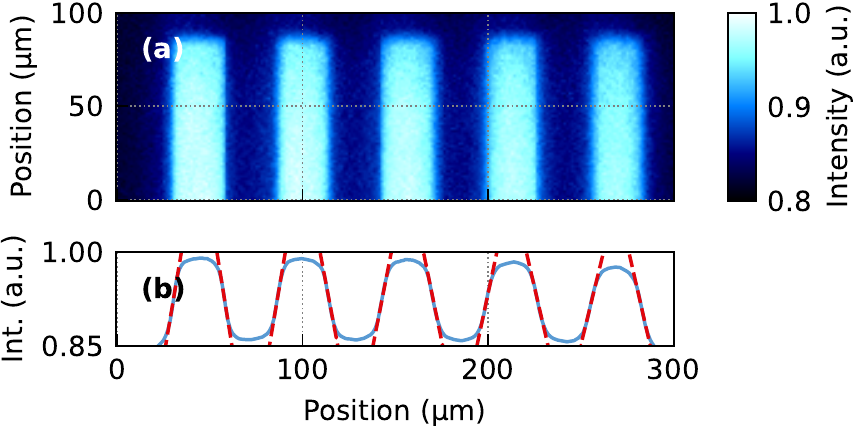}
\caption{\label{fig:optics_resolution}%
Knife-edge spatial resolution test of the lens system at $10.4\times$ magnification at room temperature.
\textbf{(a)} Cut-out from an in-air EMCCD image of an NBS~1963A resolution-test target with 18\;line~pairs~per~\si{\milli\meter} illuminated by \SI{313}{\nano\meter} light.
\textbf{(b)} Projection across the lines; the mean 10--90\% intensity rise distance is~$d_{10-90\%}=\SI{6.9 +- 0.3}{\micro\meter}$.
}
\end{figure}

Simulations performed with the software OSLO indicate a spatial resolution of the lens stack of about 110\;line pairs per \si{\milli\meter} at the \SI{20}{\percent} point of the modulation transfer function. This is compatible with the in-air characterization with a resolution test target illuminated by \SI{313}{\nano\meter}~light through a diffusive element mounted close to the target, which demonstrates a 10--90\% intensity rise distance of $d_{10-90\%} = \SI{6.9 +- 0.3}{\micro\meter}$ in object space (see Fig.~\ref{fig:optics_resolution}).
This is sufficient to resolve two trapped Be$^+$ ions at an axial secular frequency of up to approximately~$\omega_{z,0} \approx 2\pi \times \SI{1.5}{\mega\hertz}$.
Even at the lowest magnification of $8\times$, this corresponds to a separation of several pixels on the camera. Depending on the material properties, the performance of the lens system deviates with temperature  from that of the carefully aligned set up at \SI{300}{\kelvin}\cite{Asfour2017}.

\section{Ion-trap characterization}
\label{sec:characterization}

\begin{figure}
  \includegraphics[width=\columnwidth]{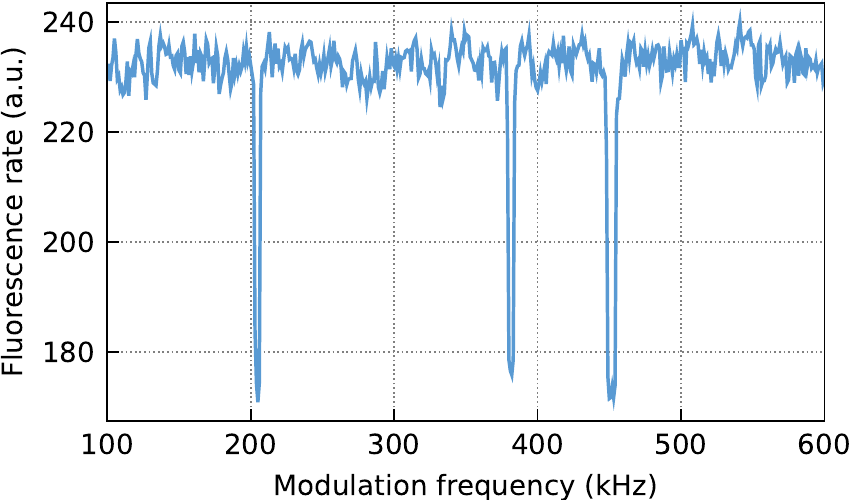}
  \caption{\label{fig:be_secular_frequencies}%
Example of a secular frequency determination. Fluorescence rate as function of tickling frequency. Three clear resonances are visible, corresponding to the axial and two radial center-of-mass secular frequencies.
}
\end{figure}

To characterize the strength of the axial and radial trapping fields, the secular frequencies of trapped $\text{Be}^\text{+}$ ions were measured by exciting the respective motional degrees of freedom.
This was achieved by applying an oscillating `tickling' electric field to the mirror electrode behind the trap.
Secular motion driven at resonance causes elongation of the ion images on the camera and a drop in fluorescence rate (see Fig.~\ref{fig:be_secular_frequencies}).
One of the methods of calibrating the magnification factor of the imaging system was using the ion separation distance on the camera image of an axial two-ion Be$^+$ crystal at a known axial secular frequency.

\begin{figure}
  \includegraphics[width=\columnwidth]{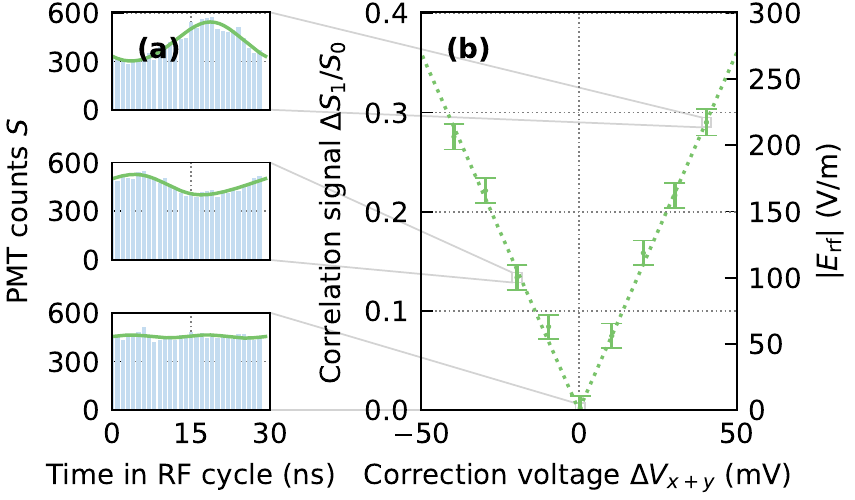}
  \caption{\label{fig:rf_photon_correlation}%
Minimization of micromotion using RF-photon correlation.
\textbf{(a)} The observed modulation~$\Delta S_1$ at the trap RF frequency~$\Omega_\text{RF}$ of the average fluorescence rate~$S_0$ is minimized by \textbf{(b)}~moving the ion along the $(\hat{x} + \hat{y})/\sqrt{2}$ axis.
Note that here the direction of induced micromotion has a component along the cooling laser propagation direction~$\hat{k}$ as it is orthogonal to the displacement.
The right-hand scale indicates the corresponding oscillating electric field amplitude~$E_\text{RF}$ at the (displaced) mean ion position.
}
\end{figure}

Micromotion caused by stray electric fields is minimized by a combination of observing the ion position along the $(\hat{x} - \hat{y})/\sqrt{2}$ axis on the camera image as function of radial trapping potential \cite{Gloger2015} and using the RF-photon correlation technique \cite{Berkeland1998,Keller2015} for reducing displacement along the $(\hat{x} + \hat{y})/\sqrt{2}$ axis (orthogonal to the image plane).
In the latter method, modulation of the Be$^+$ fluorescence rate in sync with the RF trapping field cycle is investigated.
We obtain the timestamp of PMT counts with nanosecond resolution from a Sinara DIO module.
Because this module and the function generator driving the trap are phase-locked to a common clock signal, finding the time relative to the RF phase is a simple modulo operation with the known trap drive frequency~$\Omega_\text{RF}$.
Figure~\ref{fig:rf_photon_correlation} shows a typical RF-photon correlation measurement, where the residual oscillating electric field component \cite{Keller2015} at the ion position was minimized to $\norm{\hat{k} \cdot \vec{E}_\text{RF}} \leq 16\;\text{V/m}$.

\subsection{Retrapping highly charged ions}
\label{sec:retrapping}

\begin{figure*}
  \includegraphics[width=\textwidth]{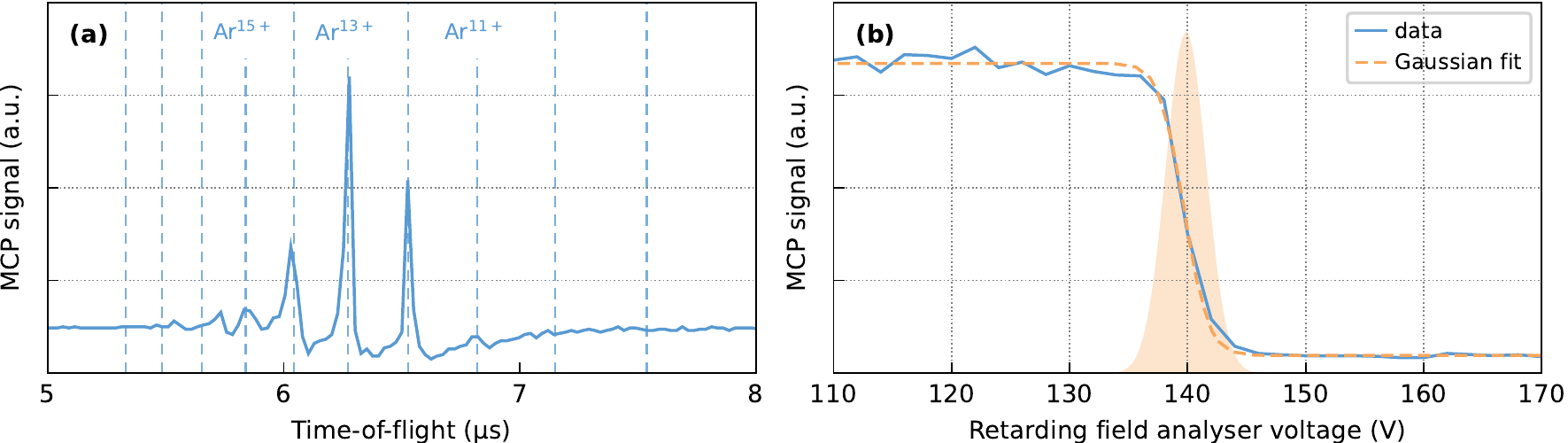}
  \caption{\label{fig:beamlinetofselection}%
Characterization of the HCI transfer beamline.
\textbf{(a)} Example of Ar HCI time-of-flight (TOF) at an EBIT extracting rate of~\SI{0.7}{\hertz}; peaks correspond to different charge states present due to the constant injection of neutral Ar into the EBIT.
Vertical dashed lines show a fit with Eq.~\eqref{eq:chargetomasstof} for argon, marking the peaks as labeled. The charge-breeding time can be used to optimize the yield of the desired charge states.
\textbf{(b)} Normalized kinetic energy distribution for $^\text{40}\text{Ar}^\text{14+}$ after optimizing the bunching in the pulsed drift tube. The Gaussian fit yields an energy spread~(1$\sigma$) of about~\SI{1.7}{\volt\times\charge}.
}
\end{figure*}

We now turn to the procedure for preparing cold highly charged ions in the spectroscopy trap.
Usual breeding times for argon ions of charge states 10$^+$ to 16$^+$ in the EBIT are on the order of hundreds of milliseconds for a \SI{10}{\milli\ampere} electron beam with an energy of about \SI{1}{\kilo\volt}. By switching the potential of the central drift tube with a high-voltage switch, the trap is inverted, and ions are emitted into the beamline. Based on results from identical miniature EBITs, a single dump contains in total several million ions in various charge states . 
They can be separated using time-of-flight (TOF) spectroscopy, and undesired species are filtered out by pulsing a kicker electrode.
The initial kinetic energy is reduced in a pulsed drift tube (PDT) before reaching the spectroscopy trap, which can be positively biased to control the final kinetic energy.

After setting the EBIT parameters for  producing the desired HCI, we iteratively optimize the voltages applied to the beamline electrodes for maximum transmission, selecting the species by TOF, and minimize the kinetic energy spread of the ion bunches. 

In general, the EBIT produces ions with a distribution of charge states depending on the continuous or pulsed influx of neutral atoms, and the ratio of the rates for electron-impact ionization and photorecombination. Isotopes and residual gas impurity ions also appear in the TOF spectra. Each ion bunch separates according to the extant charge-to-mass ratios (see Fig.~\ref{fig:beamlinetofselection}a):
For an ion of charge~$q$ and mass~$m$,
\begin{equation} \label{eq:chargetomasstof}
  t \propto \frac{1}{\sqrt{q/m}} \text{.}
\end{equation}
For selecting a HCI species of interest, one electrode of~SL3 is pulsed to let only ions within a time window of a fraction of a~\si{\micro\second} pass, while at other times deflecting unwanted ions. 

After beamline optimization, the ions are bunched and decelerated with a pulsed drift tube \cite{Schmoeger2015a}. It reduces the mean kinetic energy of the ions as well as their energy spread. Its timing is very critical, and in most cases only a single charge state is decelerated. Downstream, a second MCP is used to measure and optimize the kinetic energy distribution (see Fig.~\ref{fig:beamlinetofselection}b) with its attached retarding-field analyzer by sweeping the retarding potential~$\Phi_\text{grid}$. At the end of the beamline, the HCIs are brought from the initial mean kinetic energy of \SI{700}{\volt\times\charge} down to about~\SI{140}{\volt\times\charge}, low enough for the following retrapping step, but keeping the ion beam from diverging too much before reaching the spectroscopy trap.

\begin{figure}
\includegraphics[width=\columnwidth]{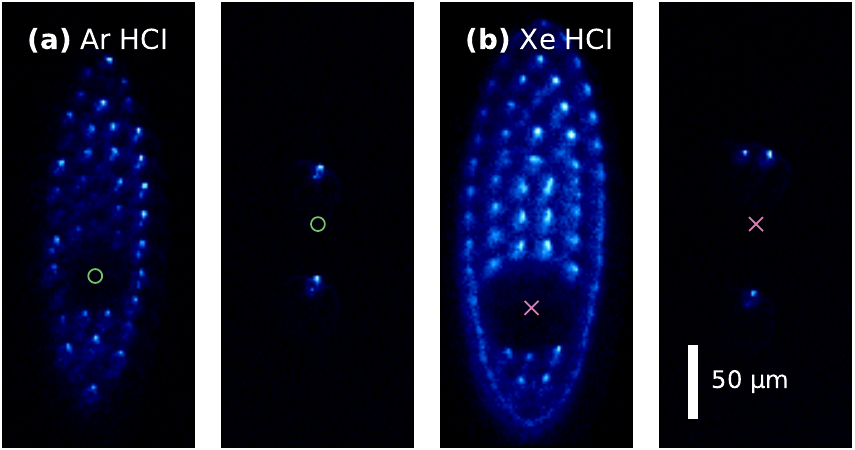}
\caption{\label{fig:hcicrystals} Examples of mixed-species crystals consisting of $\text{Be}^+$ ions and \textbf{(a)} one argon or \textbf{(b)} one xenon HCI.
The left-hand image displays a larger crystal immediately after (re-)capturing one HCI.
Such crystals are exposed to excitation of $\text{Be}^+$ secular motion while suspending Doppler cooling for reducing the number of $\text{Be}^+$ as needed.
}
\end{figure}

After passing through two Einzel lenses mounted on the heat shields, the HCI enter the superconducting linear Paul trap. The entire trap is biased to a positive potential set about~\SI{1}{\volt} below the HCI kinetic energy to further slow them down on entering the trap.
A cylindrical mirror electrode is mounted at each end of the about~\SI{15}{\centi\meter} long quadrupole trap (see Fig.~\ref{fig:schematic_overview}). After briefly pulsing down the one facing the beamline to allow HCI to enter, both mirror electrodes stay at a higher potential reflecting the HCI back and forth under radial RF confinement.
The closing timing is carefully optimized to prevent HCI leaving the trap after the first reflection.

At this point, the residual kinetic energy of the HCI is still larger than the axial trap depth in the section in the middle of the quadrupole electrodes. Following the established technique \cite{Schmoeger2015,Schmoeger2015a}, an ion crystal consisting of several dozen laser-cooled $\text{Be}^\text{+}$ ions is prepared there beforehand.
On each pass through the crystal while oscillating between the mirror electrodes, the HCI dissipate energy, transferring it to the laser-cooled ion crystal though the Coulomb interaction.
They eventually reach standstill after some hundreds of reflections, forming a mixed-species crystal with the $\text{Be}^\text{+}$ ions. Then, sympathetic cooling by the $\text{Be}^\text{+}$ ions ensures reaching mK temperatures.

Since HCI do not interact with the Doppler-cooling laser, their signature on the crystal images is a dark spot with a radius larger than the distance between individual Be$^+$ ions due to the stronger Coulomb repulsion.
Examples of mixed crystals containing one argon or one xenon HCI are displayed in Fig.~\ref{fig:hcicrystals}.
Charge exchange reactions with residual gas molecules, primarily hydrogen in the cryogenic trap, reduce the lifetime of the HCI to about 5~to~10~min.
We plan to reduce the vacuum conductance to the inner trap region by closing off ports with windows wherever possible, and lowering the operating temperature of the trap to increase this lifetime.

\subsection{Charge-state determination}

For verifying the charge state of a re-trapped HCI, we examine the enclosing ion crystal.
The HCI is strongly coupled to the Be$^+$ ions, and its effect depends on its charge-to-mass ratio.
Two sensitive ion crystal properties are its structure and secular mode frequencies. To calculate their dependence on the co-trapped HCI species, we consider the potential energy of an ion crystal in the harmonic pseudopotentials of the trap, and the Coulomb repulsion.
For a crystal of~$N$ ions with charges~$q_i$ and masses~$m_i$ at positions~$\vec{r}_i = (x_i, y_i, z_i)$, the potential energy is given by\cite{Thompson2015}
\begin{eqnarray}
  V(\vec{r}_1, ..., \vec{r}_N) = \sum_i { \frac{1}{2} m_i \left( \omega_{x,i}^2 x_i^2 + \omega_{y,i}^2 y_i^2 + \omega_{z,i}^2 z_i^2 \right) } \nonumber\\
  {} + \frac{1}{4 \pi \epsilon_0} \sum_{i\neq j} {\frac{q_i q_j}{\norm{\vec{r}_i - \vec{r}_j}} } \text{.} \label{eq:potential}
\end{eqnarray}
Here, the trap depth is parametrized by the secular frequencies~$\omega_{k,i}$, which depend on the charge and mass of ion~$i$.
Defining the secular frequencies for a single Be$^+$ ion with $q_0 = \SI[retain-explicit-plus=true]{+1}{\elementarycharge}$ and $m_0 = \SI{9.01}{\atomicmassunit}$ as nominal frequencies~$\omega_{k,0}$, the secular frequencies for a HCI of charge~$q_i = \theta_i q_0$ and mass~$m_i = \mu_i m_0$ are related to the nominal frequencies as
\begin{eqnarray}
  \omega_{x,i}^2 &=& \frac{\theta_i}{\mu_i} \frac{\omega_{x,0}^2 - \omega_{y,0}^2 - \omega_{z,0}^2}{2} + \frac{\theta_i^2}{\mu_i^2} \frac{\omega_{x,0}^2 + \omega_{y,0}^2 + \omega_{z,0}^2}{2} \nonumber\\
  \omega_{y,i}^2 &=& \frac{\theta_i}{\mu_i} \frac{\omega_{y,0}^2 - \omega_{x,0}^2 - \omega_{z,0}^2}{2} + \frac{\theta_i^2}{\mu_i^2} \frac{\omega_{x,0}^2 + \omega_{y,0}^2 + \omega_{z,0}^2}{2} \\
  \omega_{z,i}^2 &=& \frac{\theta_i}{\mu_i} \omega_{z,0}^2 \text{.} \nonumber
\end{eqnarray}
The equilibrium positions~$\vec{r}_{\text{eq},i}$ of an ion crystal for given confinement strengths~$\omega_{k,0}$ can be found by minimizing~Eq.~\eqref{eq:potential}.

As an example, we calculate the equilibrium positions for the type of ion crystal needed for quantum logic, consisting of one Be$^+$ ion and one HCI.
Assuming sufficiently strong radial confinement~$\omega_{z,0} \ll \omega_{x,0},  \omega_{y,0}$, the crystal will be oriented along the trap axis with $x_{\text{eq},i} = y_{\text{eq},i} = 0$.
The equilibrium positions of the two ions are the solution to
\begin{equation} \label{eq:equipos_axial}
  \left. \frac{\partial V}{\partial z_i} \right|_{z_i = z_{\text{eq},i}} = 0 \text{.}
\end{equation}
This can be solved analytically to yield
\begin{equation}
  z_{\text{eq},\text{Be$^+$}} = \mp \frac{\theta_\text{HCI}}{(1 + \theta_\text{HCI})^{2/3}} \ell_0 \text{,} \quad z_{\text{eq},\text{HCI}} = \pm \frac{1}{(1 + \theta_\text{HCI})^{2/3}} \ell_0 \text{,}
\end{equation}
where~$\ell_0$ is the characteristic length scale defined as
\begin{equation} \label{eq:one}
  \ell_0 = \sqrt[3]{\frac{q_0^2}{4 \pi \varepsilon_0 m_0 \omega_{z,0}^2}} \text{.}
\end{equation}
Note that the positions only depend on the charge state of the HCI, and not on its mass.
For larger ion ensembles, numerical minimization may be used in Eq.~\eqref{eq:equipos_axial}.
A small modification of this configuration may be caused by light forces acting on the Doppler-cooled Be$^+$ ions but not on the HCI, which is neglected in Eq.~\eqref{eq:potential}.

\begin{figure}
\includegraphics[width=\columnwidth]{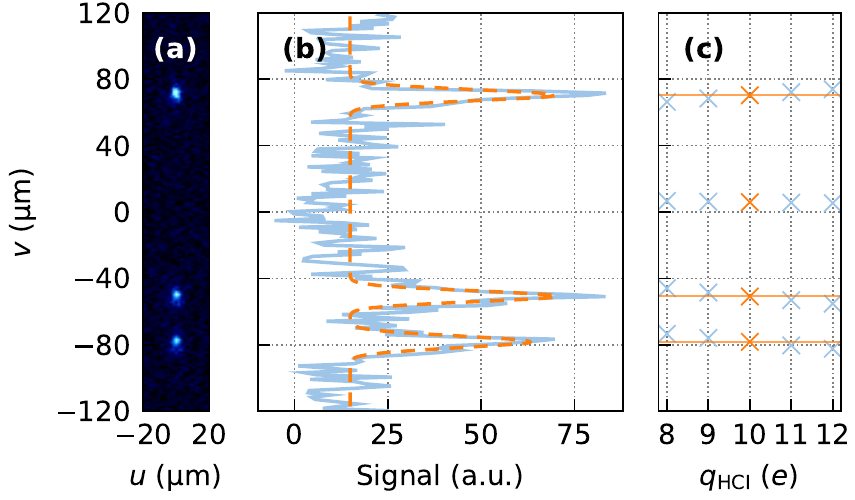}
\caption{\label{fig:ar10_charge_state}%
Example charge-state determination for an axial linear crystal consisting of a $^{40}$Ar HCI co-trapped with three $^{9}$Be$^+$ ions. \textbf{(a)}~ Crystal image showing Be$^+$ fluorescence.
\textbf{(b)}~Ion positions as fitted with three two-dimensional Gaussian functions and a constant background to the image (dashed line).
\textbf{(c)}~ Comparison of calculated (crosses) at $\omega_{z,0} = 2\pi \times \SI{118.3}{\kilo\hertz}$ to fitted equilibrium positions (horizontal lines, uncertainty smaller than line thickness). The position of the inner Be$^+$ ion indicates a HCI charge state of $q_\text{HCI} = \SI[retain-explicit-plus=true]{+9.8 +- 0.2}{\elementarycharge}$.}
\end{figure}

To determine the charge state of re-trapped HCI, we analyze the ion positions in axial linear mixed crystals.
An example with one Ar HCI and three Be$^+$ ions is depicted in Fig.~\ref{fig:ar10_charge_state}.
Observed and calculated equilibrium positions are compared by minimizing Eq.~\eqref{eq:potential} values for different assumed charge states~$q_\text{HCI}$.
We thus conclude that the dark spot in Fig.~\ref{fig:ar10_charge_state} contains an $^{40}\text{Ar}^{10+}$ ion.
Note that in this ion configuration, it is in fact possible to obtain the charge state of the HCI independent of the secular frequency from the position of the inner Be$^+$ ion relative to the the outer Be$^+$ ions.

\subsection{Spatial thermometry}
\label{sec:spatial_thermometry}

After establishing the HCI charge state, recorded images of mixed crystals can be used to determine ion temperatures. Ions in thermal equilibrium randomly move in the trapping potential, and their spatial probability distribution is visible on camera images. For a single Be$^+$ ion at temperature~$T$ in a harmonic potential in the weak-binding limit, the time-averaged spatial distribution in each of the three principal trap directions~$k$ (averaged over time) is Gaussian with an approximate width of \cite{Knuenz2012,Rajagopal2016}
\begin{equation} \label{eq:spatial_extent_beion}
  \sigma_{\text{th},k,0} = \sqrt{ \frac{k_B T}{m_0 \omega_{k,0}^2} } \text{.}
\end{equation}
Here, we neglect micromotion, which would slightly increase the width in both radial directions\cite{Blatt1986}, but remains below the level of a few percent for small values of the Matthieu stability parameter $q\ll 1$.

In a mixed-species ion crystal, all motional modes are thermally excited following the kinetic energy distribution.
The width of the spatial distribution in direction~$k$ for ion~$i$ is then given by \cite{Rajagopal2016}
\begin{equation} \label{eq:thermometry_gamma_def}
  \sigma_{\text{th},k,i} = \sqrt{ \sum_p { S_{k,i,p} \frac{k_B T}{m_i \omega_{k,p}^2 } } } = \sqrt{ \gamma_{k,i}^2 \frac{k_B T}{m_0 \omega_{z,0}^2} } \text{,}
\end{equation}
where the sum runs over all normal modes~$p$ of the crystal, with mode frequencies~$\omega_{k,p}$ and eigenvectors~$S_{k,i,p}$.
In general, at a given temperature~$T$ the spatial extent for each ion is different; by normalizing all values to the one of a given Be$^+$ ion from Eq.~\eqref{eq:spatial_extent_beion}, the scaling for each ion can be expressed in a set of weighting factors~$\gamma_{k,i}$ that depend only on the crystal configuration and the ratios of the radial secular frequencies to the axial secular frequency. Note that for a crystal consisting purely of Be$^+$ ions, $\omega_{z,0}$ is the frequency of the axial center-of-mass mode.

\begin{table}
\caption{\label{tab:gamma_factors}%
Numerically calculated spatial thermometry weighting factors for an axial linear crystal consisting of a single Ar$^{10+}$ ion co-trapped with 3 Be$^+$ ions . Horizontal axis ($u$) corresponds to radial trap coordinate; vertical axis ($v$), to axial trap coordinate (see main text for details).
}
\begin{ruledtabular}
\begin{tabular}{lldd}
\multicolumn{2}{l}{Ion} & \multicolumn{1}{c}{\textrm{Image horizontal~$\gamma_{u,i}$}} &
\multicolumn{1}{c}{\textrm{Image vertical~$\gamma_{v,i}$}} \\
\colrule
1 & $\text{Be}^\text{+}$ & 0.291 & 0.665 \\
2 & $\text{Ar}^\text{10+}$ & 0.123 & 0.358 \\
3 & $\text{Be}^\text{+}$ & 0.319 & 0.536 \\
4 & $\text{Be}^\text{+}$ & 0.304 & 0.640 \\
\end{tabular}
\end{ruledtabular}
\end{table}

We obtain the motional modes of a crystal by numerically solving the eigensystem describing secular ion motion as perturbation of its equilibrium position, defining $\vec{r}_i(t) = \vec{r}_{\text{eq},i} + \vec{\Delta r}_i (t)$.
When an ion is cold enough, the local potential around it can be approximated as harmonic by linearizing all forces. The Lagrangian for the axial secular motion is written in terms of the displacements~${\Delta z}_i$ and corresponding velocities~${\Delta \dot{z}}_i$ as:
\begin{equation}
  L_z = \frac{1}{2} \sum_i { m_i {\Delta \dot{z}}_i^2 } - \frac{1}{2} \sum_{i,j}{  \left. \frac{\partial^2 V}{\partial z_i \partial z_j} \right|_{{\Delta z}_i = 0} {\Delta z}_i {\Delta z}_j } \text{,}
\end{equation}
taking the potential~$V$ from Eq.~\eqref{eq:potential} and equilibrium positions from Eq.~\eqref{eq:equipos_axial}.
The axial eigenmodes ($\omega_{z,p}$, $S_{z,i,p}$) of the crystal are then numerically obtained from the corresponding equations of motion, and the calculation is repeated for the two radial directions.

The spatial extent of ion~$i$ on the camera image is the convolution of its projected spatial probability distribution and the point-spread function (PSF) of the imaging system.
Assuming an approximately Gaussian PSF, this convolution results in
\begin{equation} \label{eq:sigma_psf_thermal}
  \sigma_{u,i} = \sqrt{ \sigma_{\text{PSF},u}^2 + \sigma_{\text{th},u,i}^2 } \quad \text{and} \quad \sigma_{v,i} = \sqrt{ \sigma_{\text{PSF},v}^2 + \sigma_{\text{th},v,i}^2 } \text{,}
\end{equation}
denoting the two image axes as $u,v$.
In our experimental setup, one image axis corresponds to the axial coordinate of the trap, while the other axis mixes both radial coordinates under a \SI{45}{\degree} angle (see Sec.~\ref{sec:setup_trap}).
We obtain the projected spatial extents by way of Eq.~\eqref{eq:thermometry_gamma_def}, defining the effective weighting factors as:
\begin{equation}
  \gamma_{u,i} = \sqrt{ \frac{\gamma_{x,i}^2 + \gamma_{y,i}^2}{2} } \quad \text{and} \quad \gamma_{v,i} = \gamma_{z,i} \text{.}
\end{equation}


\begin{figure}
\includegraphics[width=\columnwidth]{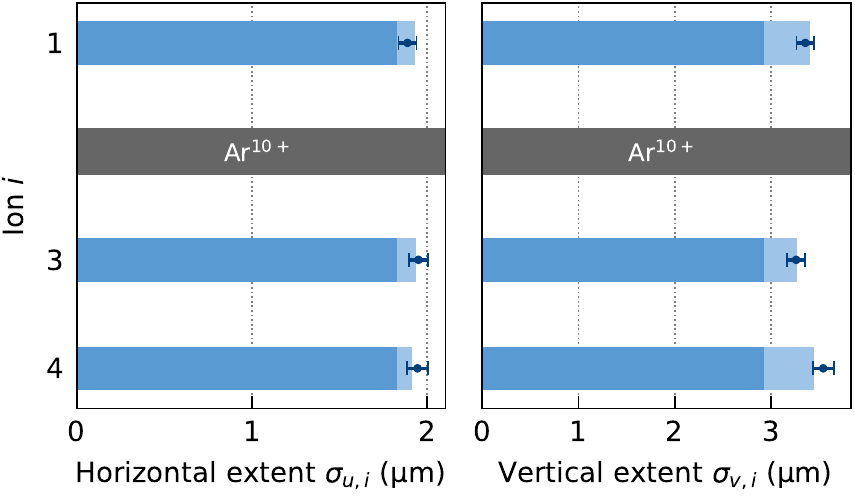}
\caption{\label{fig:ar10_spatial_thermometry}%
Spatial thermometry of mixed species crystal in Fig.~\ref{fig:ar10_charge_state}. The data points with error bars represent the spatial extent of Be$^+$ ions on the image horizontal ($u$) and vertical ($v$) coordinates, corresponding to radial and axial trap coordinates, respectively.
Bars show a fit with Eq.~\eqref{eq:sigma_psf_thermal}.
The (small) difference in ion spatial extent following the relative factors given in Table~\ref{tab:gamma_factors} is signature of a thermal component; the fit yields a temperature of $T = \SI{4 +- 3}{\milli\kelvin}$. The darker part of the bars shows the PSF contribution, taken to be identical for all ions.
}
\end{figure}

For the trapping parameters of the crystal in Fig.~\ref{fig:ar10_charge_state}, the calculated weighting factors~$\gamma_{u,i}$ and $\gamma_{v,i}$ are given in Table~\ref{tab:gamma_factors}.
Note that normalizing to the nominal axial secular frequency in Eq.~\eqref{eq:thermometry_gamma_def} means that the $\gamma_{v,i}$~values in this table are independent of the axial secular frequency, whereas the $\gamma_{u,i}$~values depend on the ratios of the two radial secular frequencies to the axial secular frequency.
The observed horizontal and vertical spatial extents of the ions in the crystal of Fig.~\ref{fig:ar10_charge_state} obtained from the two-dimensional Gaussian fits to the image data are shown in Fig.~\ref{fig:ar10_spatial_thermometry}.
From Eq.~\eqref{eq:sigma_psf_thermal} it is clear that a thermal contribution would show as a relative difference in spatial extent between the ions that follows the pattern of the weighting factors in Table~\ref{tab:gamma_factors}.

The results of a global fit to the observed spatial extents is shown as colored bars in Fig.~\ref{fig:ar10_spatial_thermometry}, taking $\sigma_{\text{PSF},u}$, $\sigma_{\text{PSF},v}$ and $T$ as free parameters.
The fitted temperature from the $\text{Be}^+$ ions is $T = \SI{4 +- 3}{\milli\kelvin}$.
We can assume thermal equilibrium due to the strong ion-ion Coulomb interaction, and thus this temperature also holds for the $\text{Ar}^\text{10+}$ ion, confirming cooling of HCI to millikelvin temperatures close to the Doppler limit.
The uncertainties from spatial thermometry are relatively large, mostly because it requires resolving a small thermal component blurred by the larger PSF contribution.
Assuming a symmetric Gaussian PSF, the relation $\sigma_{\text{PSF}} = d_{10-90\%} / (2\sqrt{2} \erf^{-1}{0.8})$ holds, which indicates general agreement between these PSF values and the in-air characterization of the cryogenic optics described in Sec.~\ref{sec:optics}.



\section{Superconducting magnetic shielding}
\label{sec:superconducting_shielding}

A unique feature of CryPTEx-SC is its ability to permanently freeze a pre-set magnetic field flux. When a hollow superconductor is cooled below its critical temperature~$T_c$ in a finite magnetic field, the onset of perfect diamagnetism expels an extant external field from the bulk material by inducing eddy currents on its outer and inner surfaces. This freezes the magnetic flux inside the hollow, as already observed by Meissner and Ochsenfeld \cite{MeissnerOchsenfeld1933}.

Niobium, a type-II superconductor, exhibits flux pinning when the expulsion of magnetic flux from the bulk of the superconductor is not complete, forming flux tubes at pinning centers such as lattice imperfections or impurities. This effect is stronger in thin layers.
Trapped flux causes finite surface resistance in superconducting radio-frequency (RF) cavities \cite{Vallet1992}.

\begin{figure}
\includegraphics[width=\columnwidth]{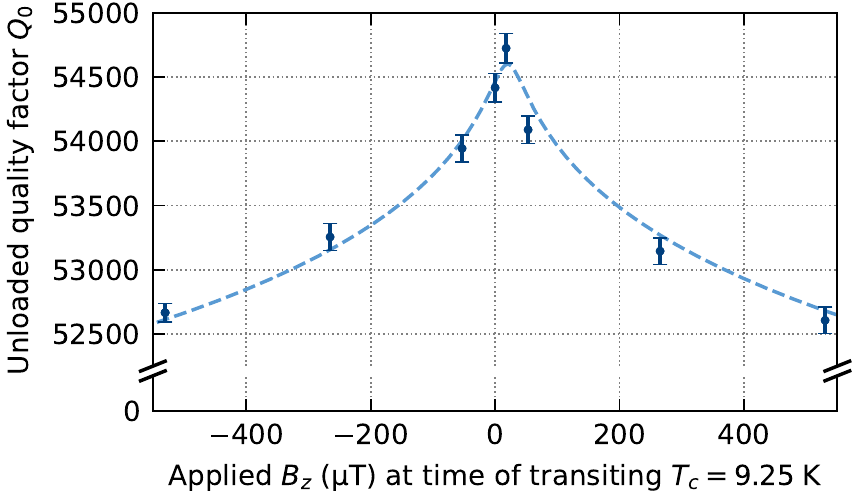}
\caption{\label{fig:qfactor_effect_bfield}%
Unloaded quality factor~$Q_0$ of the~\SI{35}{\mega\hertz} quadrupole resonance at $T=\SI{4.2}{\kelvin}$ for high input power as function of the vertical magnetic flux density applied at the time the superconductor was cooled below~$T_c$.
The dashed line is an empirical fit assuming a power law for the surface resistance~$R_s \propto \norm{\vec{B}}^n$ yielding $n \approx \num{0.3}$.
Maximum $Q_0$ is reached at a finite positive $B_z$ that cancels the background magnetic field.
}
\end{figure}

We observe some influence of the magnetic field strength at the time of cooling down below~$T_c=\SI{9.2}{\kelvin}$ on the quality factor. When operating it in superconducting state at higher RF input powers, a reduction depending on the value of the magnetic field trapped during the last cooldown is seen (see Fig.~\ref{fig:qfactor_effect_bfield}).
For this measurement, we repeatedly warmed up the resonator above~$T_c$ using a resistive heater, applied an external magnetic field and let the resonator cool down to~$T=\SI{4.2}{\kelvin}$ again.
The maximum applied magnetic field magnitude of~\SI{0.6}{\milli\tesla} is far below the critical field strength~$H_c$ of order~\SI{100}{\milli\tesla} for niobium, thus the material always reached the Meissner state.

Reflective measurements of the unloaded quality factor~$Q_0$ of the quadrupole resonance at~\SI{35}{\mega\hertz} were done with a network analyzer at~\SI{20}{\dBm} input power connected to the inductive coupler.
An increase of~\SI{0.5}{\milli\tesla} in $\vec{B}$-field magnitude reduces~$Q_0$ by about~\SI{5}{\percent} with also an associated~$10^{-7}$ fractional shift of the resonance frequency.
The highest quality factor is reached at a finite vertical magnetic flux density compensating the background $\vec{B}$-field component.
Note that the magnetic field from the field coils might not be homogeneous throughout all of the resonator material.

The resonator quality factor $Q_0$ can be written in terms of the geometry factor~$G$ and the surface resistance~$R_s$ at the relevant RF frequency~$\omega_0$ as \cite{Padamsee2001}
\begin{equation}
  Q_0 = \frac{G}{R_s} \text{.}
\end{equation}
The geometry factor is the ratio of the magnetic field strength integrated over the cavity volume to that integrated over the cavity surface, \cite{Padamsee2001}
\begin{equation}
  G = \frac{\omega_0 \mu_0 \int_V {\norm{\vec{H}}^2 \mathrm{d}v}}{\int_S {\norm{\vec{H}}^2 \mathrm{d}s}} \text{,}
\end{equation}
which only depends on the resonator geometry. For our cavity we calculate it with the finite-element method (FEM) to be $G=\SI{1.7}{\ohm}$.
If we assign the observed change in~$Q_0$ to an increase in the surface resistance related to flux trapping, our data appears to suggest a nonlinear dependence rather than the typical linear dependence at low $\vec{B}$-field strength \cite{Kim1965}.
Fig.~\ref{fig:qfactor_effect_bfield} shows an empirical least-squares fit assuming a power-law behavior for $R_s \propto \norm{\vec{B}}^n$, yielding $n \approx \num{0.3}$.
However, we note that we only observe this $\vec{B}$-field dependent effect at high in-coupled power.
At low power, $Q_0$ is higher, reaching~$Q_0\approx \num{200000}$, and shows no dependence on~$\norm{\vec{B}}$ over the studied range.
Similar interplay with the intra-cavity electric field magnitude has been associated with hydride-formation in Nb superconducting cavities \cite{Bonin1992}, suggesting that this process may also play a role in the overall performance of our cavity.
In any case, under current operating conditions, the effect of the trapped flux used as quantization axis for the ions does not significantly limit trap performance.

\subsection{Microwave hfs spectroscopy}
\label{sec:beion_microwave_spectroscopy}

Generally, magnetic-field insensitive hyperfine transitions are chosen as qubits for quantum information processing, to limit decoherence caused by fluctuating external fields.
By operating at a specific (`magic') magnetic field magnitude, one ground state hyperfine transition component in Be$^+$ can be made first-order insensitive to magnetic field fluctuations at a finite field magnitude \cite{Gaebler2016}.
However, here we investigate magnetically-sensitive hyperfine-structure (hfs) transitions in order to benchmark the superconducting shielding performance of our trap.

\begin{figure}
\includegraphics[width=\columnwidth]{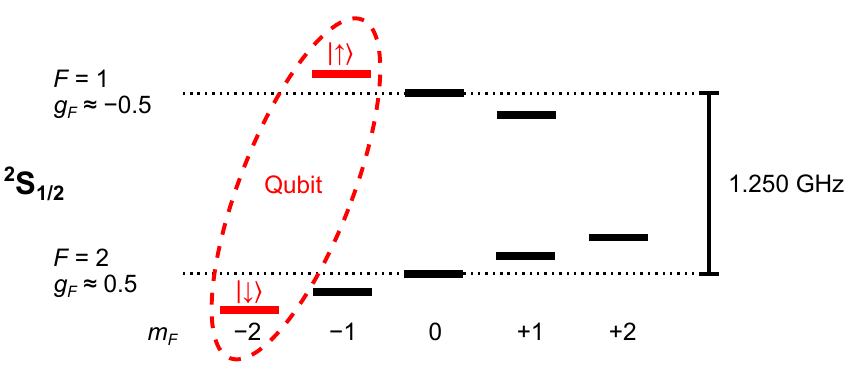}
\caption{\label{fig:be_qubit}%
Hyperfine structure of the $^\text{2}\text{S}_\text{1/2}$ ground state of $^9$Be$^+$. The transition component $(F, m_F) = (2, -2) \leftrightarrow (1, -1)$ with maximal first-order Zeeman shift is chosen as qubit for most of this work.
}
\end{figure}

The magnetic field in the trap was characterized by microwave spectroscopy on single Be$^+$ ions (see Fig.~\ref{fig:be_qubit}) using the transition $\ket{\downarrow} = \ket{F=2,\; m_F=-2}$ to $\ket{\uparrow} = \ket{1, -1}$ with an on-resonance Rabi time of~\SI{33}{\micro\second}.
We apply the Ramsey method with time-separated oscillatory fields to obtain good frequency resolution \cite{Ramsey1950}.
Pulse sequences are implemented using the ARTIQ experiment control system \cite{ARTIQ} to run the RF signal generators for the AOMs, the microwave source, and for recording the signal from the PMT.

\begin{figure}
\includegraphics[width=\columnwidth]{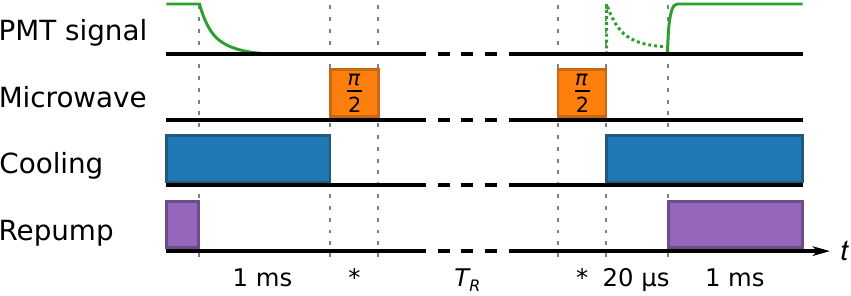}
\caption{\label{fig:ramsey_sequence}%
Ramsey-pulse sequence: switching scheme of microwave radiation, cooling and repumper laser.
Sketched is also the fluorescence signal measured with the PMT.
See main text for details.
}
\end{figure}

The sequence starts with both the cooling and repumper \SI{313}{\nano\meter} lasers switched on for Doppler cooling (see Fig.~\ref{fig:ramsey_sequence}).
State preparation consists of turning off the repumper laser, so that the cooling laser, which is not entirely free of a $\pi$-polarized component, pumps the Be$^+$ ion with highest probability into the upper qubit state~$\ket{\uparrow}$.
Subsequently, the cooling laser is also turned off, and a microwave pulse sequence comprised of a $\pi/2$ pulse, a variable wait time~$T_R$ and a second $\pi/2$ pulse is executed.
For state detection, the cooling laser is unblocked, and fluorescence recorded for a set time matching the timescale of optical pumping.
If the sequence left the ion in the $\ket{\uparrow}$ state, no fluorescence appears; if instead the microwave radiation successfully flips the qubit to the lower state~$\ket{\downarrow}$, fluorescence will be observed until the ion is pumped into the dark qubit state $\ket{\uparrow}$ again.
This timescale, which depends on the purity of circular polarization of the laser, its intensity, and the alignment between its propagation direction and the quantization axis, limits the signal-to-noise ratio for a single detection trial.
After the detection time window, both lasers are unblocked for \SI{1}{\milli\second} to (re-)cool the ion, and then the sequence can be repeated.

\begin{figure*}
\includegraphics[width=\textwidth]{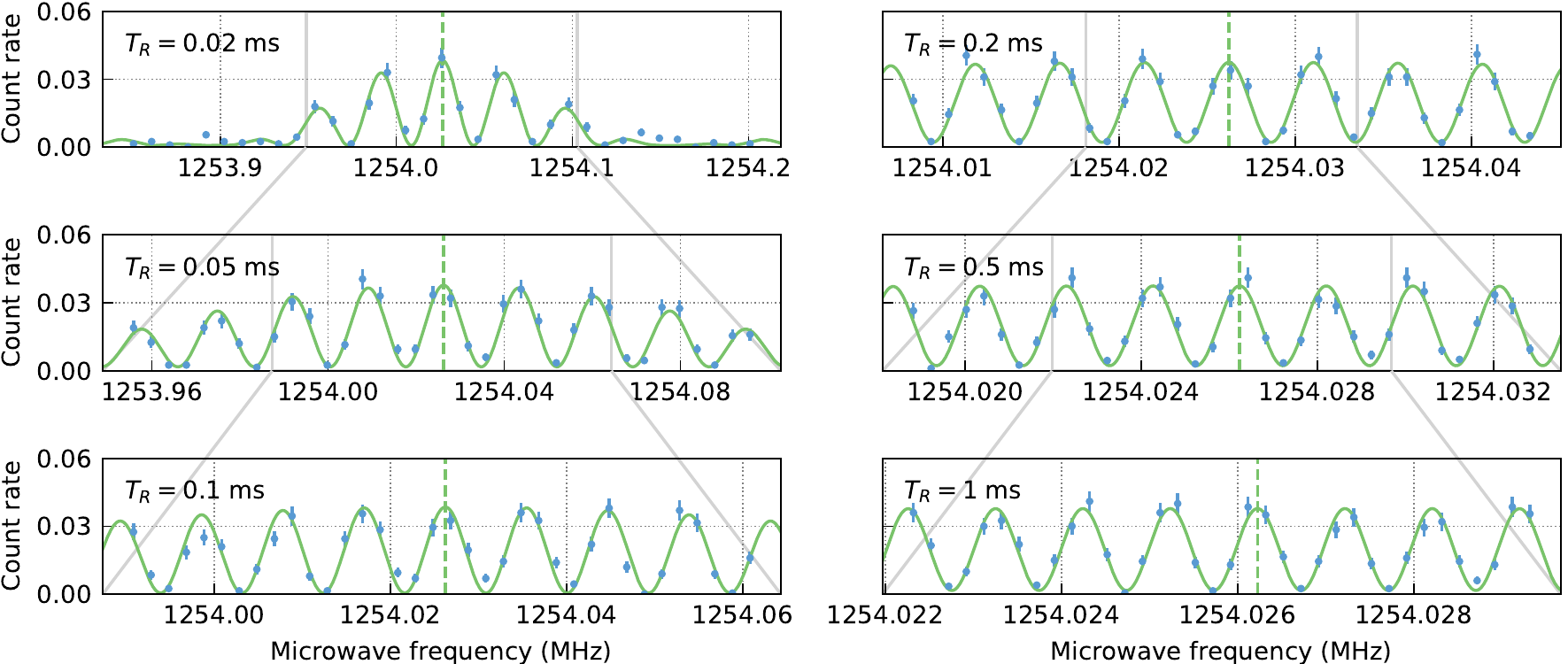}
\caption{\label{fig:ramsey_example}%
Example Ramsey measurement series showing the observed fluorescence rate versus the microwave frequency near the qubit hfs transition. Data are for wait times~$T_R$ between \SI{10}{\micro\second} and \SI{1}{\milli\second}, adapting the frequency range and step size at each $T_R$ for keeping the number of measured fringes roughly constant.
The line shows a global fit of the excitation probability; a dashed vertical line marks fitted resonance frequency.
Each data point is averaged over 2000 sequences; uncertainties are derived from Poisson statistics. The $\vec{B}$-field magnitude at the ion position determined from this series is \SI{190.5026 +- 0.0003}{\micro\tesla}. At this time, the magnetic field was not well-aligned with the cooling laser propagation direction, resulting in few signal photons per repeat due to rapid optical pumping into dark states.
}
\end{figure*}

As function of the frequency detuning, the observed transition probability shows the characteristic Ramsey pattern (see Fig.~\ref{fig:ramsey_example}), where the width of its envelope depends on the Rabi time and the fringe spacing on $1/T_R$. While accurate resonance frequency determination requires long wait times, finding among the many fringes the central one becomes cumbersome. For this, we combine a series of frequency scans with different frequency ranges and values of $T_R$. We identify the central peak at each $T_R$ directly \cite{Chuchelov2019} using a global least-squares fit of the Ramsey pattern \cite{Ramsey1950} yielding the transition frequency.

\subsection{Decay of trapped magnetic flux}

\begin{figure}
\includegraphics[width=\columnwidth]{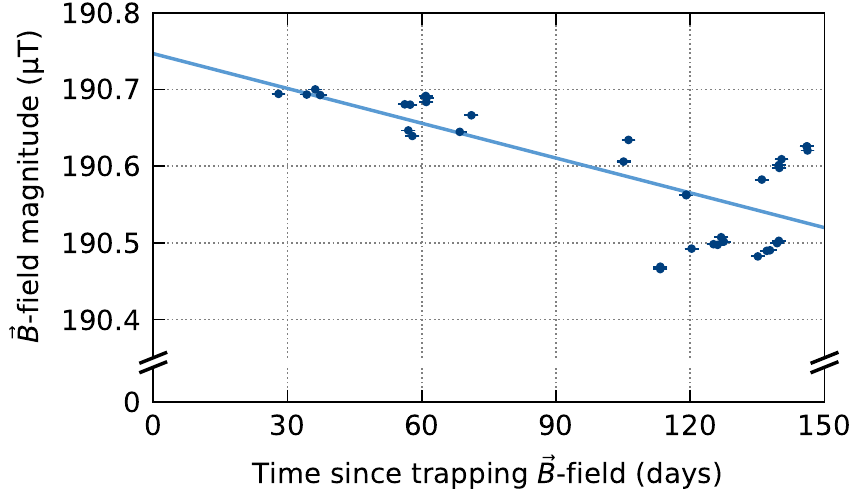}
\caption{\label{fig:frozen_bfield_decay}%
Long-term stability of the magnetic flux trapped inside the superconducting resonator, with its magnitude decreasing less than 0.2\% over several months. The fitted exponential decay would yield an effective lifetime of about 300~years.
Uncertainties are smaller than the symbols; variation in chosen trapping parameters causes scatter in the data.
}
\end{figure}

Using the method described above, the frequency of the magnetic-field-sensitive transition was measured, yielding the Zeeman splitting and the $\vec{B}$-field magnitude at the ion position. After trapping a magnetic flux in the superconductor of about~\SI{190}{\micro\tesla} at the trap center, the external field coils were switched off and measurements were performed over several months (see Fig.~\ref{fig:frozen_bfield_decay}) during which the resonator was kept below its critical temperature.

From the measured frequency of the qubit hfs transition, we subtract the zero-field hyperfine splitting, known to \num{e-11} precision \cite{Shiga2011}, and obtain the Zeeman shift~$\Delta\nu_Z$.
To first order, the magnetic field magnitude is then
\begin{equation}
  \norm{\vec{B}} = \frac{h \Delta\nu_Z}{\mu_B \Delta(g_F m_F)} \text{,}
\end{equation}
where the numerical factor for the qubit transition is given by
$\Delta(g_F m_F) \approx \num{1.5018}$,
corresponding to a magnetic field sensitivity of \SI{21}{\kilo\hertz/\micro\tesla}.
To include quadratic and higher order coupling between the ground state hyperfine levels, we use the Breit--Rabi formula \cite{Shiga2011}.
Fig.~\ref{fig:frozen_bfield_decay} shows how the trapped magnetic field magnitude decreased by only about~\SI{0.1}{\percent} over 100 days.
Assuming a continuous exponential decay, the least-squares-fitted lifetime of the trapped $\vec{B}$-field is about~300~years, which would correspond to a continuous fractional change in magnitude of
\begin{equation}
  \frac{1}{\norm{\vec{B}}} \frac{\mathrm{d} \norm{\vec{B}}}{\mathrm{d} t} \approx \SI{1e-10}{\per\second} \text{.}
\end{equation}
During longer measurement series, we observe frequency drifts consistent with $\vec{B}$-field changes at this order of magnitude.
However, these drifts are non-monotonic and at least in part appear to correlate with temperature fluctuations of the cryogenic system.
Small changes of ion position in a $\vec{B}$-field gradient could also cause frequency shifts, which we explore further in Section~\ref{sec:single_qubit_addressing}. The scatter in right half of Fig.~\ref{fig:frozen_bfield_decay} corresponds to a greater variation in trap operating parameters chosen during this period.

Presently we cannot establish whether the $\vec{B}$-field decay is continuous (aside from flux quantization) or possibly activated by operation at high RF input power or superconductor temperature.
The lifetime may thus represent an effective value for typical trap operation.
In any case, the long-term stability of the pre-set quantization $\vec{B}$-field demonstrated here is a great advantage for quantum experiments, comparable to operation with permanent magnets \cite{Ruster2016}.
There, thermal drifts determine the long term field stability; we similarly find that stabilizing the resonator temperature by a PID-loop controlling current through a heating resistor mounted on the resonator body yields the most stable conditions for qubit operation.

\subsection{Shielding of external magnetic fields}

Fluctuating external magnetic fields can induce decoherence. We now investigate how efficiently the superconducting resonator shields the trapped ions from environmental magnetic noise. Our outermost shielding within the vacuum chamber consists of nested OFHC (oxygen-free high conductivity) copper heat shields \cite{Stark2021} that are nearly identical to those of the sister experiment \cite{Leopold2019,Micke2020} at PTB-Braunschweig. These shields filter magnetic field noise above corner frequencies of \SIrange{0.14}{0.30}{\hertz} along different axes by induced counteracting eddy currents, as characterized in the PTB trap \cite{Leopold2019}.

\begin{figure}
\includegraphics[width=\columnwidth]{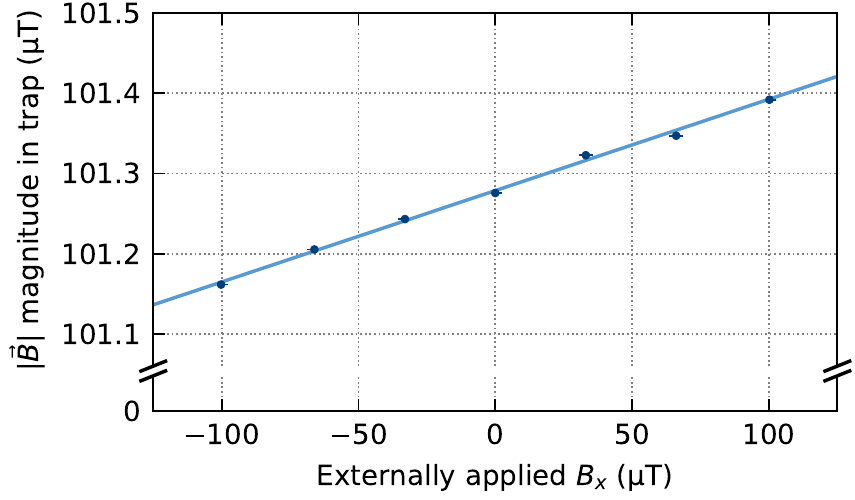}
\caption{\label{fig:bfield_dc_shielding}%
DC magnetic shielding: $\vec{B}$-field flux density at the ion position from qubit resonance frequency versus externally applied $\vec{B}$-field (under a \SI{30}{\degree} angle). Note the different axis scales.
Uncertainties are smaller than the symbol size.
}
\end{figure}

Moreover, our superconducting resonator also shields magnetic field changes down to zero frequency. We determine the effective DC shielding factor by observing the qubit resonance frequency while applying external fields using each pair of coils in turn. Figure~\ref{fig:bfield_dc_shielding} shows the results for a variable DC external field along the trap axis, which is under a \SI{30}{\degree} angle to the direction of the previously trapped $\vec{B}$-field. A linear component as function of the externally applied field is apparent here. We calculate the shielding factor for the trapped ion as the ratio of the observed change to the expected change without the superconductor by a linear least-squares fit, resulting in~\num{761 +- 18}, with the uncertainty reflecting the spread in the data.
The fact that the shielding is not complete could be caused by the various openings in the resonator housing, where magnetic field lines may be able to slowly penetrate the inner volume. Changes in qubit frequency using the other two pairs of coils are consistent with the same shielding factor, taking into account the relative angles of applied fields.
This demonstrates a passive magnetic shielding
of about \SI{57}{\deci\bel}  
at frequencies down to DC, comparable to some dedicated magnetically-shielded rooms \cite{Altarev2014} employing nested shields of \textmu-metal. We can assume that for higher frequencies the total shielding factor of our setup is at least as good as the product of this and the component from the OFHC heat shields \cite{Leopold2019}. 

\subsection{Qubit coherence time}

Magnetic noise, as well as power and frequency fluctuations of the microwave source driving the Be$^+$ qubit reduce its coherence time.
We reference the microwave function generator to a clock signal with a stability of~$10^{-10}\sqrt{\text{s}}/\sqrt{\tau}$ over time~$\tau$.
We investigate the qubit decoherence by two methods: observing the Ramsey fringe contrast, which is sensitive also to low-frequency noise \cite{Ladd2010} ($T_2^*$), and spin-echo sequences sensitive to high-frequency noise ($T_2$).

\begin{figure}
\includegraphics[width=\columnwidth]{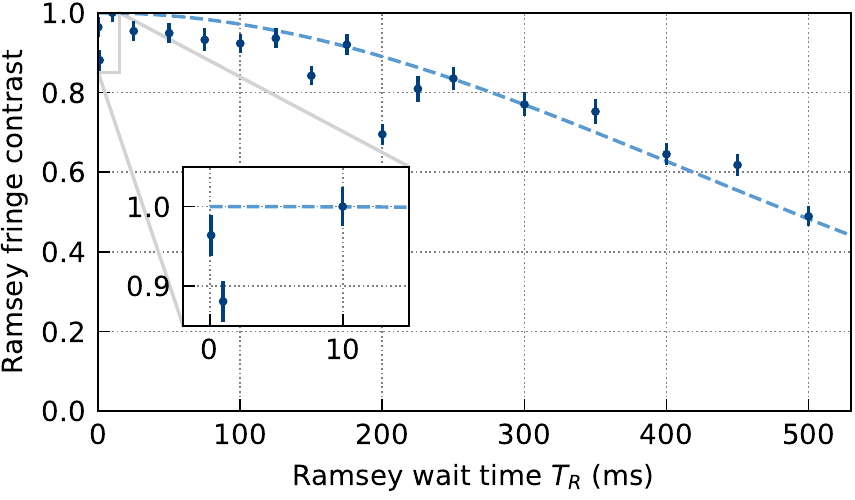}
\caption{\label{fig:ramsey_contrast}%
Measurement of Ramsey fringe contrast as function of wait time~$T_R$
up to \SI{500}{\milli\second}. 
The reduction in contrast is consistent with non-Markovian noise; a Gaussian fit (dashed line) 
indicates a $1/\sqrt{e}$ coherence time of $\tau = \SI{414 +- 22}{\milli\second}$. 
}
\end{figure}

Decoherence manifests in a contrast reduction of the Ramsey pattern with increasing wait time.
A series with wait times $T_R$
of up to~\SI{500}{\milli\second} 
are shown in Fig.~\ref{fig:ramsey_contrast}.
We determine the contrast by varying the detuning of the microwave pulses. The functional shape of the decoherence as function of wait time depends on the noise spectrum. If Gaussian phase noise is the dominant process, the coherence depends on the relative length of the noise correlation time~$\tau$ and the experimental cycle \cite{Monz2011}. For $T_R$ shorter than the noise correlation time, the coherence exhibits a Gaussian decay~$\propto e^{-t^2 / (2\tau^2)}$, which at longer wait times turns into an exponential decay~$\propto e^{-t/{T}}$.
The contrast in Fig.~\ref{fig:ramsey_contrast} shows Gaussian reduction pointing to a noise-correlation time on the order of the maximum wait time.
A least-squares fit 
yields~$\tau = \SI{414 +- 22}{\milli\second}$. 

\begin{figure}
\includegraphics[width=\columnwidth]{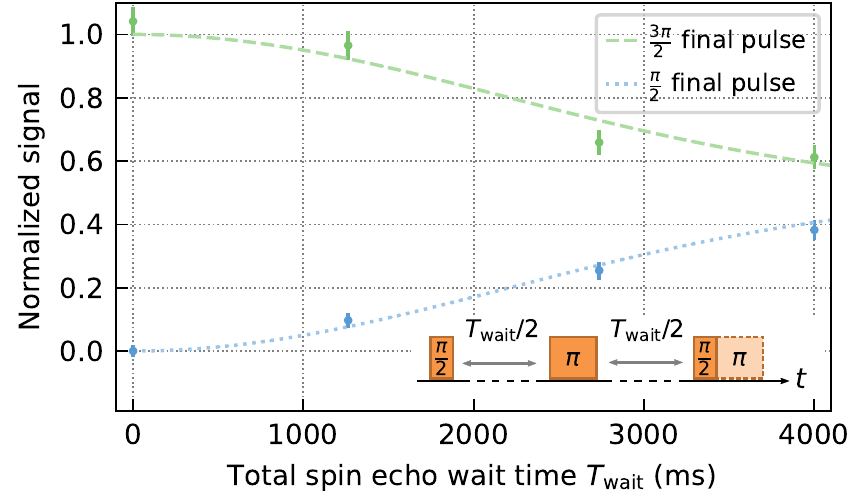}
\caption{\label{fig:spin_echo}%
Spin-echo measurement with a total wait time of up to~\SI{4}{\second}.
Inset shows the pulse sequence, where the final pulse controls whether the~$\ket{\uparrow}$ or $\ket{\downarrow}$ state is reached in absence of decoherence.
A Gaussian fit (dashed line) yields a $1/\sqrt{e}$ coherence time of~$\tau = \SI{2200 +- 100}{\milli\second}$.
}
\end{figure}

In the Hahn spin-echo sequence \cite{Vandersypen2005}, a $\pi$-pulse is added at the midpoint of the wait time in a Ramsey sequence (see inset Fig.~\ref{fig:spin_echo}). 
This reverses the low-frequency phase evolution of the spin, refocusing in the second half a phase rotation accrued in the first half.
This cancels the effect of slow drifts of the qubit frequency, leaving dephasing by random noise dominant.
Contrary to the Ramsey case, the refocusing pulse causes the qubit to end up in the same state it started from when driven on resonance. Decoherence will increase the probability of ending in the other state, with both probabilities becoming equal when coherence is completely lost.
To independently measure signal and background, we interleave sequences where the final pulse effects a $3\pi/2$ spin rotation, thereby inverting the probabilities of reaching either state.

Results of spin-echo measurements where~$T_\text{wait}$ was varied between~\SI{100}{\micro\second} and \SI{4}{\second} are shown in Fig.~\ref{fig:spin_echo}. The probability of finding the correct state decreases with~$T_\text{wait}$, becoming close to random at the longest wait time.
A least-squares Gaussian fit yields $\tau = \SI{2200 +- 100}{\milli\second}$.
This coherence time is significantly longer than in the Ramsey case, and shows the refocusing effect of the spin-echo method.
Assuming the decoherence to be caused by Gaussian $\vec{B}$-field noise, its root-mean-square fluctuations can be quantified as \cite{Monz2011,Leopold2019}
\begin{equation}
  \sqrt{\expect{\Delta B}^2} \approx \frac{\hbar}{\mu_B \Delta(g_F m_F)} \frac{1}{\tau} \text{,}
\end{equation}
which gives an root-mean-square (rms) noise amplitude of \SI{3.5 +- 0.2}{\pico\tesla}.

\subsection{Single qubit addressing}
\label{sec:single_qubit_addressing}

\begin{figure}
\includegraphics[width=\columnwidth]{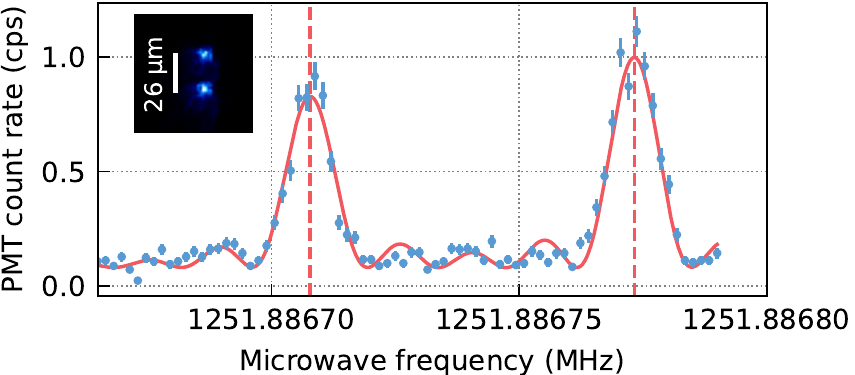}
\caption{\label{fig:individual_addressing}%
Individual addressing of two Be$^+$ simultaneously trapped qubits.
A small $\vec{B}$-field gradient splits the two transition frequencies apart
by~\SI{60}{Hz},
resolvable when attenuating the microwave intensity
by about~\SI{70}{\deci\bel}.
The fit comprises two Rabi line shapes where the widths are consistent with the Rabi time of~\SI{150}{\milli\second}.
Inset displays an EMCCD image of the two-ion axial crystal.
}
\end{figure}

These long coherence times allow us to greatly reduce the intensity of the microwave radiation while still coherently exciting the qubits, thus increasing the Rabi time and reducing the frequency width of the excitation. By measuring the qubit resonance frequency of a single ion moved along the trap axis near its center, we observed in the trapped $\vec{B}$-field an undesired gradient
of \SI{0.1}{\micro\tesla/\milli\meter}.
While this can be eliminated with the external field coils in the next cooldown, its presence allows to distinguish two simultaneously trapped Be$^+$ ion based on the difference in resonance frequencies.
Figure~\ref{fig:individual_addressing} shows that by increasing the Rabi time to \SI{150}{\milli\second} we are able to independently excite two Be$^+$ qubits
separated by about~\SI{60}{\hertz}.
The small difference in peak height is explained by a position-dependent light collection efficiency; we see the same effect when moving a single ion along the trap axis.

Magnetic field gradients in combination with ion motion are also a potential source of decoherence.
The ion motion is described by a thermal distribution of quantized harmonic oscillations, treated in more detail in Section~\ref{sec:spatial_thermometry}.
Using Eq.~\eqref{eq:spatial_extent_beion} with a typical axial secular frequency of $\omega_{z,0} \approx 2\pi \times \SI{215}{\kilo\hertz}$ and an ion temperature of $T \approx \SI{0.5}{\milli\kelvin}$ close to the Be$^+$ Doppler temperature\cite{Leibfried2003}, we find an rms length scale of $\sigma_z \approx \SI{0.5}{\micro\meter}$.
In the $\vec{B}$-field gradient, this corresponds to an rms $\vec{B}$-field magnitude
of about $\sigma_B \approx \SI{50}{\pico\tesla}$.
However, during the Ramsey and spin-echo measurements, the cooling lasers are blocked and no thermalization takes place, which results in purely harmonic ion motion absent heating.
Such motion would lead to an effect that is periodic in the wait time at the oscillation period and not the observed decoherence.

\section{Discussion}

We have demonstrated the primary feature of our CryPTEx-SC experiment: preparing cold highly charged ions in a low-noise trap for precision spectroscopy.
The HCI are produced in an electron beam ion trap, before re-trapping and sympathetic cooling in a crystal of $^9$Be$^+$ ions pre-loaded in the cryogenic radio-frequency trap for spectroscopy.
The trap features a multi-lens cryogenic objective with large numerical aperture and working distance designed to minimize black-body radiation in the trap region.
Analysis of mixed ion crystal images enabled identification of the HCI charge state and confirmation of reaching mK temperatures.

The setup is similar to its sister experiment \cite{King2022} at PTB, Braunschweig but has the unique feature of integrating the linear RF ion trap with a superconducting resonator \cite{Stark2021}, which minimizes magnetic field noise without requiring active field stabilization.
This is beneficial for coherent manipulations of ion qubits, including quantum logic spectroscopy techniques.
Flux trapping in the superconducting niobium preserves the $\vec{B}$-field present at the time of cooldown, making stable external quantization fields unnecessary and eliminating associated technical noise.
By performing microwave spectroscopy on the ground state hyperfine structure of Be$^+$ ions, we found the fractional decay of the stored $\vec{B}$-field to be on the order of \SI{e-10}{\per\second}.
Presence of a small $\vec{B}$-field gradient along the trap axis optionally enables individual qubit addressing.

Moreover, the superconductor passively attenuates external magnetic field fluctuations down to zero frequency with a measured shielding factor at DC of~\SI{57}{\deci\bel}.
At finite frequencies (above about~\SI{0.1}{\hertz}), the two copper thermal shields enclosing the spectroscopy trap provide additional shielding \cite{Leopold2019}.
Using Ramsey measurements,
we observe coherence times of~\SI{>400}{\milli\second}
on the maximally $\vec{B}$-field sensitive Be$^+$ qubit without active field stabilization or AC line triggering, over an order of magnitude longer than reported for the PTB trap without superconducting shielding \cite{Leopold2019} and comparable to experiments implementing permanent magnets and double \textmu{}-metal shielding \cite{Ruster2016}.
In spin-echo measurements, where slow drifts are canceled, the observed coherence time of \SI{2.2 +- 0.1}{\second} corresponds to Gaussian $\vec{B}$-field noise with a root-mean-square amplitude of $\sqrt{\expect{\Delta B}^2} = \SI{3.5 +- 0.2}{\pico\tesla}$.
Potential remaining sources of decoherence include external $\vec{B}$-field noise intruding through the openings in the resonator, noise on the DC trap electrodes, ion motion in the $\vec{B}$-field gradient, ion heating and microwave noise.
Quantum lock-in techniques could be implemented \cite{Kotler2011} to further investigate the noise spectrum. 

Longer coherence times in trapped ion qubits have been demonstrated for transitions intrinsically insensitive to $\vec{B}$-field fluctuations \cite{Harty2014}, using decoherence-free subspaces \cite{Kotler2014} or dynamic decoupling \cite{Wang2021}.
The latter techniques add overhead to experimental procedures and increase demands on gate fidelity.
Our results show that in a cryogenic ion trap that is passively shielded by the combination of enclosing superconducting material and copper thermal shields, reaching long coherence times is not limited to magnetically-insensitive trapped ion qubits.
This is advantageous for direct extreme-ultraviolet frequency-comb spectroscopy \cite{Lyu2020}, optical clocks \cite{Ludlow2015,King2022}, as well as quantum computing \cite{Haffner2008} and simulation \cite{Blatt2012} with trapped ions.

Currently, we are setting up a laser system for driving Raman transitions in Be$^+$ for sideband cooling and quantum logic spectroscopy of HCI. Planned improvements in the vacuum system and the RF coupling will lead to lower operating pressure and longer lifetimes of the trapped HCI. The present setup is the blueprint for an upgraded superconducting trap (project VAUQSI), which will soon be assembled.

Improved trapping and cooling of HCI will bring opportunities for fundamental physics research with this large class of atomic systems.
The demonstrated low $\vec{B}$-field noise will be beneficial not only for HCI research, but also for other atomic and molecular ions.
In particular, the decoherence rate in our trap will be a great advantage for future quantum logic studies involving tens of ion qubits, where fidelity has to be boosted to reduce error rates.
By constructing decoherence-free subspaces \cite{Lidar1998} in ensembles of trapped ions, the present insensitivity to magnetic noise could be further enhanced, making the qubits better probes of, e.g., spin-spin interactions \cite{Kotler2014}, isotopic shifts \cite{Manovitz2019}, violations of Lorentz invariance \cite{Pruttivarasin2015,Dreissen2022}, and in searches for physics beyond the Standard Model \cite{Porsev2020,Kozlov2018}. 

\begin{acknowledgments}
We thank the MPIK workshops with T.~Spranz and his team for the outstanding quality of their work in manufacturing hardware for the experiment, and the engineering design group of F.~Müller for their excellent support.
We thank S.~Schiller and S.~Sturm for providing access to reference signals from a maser and GPS clock.
We thank our collaborators at Physikalisch-Technische Bundesanstalt for providing us with a bi-aspheric lens and laser components.
This project received funding from the Max-Planck Society;
the Max-Planck–Riken–PTB-Center for Time, Constants and Fundamental Symmetries;
the European Metrology Programme for Innovation and Research (EMPIR), which is co-financed by the Participating States and from the European Union's Horizon 2020 research and innovation program (Project No.~17FUN07~CC4C and 20FUN01~TSCAC);
the European Research Council (ERC) under the European Union's Horizon 2020 research and innovation programme (grant agreement No.~101019987);
and the Deutsche Forschungsgemeinschaft (DFG, German Research Foundation) through the collaborative research center SFB~1225~ISOQUANT, through Germany's Excellence Strategy–EXC-2123 QuantumFrontiers–390837967, and through SCHM2678/5-1.
We thank for generous funding by the German Federal Ministry of Education and Research (BMBF) within the program 13N15973 "Quantum technologies -- from basic research to market" (Projekt VAUQSI -- Viel-Frequenz-Ansteuerung Ultrastabiler Qubits in Supraleitenden Ionenfallen).
\end{acknowledgments}

\section*{Author declarations}

\subsection*{Conflict of interest}

The authors have no conflicts to disclose.

\subsection*{Author contributions}

\textbf{Elwin~A. Dijck:} Conceptualization (supporting); Formal analysis (lead); Investigation (lead); Methodology (lead); Supervision (supporting); Writing -- original draft (lead); Writing -- review \& editing (lead).
\textbf{Christian Warnecke:} Conceptualization (supporting); Formal analysis (supporting); Investigation (equal); Methodology (supporting); Supervision (supporting); Writing -- original draft (supporting); Writing -- review \& editing (equal).
\textbf{Malte Wehrheim:} Formal analysis (supporting); Investigation (equal); Methodology (supporting); Writing -- review \& editing (equal).
\textbf{Ruben~B. Henninger:} Formal analysis (supporting); Investigation (equal); Methodology (supporting); Writing -- review \& editing (supporting).
\textbf{Julia Eff:} Investigation (supporting); Writing -- review \& editing (supporting).
\textbf{Kostas Georgiou:} Investigation (supporting); Writing -- review \& editing (supporting).
\textbf{Andrea Graf:} Resources (lead).
\textbf{Stepan Kokh:} Investigation (equal); Writing -- review \& editing (supporting).
\textbf{Lakshmi~P. {Kozhiparambil~Sajith}:} Investigation (supporting); Writing -- review \& editing (supporting).
\textbf{Christopher Mayo:} Investigation (supporting); Writing -- review \& editing (supporting).
\textbf{Vera~M. Sch\"{a}fer:} Investigation (supporting); Writing -- review \& editing (equal).
\textbf{Claudia Volk:} Investigation (supporting); Writing -- review \& editing (supporting).
\textbf{Piet~O. Schmidt:} Conceptualization (supporting); Funding acquisition (supporting); Writing -- review \& editing (equal).
\textbf{Thomas Pfeifer:} Conceptualization (supporting); Funding acquisition (equal); Project administration (equal); Supervision (equal); Writing -- review \& editing (equal).
\textbf{Jos\'{e}~R. {Crespo~L\'{o}pez-Urrutia}:} Conceptualization (lead); Funding acquisition (lead); Project administration (lead); Supervision (lead); Writing -- original draft (supporting); Writing -- review \& editing (equal).

\section*{Data availability}

The data that support the findings of this study are available from the corresponding author upon reasonable request.

\bibliography{firsthcis}

\begin{thebibliography}{82}%
\makeatletter
\providecommand \@ifxundefined [1]{%
 \@ifx{#1\undefined}
}%
\providecommand \@ifnum [1]{%
 \ifnum #1\expandafter \@firstoftwo
 \else \expandafter \@secondoftwo
 \fi
}%
\providecommand \@ifx [1]{%
 \ifx #1\expandafter \@firstoftwo
 \else \expandafter \@secondoftwo
 \fi
}%
\providecommand \natexlab [1]{#1}%
\providecommand \enquote  [1]{``#1''}%
\providecommand \bibnamefont  [1]{#1}%
\providecommand \bibfnamefont [1]{#1}%
\providecommand \citenamefont [1]{#1}%
\providecommand \href@noop [0]{\@secondoftwo}%
\providecommand \href [0]{\begingroup \@sanitize@url \@href}%
\providecommand \@href[1]{\@@startlink{#1}\@@href}%
\providecommand \@@href[1]{\endgroup#1\@@endlink}%
\providecommand \@sanitize@url [0]{\catcode `\\12\catcode `\$12\catcode
  `\&12\catcode `\#12\catcode `\^12\catcode `\_12\catcode `\%12\relax}%
\providecommand \@@startlink[1]{}%
\providecommand \@@endlink[0]{}%
\providecommand \url  [0]{\begingroup\@sanitize@url \@url }%
\providecommand \@url [1]{\endgroup\@href {#1}{\urlprefix }}%
\providecommand \urlprefix  [0]{URL }%
\providecommand \Eprint [0]{\href }%
\providecommand \doibase [0]{http://dx.doi.org/}%
\providecommand \selectlanguage [0]{\@gobble}%
\providecommand \bibinfo  [0]{\@secondoftwo}%
\providecommand \bibfield  [0]{\@secondoftwo}%
\providecommand \translation [1]{[#1]}%
\providecommand \BibitemOpen [0]{}%
\providecommand \bibitemStop [0]{}%
\providecommand \bibitemNoStop [0]{.\EOS\space}%
\providecommand \EOS [0]{\spacefactor3000\relax}%
\providecommand \BibitemShut  [1]{\csname bibitem#1\endcsname}%
\let\auto@bib@innerbib\@empty
\bibitem [{\citenamefont {Kozlov}\ \emph {et~al.}(2018)\citenamefont {Kozlov},
  \citenamefont {Safronova}, \citenamefont {Crespo L{\'o}pez-Urrutia},\ and\
  \citenamefont {Schmidt}}]{Kozlov2018}%
  \BibitemOpen
  \bibfield  {author} {\bibinfo {author} {\bibfnamefont {M.~G.}\ \bibnamefont
  {Kozlov}}, \bibinfo {author} {\bibfnamefont {M.~S.}\ \bibnamefont
  {Safronova}}, \bibinfo {author} {\bibfnamefont {J.~R.}\ \bibnamefont {Crespo
  L{\'o}pez-Urrutia}}, \ and\ \bibinfo {author} {\bibfnamefont {P.~O.}\
  \bibnamefont {Schmidt}},\ }\bibfield  {title} {\enquote {\bibinfo {title}
  {Highly charged ions: {O}ptical clocks and applications in fundamental
  physics},}\ }\href {\doibase 10.1103/RevModPhys.90.045005} {\bibfield
  {journal} {\bibinfo  {journal} {Rev. Mod. Phys.}\ }\textbf {\bibinfo {volume}
  {90}},\ \bibinfo {pages} {045005} (\bibinfo {year} {2018})}\BibitemShut
  {NoStop}%
\bibitem [{\citenamefont {Rehbehn}\ \emph {et~al.}(2021)\citenamefont
  {Rehbehn}, \citenamefont {Rosner}, \citenamefont {Bekker}, \citenamefont
  {Berengut}, \citenamefont {Schmidt}, \citenamefont {King}, \citenamefont
  {Micke}, \citenamefont {Gu}, \citenamefont {M{\"u}ller}, \citenamefont
  {Surzhykov},\ and\ \citenamefont {Crespo L{\'o}pez-Urrutia}}]{Rehbehn2021}%
  \BibitemOpen
  \bibfield  {author} {\bibinfo {author} {\bibfnamefont {N.-H.}\ \bibnamefont
  {Rehbehn}}, \bibinfo {author} {\bibfnamefont {M.~K.}\ \bibnamefont {Rosner}},
  \bibinfo {author} {\bibfnamefont {H.}~\bibnamefont {Bekker}}, \bibinfo
  {author} {\bibfnamefont {J.~C.}\ \bibnamefont {Berengut}}, \bibinfo {author}
  {\bibfnamefont {P.~O.}\ \bibnamefont {Schmidt}}, \bibinfo {author}
  {\bibfnamefont {S.~A.}\ \bibnamefont {King}}, \bibinfo {author}
  {\bibfnamefont {P.}~\bibnamefont {Micke}}, \bibinfo {author} {\bibfnamefont
  {M.~F.}\ \bibnamefont {Gu}}, \bibinfo {author} {\bibfnamefont
  {R.}~\bibnamefont {M{\"u}ller}}, \bibinfo {author} {\bibfnamefont
  {A.}~\bibnamefont {Surzhykov}}, \ and\ \bibinfo {author} {\bibfnamefont
  {J.~R.}\ \bibnamefont {Crespo L{\'o}pez-Urrutia}},\ }\bibfield  {title}
  {\enquote {\bibinfo {title} {Sensitivity to new physics of isotope-shift
  studies using the coronal lines of highly charged calcium ions},}\ }\href
  {\doibase 10.1103/PhysRevA.103.L040801} {\bibfield  {journal} {\bibinfo
  {journal} {Phys. Rev. A}\ }\textbf {\bibinfo {volume} {103}},\ \bibinfo
  {pages} {L040801} (\bibinfo {year} {2021})}\BibitemShut {NoStop}%
\bibitem [{\citenamefont {Liang}\ \emph {et~al.}(2021)\citenamefont {Liang},
  \citenamefont {Zhang}, \citenamefont {Guan}, \citenamefont {Lu},
  \citenamefont {Xiao}, \citenamefont {Chen}, \citenamefont {Huang},
  \citenamefont {Zhang}, \citenamefont {Li}, \citenamefont {Zou}, \citenamefont
  {Li}, \citenamefont {Yan}, \citenamefont {Derevianko}, \citenamefont {Zhan},
  \citenamefont {Shi},\ and\ \citenamefont {Gao}}]{Liang2021}%
  \BibitemOpen
  \bibfield  {author} {\bibinfo {author} {\bibfnamefont {S.-Y.}\ \bibnamefont
  {Liang}}, \bibinfo {author} {\bibfnamefont {T.-X.}\ \bibnamefont {Zhang}},
  \bibinfo {author} {\bibfnamefont {H.}~\bibnamefont {Guan}}, \bibinfo {author}
  {\bibfnamefont {Q.-F.}\ \bibnamefont {Lu}}, \bibinfo {author} {\bibfnamefont
  {J.}~\bibnamefont {Xiao}}, \bibinfo {author} {\bibfnamefont {S.-L.}\
  \bibnamefont {Chen}}, \bibinfo {author} {\bibfnamefont {Y.}~\bibnamefont
  {Huang}}, \bibinfo {author} {\bibfnamefont {Y.-H.}\ \bibnamefont {Zhang}},
  \bibinfo {author} {\bibfnamefont {C.-B.}\ \bibnamefont {Li}}, \bibinfo
  {author} {\bibfnamefont {Y.-M.}\ \bibnamefont {Zou}}, \bibinfo {author}
  {\bibfnamefont {J.-G.}\ \bibnamefont {Li}}, \bibinfo {author} {\bibfnamefont
  {Z.-C.}\ \bibnamefont {Yan}}, \bibinfo {author} {\bibfnamefont
  {A.}~\bibnamefont {Derevianko}}, \bibinfo {author} {\bibfnamefont {M.-S.}\
  \bibnamefont {Zhan}}, \bibinfo {author} {\bibfnamefont {T.-Y.}\ \bibnamefont
  {Shi}}, \ and\ \bibinfo {author} {\bibfnamefont {K.-L.}\ \bibnamefont
  {Gao}},\ }\bibfield  {title} {\enquote {\bibinfo {title} {Probing multiple
  electric-dipole-forbidden optical transitions in highly charged nickel
  ions},}\ }\href {\doibase 10.1103/PhysRevA.103.022804} {\bibfield  {journal}
  {\bibinfo  {journal} {Phys. Rev. A}\ }\textbf {\bibinfo {volume} {103}},\
  \bibinfo {pages} {022804} (\bibinfo {year} {2021})}\BibitemShut {NoStop}%
\bibitem [{\citenamefont {Berengut}, \citenamefont {Dzuba},\ and\ \citenamefont
  {Flambaum}(2010)}]{Berengut2010}%
  \BibitemOpen
  \bibfield  {author} {\bibinfo {author} {\bibfnamefont {J.~C.}\ \bibnamefont
  {Berengut}}, \bibinfo {author} {\bibfnamefont {V.~A.}\ \bibnamefont {Dzuba}},
  \ and\ \bibinfo {author} {\bibfnamefont {V.~V.}\ \bibnamefont {Flambaum}},\
  }\bibfield  {title} {\enquote {\bibinfo {title} {Enhanced laboratory
  sensitivity to variation of the fine-structure constant using highly charged
  ions},}\ }\href {\doibase 10.1103/PhysRevLett.105.120801} {\bibfield
  {journal} {\bibinfo  {journal} {Phys. Rev. Lett.}\ }\textbf {\bibinfo
  {volume} {105}},\ \bibinfo {pages} {120801} (\bibinfo {year}
  {2010})}\BibitemShut {NoStop}%
\bibitem [{\citenamefont {Berengut}\ \emph {et~al.}(2012)\citenamefont
  {Berengut}, \citenamefont {Dzuba}, \citenamefont {Flambaum},\ and\
  \citenamefont {Ong}}]{Berengut2012}%
  \BibitemOpen
  \bibfield  {author} {\bibinfo {author} {\bibfnamefont {J.~C.}\ \bibnamefont
  {Berengut}}, \bibinfo {author} {\bibfnamefont {V.~A.}\ \bibnamefont {Dzuba}},
  \bibinfo {author} {\bibfnamefont {V.~V.}\ \bibnamefont {Flambaum}}, \ and\
  \bibinfo {author} {\bibfnamefont {A.}~\bibnamefont {Ong}},\ }\bibfield
  {title} {\enquote {\bibinfo {title} {Highly charged ions with {E1}, {M1}, and
  {E2} transitions within laser range},}\ }\href {\doibase
  10.1103/PhysRevA.86.022517} {\bibfield  {journal} {\bibinfo  {journal} {Phys.
  Rev. A}\ }\textbf {\bibinfo {volume} {86}},\ \bibinfo {pages} {022517}
  (\bibinfo {year} {2012})}\BibitemShut {NoStop}%
\bibitem [{\citenamefont {Bekker}\ \emph {et~al.}(2019)\citenamefont {Bekker},
  \citenamefont {Borschevsky}, \citenamefont {Harman}, \citenamefont {Keitel},
  \citenamefont {Pfeifer}, \citenamefont {Schmidt}, \citenamefont {Crespo
  L{\'o}pez-Urrutia},\ and\ \citenamefont {Berengut}}]{Bekker2019}%
  \BibitemOpen
  \bibfield  {author} {\bibinfo {author} {\bibfnamefont {H.}~\bibnamefont
  {Bekker}}, \bibinfo {author} {\bibfnamefont {A.}~\bibnamefont {Borschevsky}},
  \bibinfo {author} {\bibfnamefont {Z.}~\bibnamefont {Harman}}, \bibinfo
  {author} {\bibfnamefont {C.~H.}\ \bibnamefont {Keitel}}, \bibinfo {author}
  {\bibfnamefont {T.}~\bibnamefont {Pfeifer}}, \bibinfo {author} {\bibfnamefont
  {P.~O.}\ \bibnamefont {Schmidt}}, \bibinfo {author} {\bibfnamefont {J.~R.}\
  \bibnamefont {Crespo L{\'o}pez-Urrutia}}, \ and\ \bibinfo {author}
  {\bibfnamefont {J.~C.}\ \bibnamefont {Berengut}},\ }\bibfield  {title}
  {\enquote {\bibinfo {title} {Detection of the 5p~--~4f orbital crossing and
  its optical clock transition in $\text{Pr}^\text{9+}$},}\ }\href {\doibase
  10.1038/s41467-019-13406-9} {\bibfield  {journal} {\bibinfo  {journal} {Nat.
  Commun.}\ }\textbf {\bibinfo {volume} {10}} (\bibinfo {year} {2019}),\
  10.1038/s41467-019-13406-9}\BibitemShut {NoStop}%
\bibitem [{\citenamefont {Porsev}\ \emph {et~al.}(2020)\citenamefont {Porsev},
  \citenamefont {Safronova}, \citenamefont {Safronova}, \citenamefont
  {Schmidt}, \citenamefont {Bondarev}, \citenamefont {Kozlov}, \citenamefont
  {Tupitsyn},\ and\ \citenamefont {Cheung}}]{Porsev2020}%
  \BibitemOpen
  \bibfield  {author} {\bibinfo {author} {\bibfnamefont {S.~G.}\ \bibnamefont
  {Porsev}}, \bibinfo {author} {\bibfnamefont {U.~I.}\ \bibnamefont
  {Safronova}}, \bibinfo {author} {\bibfnamefont {M.~S.}\ \bibnamefont
  {Safronova}}, \bibinfo {author} {\bibfnamefont {P.~O.}\ \bibnamefont
  {Schmidt}}, \bibinfo {author} {\bibfnamefont {A.~I.}\ \bibnamefont
  {Bondarev}}, \bibinfo {author} {\bibfnamefont {M.~G.}\ \bibnamefont
  {Kozlov}}, \bibinfo {author} {\bibfnamefont {I.~I.}\ \bibnamefont
  {Tupitsyn}}, \ and\ \bibinfo {author} {\bibfnamefont {C.}~\bibnamefont
  {Cheung}},\ }\bibfield  {title} {\enquote {\bibinfo {title} {Optical clocks
  based on the {C}f$^{15+}$ and {C}f$^{17+}$ ions},}\ }\href {\doibase
  10.1103/PhysRevA.102.012802} {\bibfield  {journal} {\bibinfo  {journal}
  {Phys. Rev. A}\ }\textbf {\bibinfo {volume} {102}},\ \bibinfo {pages}
  {012802} (\bibinfo {year} {2020})}\BibitemShut {NoStop}%
\bibitem [{\citenamefont {Shabaev}\ \emph {et~al.}(1997)\citenamefont
  {Shabaev}, \citenamefont {Tomaselli}, \citenamefont {K\"uhl}, \citenamefont
  {Artemyev},\ and\ \citenamefont {Yerokhin}}]{Shabaev1997}%
  \BibitemOpen
  \bibfield  {author} {\bibinfo {author} {\bibfnamefont {V.~M.}\ \bibnamefont
  {Shabaev}}, \bibinfo {author} {\bibfnamefont {M.}~\bibnamefont {Tomaselli}},
  \bibinfo {author} {\bibfnamefont {T.}~\bibnamefont {K\"uhl}}, \bibinfo
  {author} {\bibfnamefont {A.~N.}\ \bibnamefont {Artemyev}}, \ and\ \bibinfo
  {author} {\bibfnamefont {V.~A.}\ \bibnamefont {Yerokhin}},\ }\bibfield
  {title} {\enquote {\bibinfo {title} {Ground-state hyperfine splitting of
  high-${Z}$ hydrogenlike ions},}\ }\href {\doibase 10.1103/PhysRevA.56.252}
  {\bibfield  {journal} {\bibinfo  {journal} {Phys. Rev. A}\ }\textbf {\bibinfo
  {volume} {56}},\ \bibinfo {pages} {252--255} (\bibinfo {year}
  {1997})}\BibitemShut {NoStop}%
\bibitem [{\citenamefont {Klaft}\ \emph {et~al.}(1994)\citenamefont {Klaft},
  \citenamefont {Borneis}, \citenamefont {Engel}, \citenamefont {Fricke},
  \citenamefont {Grieser}, \citenamefont {Huber}, \citenamefont {K\"uhl},
  \citenamefont {Marx}, \citenamefont {Neumann}, \citenamefont {Schr\"oder},
  \citenamefont {Seelig},\ and\ \citenamefont {V\"olker}}]{Klaft1994}%
  \BibitemOpen
  \bibfield  {author} {\bibinfo {author} {\bibfnamefont {I.}~\bibnamefont
  {Klaft}}, \bibinfo {author} {\bibfnamefont {S.}~\bibnamefont {Borneis}},
  \bibinfo {author} {\bibfnamefont {T.}~\bibnamefont {Engel}}, \bibinfo
  {author} {\bibfnamefont {B.}~\bibnamefont {Fricke}}, \bibinfo {author}
  {\bibfnamefont {R.}~\bibnamefont {Grieser}}, \bibinfo {author} {\bibfnamefont
  {G.}~\bibnamefont {Huber}}, \bibinfo {author} {\bibfnamefont
  {T.}~\bibnamefont {K\"uhl}}, \bibinfo {author} {\bibfnamefont
  {D.}~\bibnamefont {Marx}}, \bibinfo {author} {\bibfnamefont {R.}~\bibnamefont
  {Neumann}}, \bibinfo {author} {\bibfnamefont {S.}~\bibnamefont {Schr\"oder}},
  \bibinfo {author} {\bibfnamefont {P.}~\bibnamefont {Seelig}}, \ and\ \bibinfo
  {author} {\bibfnamefont {L.}~\bibnamefont {V\"olker}},\ }\bibfield  {title}
  {\enquote {\bibinfo {title} {Precision laser spectroscopy of the ground state
  hyperfine splitting of hydrogenlike $^{209}\mathrm{Bi}^{82+}$},}\ }\href
  {\doibase 10.1103/PhysRevLett.73.2425} {\bibfield  {journal} {\bibinfo
  {journal} {Phys. Rev. Lett.}\ }\textbf {\bibinfo {volume} {73}},\ \bibinfo
  {pages} {2425--2427} (\bibinfo {year} {1994})}\BibitemShut {NoStop}%
\bibitem [{\citenamefont {Crespo L\'opez-Urrutia}\ \emph
  {et~al.}(1998)\citenamefont {Crespo L\'opez-Urrutia}, \citenamefont
  {Beiersdorfer}, \citenamefont {Widmann}, \citenamefont {Birkett},
  \citenamefont {M\aa{}rtensson-Pendrill},\ and\ \citenamefont
  {Gustavsson}}]{Crespo1998}%
  \BibitemOpen
  \bibfield  {author} {\bibinfo {author} {\bibfnamefont {J.~R.}\ \bibnamefont
  {Crespo L\'opez-Urrutia}}, \bibinfo {author} {\bibfnamefont {P.}~\bibnamefont
  {Beiersdorfer}}, \bibinfo {author} {\bibfnamefont {K.}~\bibnamefont
  {Widmann}}, \bibinfo {author} {\bibfnamefont {B.~B.}\ \bibnamefont
  {Birkett}}, \bibinfo {author} {\bibfnamefont {A.-M.}\ \bibnamefont
  {M\aa{}rtensson-Pendrill}}, \ and\ \bibinfo {author} {\bibfnamefont
  {M.~G.~H.}\ \bibnamefont {Gustavsson}},\ }\bibfield  {title} {\enquote
  {\bibinfo {title} {Nuclear magnetization distribution radii determined by
  hyperfine transitions in the $1s$ level of {H}-like ions
  ${}^{185}${R}e$^{74+}$ and ${}^{187}${R}e$^{74+}$},}\ }\href {\doibase
  10.1103/PhysRevA.57.879} {\bibfield  {journal} {\bibinfo  {journal} {Phys.
  Rev. A}\ }\textbf {\bibinfo {volume} {57}},\ \bibinfo {pages} {879--887}
  (\bibinfo {year} {1998})}\BibitemShut {NoStop}%
\bibitem [{\citenamefont {Beiersdorfer}\ \emph {et~al.}(2001)\citenamefont
  {Beiersdorfer}, \citenamefont {Utter}, \citenamefont {Wong}, \citenamefont
  {Crespo L\'opez-Urrutia}, \citenamefont {Britten}, \citenamefont {Chen},
  \citenamefont {Harris}, \citenamefont {Thoe}, \citenamefont {Thorn},
  \citenamefont {Tr\"abert}, \citenamefont {Gustavsson}, \citenamefont
  {Forss\'en},\ and\ \citenamefont
  {M\aa{}rtensson-Pendrill}}]{Beiersdorfer2001}%
  \BibitemOpen
  \bibfield  {author} {\bibinfo {author} {\bibfnamefont {P.}~\bibnamefont
  {Beiersdorfer}}, \bibinfo {author} {\bibfnamefont {S.~B.}\ \bibnamefont
  {Utter}}, \bibinfo {author} {\bibfnamefont {K.~L.}\ \bibnamefont {Wong}},
  \bibinfo {author} {\bibfnamefont {J.~R.}\ \bibnamefont {Crespo
  L\'opez-Urrutia}}, \bibinfo {author} {\bibfnamefont {J.~A.}\ \bibnamefont
  {Britten}}, \bibinfo {author} {\bibfnamefont {H.}~\bibnamefont {Chen}},
  \bibinfo {author} {\bibfnamefont {C.~L.}\ \bibnamefont {Harris}}, \bibinfo
  {author} {\bibfnamefont {R.~S.}\ \bibnamefont {Thoe}}, \bibinfo {author}
  {\bibfnamefont {D.~B.}\ \bibnamefont {Thorn}}, \bibinfo {author}
  {\bibfnamefont {E.}~\bibnamefont {Tr\"abert}}, \bibinfo {author}
  {\bibfnamefont {M.~G.~H.}\ \bibnamefont {Gustavsson}}, \bibinfo {author}
  {\bibfnamefont {C.}~\bibnamefont {Forss\'en}}, \ and\ \bibinfo {author}
  {\bibfnamefont {A.-M.}\ \bibnamefont {M\aa{}rtensson-Pendrill}},\ }\bibfield
  {title} {\enquote {\bibinfo {title} {Hyperfine structure of hydrogenlike
  thallium isotopes},}\ }\href {\doibase 10.1103/PhysRevA.64.032506} {\bibfield
   {journal} {\bibinfo  {journal} {Phys. Rev. A}\ }\textbf {\bibinfo {volume}
  {64}},\ \bibinfo {pages} {032506} (\bibinfo {year} {2001})}\BibitemShut
  {NoStop}%
\bibitem [{\citenamefont {Schiller}(2007)}]{Schiller2007}%
  \BibitemOpen
  \bibfield  {author} {\bibinfo {author} {\bibfnamefont {S.}~\bibnamefont
  {Schiller}},\ }\bibfield  {title} {\enquote {\bibinfo {title} {Hydrogenlike
  highly charged ions for tests of the time independence of fundamental
  constants},}\ }\href {\doibase 10.1103/PhysRevLett.98.180801} {\bibfield
  {journal} {\bibinfo  {journal} {Phys. Rev. Lett.}\ }\textbf {\bibinfo
  {volume} {98}},\ \bibinfo {pages} {180801} (\bibinfo {year}
  {2007})}\BibitemShut {NoStop}%
\bibitem [{\citenamefont {Oreshkina}\ \emph {et~al.}(2017)\citenamefont
  {Oreshkina}, \citenamefont {Cavaletto}, \citenamefont {Michel}, \citenamefont
  {Harman},\ and\ \citenamefont {Keitel}}]{Oreshkina2017}%
  \BibitemOpen
  \bibfield  {author} {\bibinfo {author} {\bibfnamefont {N.~S.}\ \bibnamefont
  {Oreshkina}}, \bibinfo {author} {\bibfnamefont {S.~M.}\ \bibnamefont
  {Cavaletto}}, \bibinfo {author} {\bibfnamefont {N.}~\bibnamefont {Michel}},
  \bibinfo {author} {\bibfnamefont {Z.}~\bibnamefont {Harman}}, \ and\ \bibinfo
  {author} {\bibfnamefont {C.~H.}\ \bibnamefont {Keitel}},\ }\bibfield  {title}
  {\enquote {\bibinfo {title} {Hyperfine splitting in simple ions for the
  search of the variation of fundamental constants},}\ }\href {\doibase
  10.1103/PhysRevA.96.030501} {\bibfield  {journal} {\bibinfo  {journal} {Phys.
  Rev. A}\ }\textbf {\bibinfo {volume} {96}},\ \bibinfo {pages} {030501}
  (\bibinfo {year} {2017})}\BibitemShut {NoStop}%
\bibitem [{\citenamefont {Skripnikov}\ \emph {et~al.}(2018)\citenamefont
  {Skripnikov}, \citenamefont {Schmidt}, \citenamefont {Ullmann}, \citenamefont
  {Geppert}, \citenamefont {Kraus}, \citenamefont {Kresse}, \citenamefont
  {N\"ortersh\"auser}, \citenamefont {Privalov}, \citenamefont {Scheibe},
  \citenamefont {Shabaev}, \citenamefont {Vogel},\ and\ \citenamefont
  {Volotka}}]{Skripnikov2018}%
  \BibitemOpen
  \bibfield  {author} {\bibinfo {author} {\bibfnamefont {L.~V.}\ \bibnamefont
  {Skripnikov}}, \bibinfo {author} {\bibfnamefont {S.}~\bibnamefont {Schmidt}},
  \bibinfo {author} {\bibfnamefont {J.}~\bibnamefont {Ullmann}}, \bibinfo
  {author} {\bibfnamefont {C.}~\bibnamefont {Geppert}}, \bibinfo {author}
  {\bibfnamefont {F.}~\bibnamefont {Kraus}}, \bibinfo {author} {\bibfnamefont
  {B.}~\bibnamefont {Kresse}}, \bibinfo {author} {\bibfnamefont
  {W.}~\bibnamefont {N\"ortersh\"auser}}, \bibinfo {author} {\bibfnamefont
  {A.~F.}\ \bibnamefont {Privalov}}, \bibinfo {author} {\bibfnamefont
  {B.}~\bibnamefont {Scheibe}}, \bibinfo {author} {\bibfnamefont {V.~M.}\
  \bibnamefont {Shabaev}}, \bibinfo {author} {\bibfnamefont {M.}~\bibnamefont
  {Vogel}}, \ and\ \bibinfo {author} {\bibfnamefont {A.~V.}\ \bibnamefont
  {Volotka}},\ }\bibfield  {title} {\enquote {\bibinfo {title} {New nuclear
  magnetic moment of $^{209}\mathrm{Bi}$: Resolving the bismuth hyperfine
  puzzle},}\ }\href {\doibase 10.1103/PhysRevLett.120.093001} {\bibfield
  {journal} {\bibinfo  {journal} {Phys. Rev. Lett.}\ }\textbf {\bibinfo
  {volume} {120}},\ \bibinfo {pages} {093001} (\bibinfo {year}
  {2018})}\BibitemShut {NoStop}%
\bibitem [{\citenamefont {N{\"o}rtersh{\"a}user}\ \emph
  {et~al.}(2019)\citenamefont {N{\"o}rtersh{\"a}user}, \citenamefont {Ullmann},
  \citenamefont {Skripnikov}, \citenamefont {Andelkovic}, \citenamefont
  {Brandau}, \citenamefont {Dax}, \citenamefont {Geithner}, \citenamefont
  {Geppert}, \citenamefont {Gorges}, \citenamefont {Hammen}, \citenamefont
  {Hannen}, \citenamefont {Kaufmann}, \citenamefont {K{\"o}nig}, \citenamefont
  {Kraus}, \citenamefont {Kresse}, \citenamefont {Litvinov}, \citenamefont
  {Lochmann}, \citenamefont {Maa{\ss}}, \citenamefont {Meisner}, \citenamefont
  {Murb{\"o}ck}, \citenamefont {Privalov}, \citenamefont {S{\'a}nchez},
  \citenamefont {Scheibe}, \citenamefont {Schmidt}, \citenamefont {Schmidt},
  \citenamefont {Shabaev}, \citenamefont {Steck}, \citenamefont {St{\"o}hlker},
  \citenamefont {Thompson}, \citenamefont {Trageser}, \citenamefont {Vogel},
  \citenamefont {Vollbrecht}, \citenamefont {Volotka},\ and\ \citenamefont
  {Weinheimer}}]{Noertershaeuser2019}%
  \BibitemOpen
  \bibfield  {author} {\bibinfo {author} {\bibfnamefont {W.}~\bibnamefont
  {N{\"o}rtersh{\"a}user}}, \bibinfo {author} {\bibfnamefont {J.}~\bibnamefont
  {Ullmann}}, \bibinfo {author} {\bibfnamefont {L.~V.}\ \bibnamefont
  {Skripnikov}}, \bibinfo {author} {\bibfnamefont {Z.}~\bibnamefont
  {Andelkovic}}, \bibinfo {author} {\bibfnamefont {C.}~\bibnamefont {Brandau}},
  \bibinfo {author} {\bibfnamefont {A.}~\bibnamefont {Dax}}, \bibinfo {author}
  {\bibfnamefont {W.}~\bibnamefont {Geithner}}, \bibinfo {author}
  {\bibfnamefont {C.}~\bibnamefont {Geppert}}, \bibinfo {author} {\bibfnamefont
  {C.}~\bibnamefont {Gorges}}, \bibinfo {author} {\bibfnamefont
  {M.}~\bibnamefont {Hammen}}, \bibinfo {author} {\bibfnamefont
  {V.}~\bibnamefont {Hannen}}, \bibinfo {author} {\bibfnamefont
  {S.}~\bibnamefont {Kaufmann}}, \bibinfo {author} {\bibfnamefont
  {K.}~\bibnamefont {K{\"o}nig}}, \bibinfo {author} {\bibfnamefont
  {F.}~\bibnamefont {Kraus}}, \bibinfo {author} {\bibfnamefont
  {B.}~\bibnamefont {Kresse}}, \bibinfo {author} {\bibfnamefont {Y.~A.}\
  \bibnamefont {Litvinov}}, \bibinfo {author} {\bibfnamefont {M.}~\bibnamefont
  {Lochmann}}, \bibinfo {author} {\bibfnamefont {B.}~\bibnamefont {Maa{\ss}}},
  \bibinfo {author} {\bibfnamefont {J.}~\bibnamefont {Meisner}}, \bibinfo
  {author} {\bibfnamefont {T.}~\bibnamefont {Murb{\"o}ck}}, \bibinfo {author}
  {\bibfnamefont {A.~F.}\ \bibnamefont {Privalov}}, \bibinfo {author}
  {\bibfnamefont {R.}~\bibnamefont {S{\'a}nchez}}, \bibinfo {author}
  {\bibfnamefont {B.}~\bibnamefont {Scheibe}}, \bibinfo {author} {\bibfnamefont
  {M.}~\bibnamefont {Schmidt}}, \bibinfo {author} {\bibfnamefont
  {S.}~\bibnamefont {Schmidt}}, \bibinfo {author} {\bibfnamefont {V.~M.}\
  \bibnamefont {Shabaev}}, \bibinfo {author} {\bibfnamefont {M.}~\bibnamefont
  {Steck}}, \bibinfo {author} {\bibfnamefont {T.}~\bibnamefont {St{\"o}hlker}},
  \bibinfo {author} {\bibfnamefont {R.~C.}\ \bibnamefont {Thompson}}, \bibinfo
  {author} {\bibfnamefont {C.}~\bibnamefont {Trageser}}, \bibinfo {author}
  {\bibfnamefont {M.}~\bibnamefont {Vogel}}, \bibinfo {author} {\bibfnamefont
  {J.}~\bibnamefont {Vollbrecht}}, \bibinfo {author} {\bibfnamefont {A.~V.}\
  \bibnamefont {Volotka}}, \ and\ \bibinfo {author} {\bibfnamefont
  {C.}~\bibnamefont {Weinheimer}},\ }\bibfield  {title} {\enquote {\bibinfo
  {title} {The hyperfine puzzle of strong-field bound-state {QED}},}\ }\href
  {\doibase 10.1007/s10751-019-1569-8} {\bibfield  {journal} {\bibinfo
  {journal} {Hyperfine Interact.}\ }\textbf {\bibinfo {volume} {240}},\
  \bibinfo {pages} {51} (\bibinfo {year} {2019})}\BibitemShut {NoStop}%
\bibitem [{\citenamefont {Yudin}, \citenamefont {Taichenachev},\ and\
  \citenamefont {Derevianko}(2014)}]{Yudin2014}%
  \BibitemOpen
  \bibfield  {author} {\bibinfo {author} {\bibfnamefont {V.~I.}\ \bibnamefont
  {Yudin}}, \bibinfo {author} {\bibfnamefont {A.~V.}\ \bibnamefont
  {Taichenachev}}, \ and\ \bibinfo {author} {\bibfnamefont {A.}~\bibnamefont
  {Derevianko}},\ }\bibfield  {title} {\enquote {\bibinfo {title}
  {Magnetic-dipole transitions in highly charged ions as a basis of
  ultraprecise optical clocks},}\ }\href {\doibase
  10.1103/PhysRevLett.113.233003} {\bibfield  {journal} {\bibinfo  {journal}
  {Phys. Rev. Lett.}\ }\textbf {\bibinfo {volume} {113}},\ \bibinfo {pages}
  {233003} (\bibinfo {year} {2014})}\BibitemShut {NoStop}%
\bibitem [{\citenamefont {Schmidt}\ \emph {et~al.}(2005)\citenamefont
  {Schmidt}, \citenamefont {Rosenband}, \citenamefont {Langer}, \citenamefont
  {Itano}, \citenamefont {Bergquist},\ and\ \citenamefont
  {Wineland}}]{Schmidt2005}%
  \BibitemOpen
  \bibfield  {author} {\bibinfo {author} {\bibfnamefont {P.~O.}\ \bibnamefont
  {Schmidt}}, \bibinfo {author} {\bibfnamefont {T.}~\bibnamefont {Rosenband}},
  \bibinfo {author} {\bibfnamefont {C.}~\bibnamefont {Langer}}, \bibinfo
  {author} {\bibfnamefont {W.~M.}\ \bibnamefont {Itano}}, \bibinfo {author}
  {\bibfnamefont {J.~C.}\ \bibnamefont {Bergquist}}, \ and\ \bibinfo {author}
  {\bibfnamefont {D.~J.}\ \bibnamefont {Wineland}},\ }\bibfield  {title}
  {\enquote {\bibinfo {title} {Spectroscopy using quantum logic},}\ }\href
  {\doibase 10.1126/science.1114375} {\bibfield  {journal} {\bibinfo  {journal}
  {Science}\ }\textbf {\bibinfo {volume} {309}},\ \bibinfo {pages} {749--752}
  (\bibinfo {year} {2005})}\BibitemShut {NoStop}%
\bibitem [{\citenamefont {Schm{\"o}ger}\ \emph
  {et~al.}(2015{\natexlab{a}})\citenamefont {Schm{\"o}ger}, \citenamefont
  {Versolato}, \citenamefont {Schwarz}, \citenamefont {Kohnen}, \citenamefont
  {Windberger}, \citenamefont {Piest}, \citenamefont {Feuchtenbeiner},
  \citenamefont {Pedregosa-Gutierrez}, \citenamefont {Leopold}, \citenamefont
  {Micke}, \citenamefont {Hansen}, \citenamefont {Baumann}, \citenamefont
  {Drewsen}, \citenamefont {Ullrich}, \citenamefont {Schmidt},\ and\
  \citenamefont {Crespo L{\'o}pez-Urrutia}}]{Schmoeger2015}%
  \BibitemOpen
  \bibfield  {author} {\bibinfo {author} {\bibfnamefont {L.}~\bibnamefont
  {Schm{\"o}ger}}, \bibinfo {author} {\bibfnamefont {O.~O.}\ \bibnamefont
  {Versolato}}, \bibinfo {author} {\bibfnamefont {M.}~\bibnamefont {Schwarz}},
  \bibinfo {author} {\bibfnamefont {M.}~\bibnamefont {Kohnen}}, \bibinfo
  {author} {\bibfnamefont {A.}~\bibnamefont {Windberger}}, \bibinfo {author}
  {\bibfnamefont {B.}~\bibnamefont {Piest}}, \bibinfo {author} {\bibfnamefont
  {S.}~\bibnamefont {Feuchtenbeiner}}, \bibinfo {author} {\bibfnamefont
  {J.}~\bibnamefont {Pedregosa-Gutierrez}}, \bibinfo {author} {\bibfnamefont
  {T.}~\bibnamefont {Leopold}}, \bibinfo {author} {\bibfnamefont
  {P.}~\bibnamefont {Micke}}, \bibinfo {author} {\bibfnamefont {A.~K.}\
  \bibnamefont {Hansen}}, \bibinfo {author} {\bibfnamefont {T.~M.}\
  \bibnamefont {Baumann}}, \bibinfo {author} {\bibfnamefont {M.}~\bibnamefont
  {Drewsen}}, \bibinfo {author} {\bibfnamefont {J.}~\bibnamefont {Ullrich}},
  \bibinfo {author} {\bibfnamefont {P.~O.}\ \bibnamefont {Schmidt}}, \ and\
  \bibinfo {author} {\bibfnamefont {J.~R.}\ \bibnamefont {Crespo
  L{\'o}pez-Urrutia}},\ }\bibfield  {title} {\enquote {\bibinfo {title}
  {{C}oulomb crystallization of highly charged ions},}\ }\href {\doibase
  10.1126/science.aaa2960} {\bibfield  {journal} {\bibinfo  {journal}
  {Science}\ }\textbf {\bibinfo {volume} {347}},\ \bibinfo {pages} {1233--1236}
  (\bibinfo {year} {2015}{\natexlab{a}})}\BibitemShut {NoStop}%
\bibitem [{\citenamefont {Schm{\"o}ger}\ \emph
  {et~al.}(2015{\natexlab{b}})\citenamefont {Schm{\"o}ger}, \citenamefont
  {Schwarz}, \citenamefont {Baumann}, \citenamefont {Versolato}, \citenamefont
  {Piest}, \citenamefont {Pfeifer}, \citenamefont {Ullrich}, \citenamefont
  {Schmidt},\ and\ \citenamefont {Crespo L{\'o}pez-Urrutia}}]{Schmoeger2015a}%
  \BibitemOpen
  \bibfield  {author} {\bibinfo {author} {\bibfnamefont {L.}~\bibnamefont
  {Schm{\"o}ger}}, \bibinfo {author} {\bibfnamefont {M.}~\bibnamefont
  {Schwarz}}, \bibinfo {author} {\bibfnamefont {T.~M.}\ \bibnamefont
  {Baumann}}, \bibinfo {author} {\bibfnamefont {O.~O.}\ \bibnamefont
  {Versolato}}, \bibinfo {author} {\bibfnamefont {B.}~\bibnamefont {Piest}},
  \bibinfo {author} {\bibfnamefont {T.}~\bibnamefont {Pfeifer}}, \bibinfo
  {author} {\bibfnamefont {J.}~\bibnamefont {Ullrich}}, \bibinfo {author}
  {\bibfnamefont {P.~O.}\ \bibnamefont {Schmidt}}, \ and\ \bibinfo {author}
  {\bibfnamefont {J.~R.}\ \bibnamefont {Crespo L{\'o}pez-Urrutia}},\ }\bibfield
   {title} {\enquote {\bibinfo {title} {Deceleration, precooling, and
  multi-pass stopping of highly charged ions in {B}e$^+$ {C}oulomb crystals},}\
  }\href {\doibase 10.1063/1.4934245} {\bibfield  {journal} {\bibinfo
  {journal} {Rev. Sci. Instrum.}\ }\textbf {\bibinfo {volume} {86}},\ \bibinfo
  {pages} {103111} (\bibinfo {year} {2015}{\natexlab{b}})}\BibitemShut
  {NoStop}%
\bibitem [{\citenamefont {Micke}\ \emph {et~al.}(2020)\citenamefont {Micke},
  \citenamefont {Leopold}, \citenamefont {King}, \citenamefont {Benkler},
  \citenamefont {Spie{\ss}}, \citenamefont {Schm{\"o}ger}, \citenamefont
  {Schwarz}, \citenamefont {Crespo L{\'o}pez-Urrutia},\ and\ \citenamefont
  {Schmidt}}]{Micke2020}%
  \BibitemOpen
  \bibfield  {author} {\bibinfo {author} {\bibfnamefont {P.}~\bibnamefont
  {Micke}}, \bibinfo {author} {\bibfnamefont {T.}~\bibnamefont {Leopold}},
  \bibinfo {author} {\bibfnamefont {S.~A.}\ \bibnamefont {King}}, \bibinfo
  {author} {\bibfnamefont {E.}~\bibnamefont {Benkler}}, \bibinfo {author}
  {\bibfnamefont {L.~J.}\ \bibnamefont {Spie{\ss}}}, \bibinfo {author}
  {\bibfnamefont {L.}~\bibnamefont {Schm{\"o}ger}}, \bibinfo {author}
  {\bibfnamefont {M.}~\bibnamefont {Schwarz}}, \bibinfo {author} {\bibfnamefont
  {J.~R.}\ \bibnamefont {Crespo L{\'o}pez-Urrutia}}, \ and\ \bibinfo {author}
  {\bibfnamefont {P.~O.}\ \bibnamefont {Schmidt}},\ }\bibfield  {title}
  {\enquote {\bibinfo {title} {Coherent laser spectroscopy of highly charged
  ions using quantum logic},}\ }\href {\doibase 10.1038/s41586-020-1959-8}
  {\bibfield  {journal} {\bibinfo  {journal} {Nature}\ }\textbf {\bibinfo
  {volume} {578}},\ \bibinfo {pages} {60--65} (\bibinfo {year}
  {2020})}\BibitemShut {NoStop}%
\bibitem [{\citenamefont {King}\ \emph {et~al.}(2021)\citenamefont {King},
  \citenamefont {Spie{\ss}}, \citenamefont {Micke}, \citenamefont {Wilzewski},
  \citenamefont {Leopold}, \citenamefont {Crespo L{\'o}pez-Urrutia},\ and\
  \citenamefont {Schmidt}}]{King2021}%
  \BibitemOpen
  \bibfield  {author} {\bibinfo {author} {\bibfnamefont {S.~A.}\ \bibnamefont
  {King}}, \bibinfo {author} {\bibfnamefont {L.~J.}\ \bibnamefont {Spie{\ss}}},
  \bibinfo {author} {\bibfnamefont {P.}~\bibnamefont {Micke}}, \bibinfo
  {author} {\bibfnamefont {A.}~\bibnamefont {Wilzewski}}, \bibinfo {author}
  {\bibfnamefont {T.}~\bibnamefont {Leopold}}, \bibinfo {author} {\bibfnamefont
  {J.~R.}\ \bibnamefont {Crespo L{\'o}pez-Urrutia}}, \ and\ \bibinfo {author}
  {\bibfnamefont {P.~O.}\ \bibnamefont {Schmidt}},\ }\bibfield  {title}
  {\enquote {\bibinfo {title} {Algorithmic ground-state cooling of weakly
  coupled oscillators using quantum logic},}\ }\href {\doibase
  10.1103/PhysRevX.11.041049} {\bibfield  {journal} {\bibinfo  {journal} {Phys.
  Rev. X}\ }\textbf {\bibinfo {volume} {11}},\ \bibinfo {pages} {041049}
  (\bibinfo {year} {2021})}\BibitemShut {NoStop}%
\bibitem [{\citenamefont {King}\ \emph {et~al.}(2022)\citenamefont {King},
  \citenamefont {Spie{\ss}}, \citenamefont {Micke}, \citenamefont {Wilzewski},
  \citenamefont {Leopold}, \citenamefont {Benkler}, \citenamefont {Lange},
  \citenamefont {Huntemann}, \citenamefont {Surzhykov}, \citenamefont
  {Yerokhin}, \citenamefont {Crespo L{\'o}pez-Urrutia},\ and\ \citenamefont
  {Schmidt}}]{King2022}%
  \BibitemOpen
  \bibfield  {author} {\bibinfo {author} {\bibfnamefont {S.~A.}\ \bibnamefont
  {King}}, \bibinfo {author} {\bibfnamefont {L.~J.}\ \bibnamefont {Spie{\ss}}},
  \bibinfo {author} {\bibfnamefont {P.}~\bibnamefont {Micke}}, \bibinfo
  {author} {\bibfnamefont {A.}~\bibnamefont {Wilzewski}}, \bibinfo {author}
  {\bibfnamefont {T.}~\bibnamefont {Leopold}}, \bibinfo {author} {\bibfnamefont
  {E.}~\bibnamefont {Benkler}}, \bibinfo {author} {\bibfnamefont
  {R.}~\bibnamefont {Lange}}, \bibinfo {author} {\bibfnamefont
  {N.}~\bibnamefont {Huntemann}}, \bibinfo {author} {\bibfnamefont
  {A.}~\bibnamefont {Surzhykov}}, \bibinfo {author} {\bibfnamefont {V.~A.}\
  \bibnamefont {Yerokhin}}, \bibinfo {author} {\bibfnamefont {J.~R.}\
  \bibnamefont {Crespo L{\'o}pez-Urrutia}}, \ and\ \bibinfo {author}
  {\bibfnamefont {P.~O.}\ \bibnamefont {Schmidt}},\ }\bibfield  {title}
  {\enquote {\bibinfo {title} {An optical atomic clock based on a highly
  charged ion},}\ }\href {\doibase 10.1038/s41586-022-05245-4} {\bibfield
  {journal} {\bibinfo  {journal} {Nature}\ }\textbf {\bibinfo {volume} {611}},\
  \bibinfo {pages} {43--47} (\bibinfo {year} {2022})}\BibitemShut {NoStop}%
\bibitem [{\citenamefont {Crespo L{\'o}pez-Urrutia}(2016)}]{Crespo2016}%
  \BibitemOpen
  \bibfield  {author} {\bibinfo {author} {\bibfnamefont {J.~R.}\ \bibnamefont
  {Crespo L{\'o}pez-Urrutia}},\ }\bibfield  {title} {\enquote {\bibinfo {title}
  {Frequency metrology using highly charged ions},}\ }\href {\doibase
  10.1088/1742-6596/723/1/012052} {\bibfield  {journal} {\bibinfo  {journal}
  {J. Phys.: Conf. Ser.}\ }\textbf {\bibinfo {volume} {723}},\ \bibinfo {pages}
  {012052} (\bibinfo {year} {2016})}\BibitemShut {NoStop}%
\bibitem [{\citenamefont {Jones}\ \emph {et~al.}(2005)\citenamefont {Jones},
  \citenamefont {Moll}, \citenamefont {Thorpe},\ and\ \citenamefont
  {Ye}}]{Jones2005}%
  \BibitemOpen
  \bibfield  {author} {\bibinfo {author} {\bibfnamefont {R.~J.}\ \bibnamefont
  {Jones}}, \bibinfo {author} {\bibfnamefont {K.~D.}\ \bibnamefont {Moll}},
  \bibinfo {author} {\bibfnamefont {M.~J.}\ \bibnamefont {Thorpe}}, \ and\
  \bibinfo {author} {\bibfnamefont {J.}~\bibnamefont {Ye}},\ }\bibfield
  {title} {\enquote {\bibinfo {title} {Phase-coherent frequency combs in the
  vacuum ultraviolet via high-harmonic generation inside a femtosecond
  enhancement cavity},}\ }\href {\doibase 10.1103/PhysRevLett.94.193201}
  {\bibfield  {journal} {\bibinfo  {journal} {Phys. Rev. Lett.}\ }\textbf
  {\bibinfo {volume} {94}},\ \bibinfo {pages} {193201} (\bibinfo {year}
  {2005})}\BibitemShut {NoStop}%
\bibitem [{\citenamefont {Gohle}\ \emph {et~al.}(2005)\citenamefont {Gohle},
  \citenamefont {Udem}, \citenamefont {Herrmann}, \citenamefont
  {Rauschenberger}, \citenamefont {Holzwarth}, \citenamefont {Schuessler},
  \citenamefont {Krausz},\ and\ \citenamefont {H{\"a}nsch}}]{Gohle2005}%
  \BibitemOpen
  \bibfield  {author} {\bibinfo {author} {\bibfnamefont {C.}~\bibnamefont
  {Gohle}}, \bibinfo {author} {\bibfnamefont {T.}~\bibnamefont {Udem}},
  \bibinfo {author} {\bibfnamefont {M.}~\bibnamefont {Herrmann}}, \bibinfo
  {author} {\bibfnamefont {J.}~\bibnamefont {Rauschenberger}}, \bibinfo
  {author} {\bibfnamefont {R.}~\bibnamefont {Holzwarth}}, \bibinfo {author}
  {\bibfnamefont {H.~A.}\ \bibnamefont {Schuessler}}, \bibinfo {author}
  {\bibfnamefont {F.}~\bibnamefont {Krausz}}, \ and\ \bibinfo {author}
  {\bibfnamefont {T.~W.}\ \bibnamefont {H{\"a}nsch}},\ }\bibfield  {title}
  {\enquote {\bibinfo {title} {A frequency comb in the extreme ultraviolet},}\
  }\href {\doibase 10.1038/nature03851} {\bibfield  {journal} {\bibinfo
  {journal} {Nature}\ }\textbf {\bibinfo {volume} {436}},\ \bibinfo {pages}
  {234--237} (\bibinfo {year} {2005})}\BibitemShut {NoStop}%
\bibitem [{\citenamefont {Pupeza}\ \emph {et~al.}(2021)\citenamefont {Pupeza},
  \citenamefont {Zhang}, \citenamefont {H{\"o}gner},\ and\ \citenamefont
  {Ye}}]{Pupeza2021}%
  \BibitemOpen
  \bibfield  {author} {\bibinfo {author} {\bibfnamefont {I.}~\bibnamefont
  {Pupeza}}, \bibinfo {author} {\bibfnamefont {C.}~\bibnamefont {Zhang}},
  \bibinfo {author} {\bibfnamefont {M.}~\bibnamefont {H{\"o}gner}}, \ and\
  \bibinfo {author} {\bibfnamefont {J.}~\bibnamefont {Ye}},\ }\bibfield
  {title} {\enquote {\bibinfo {title} {Extreme-ultraviolet frequency combs for
  precision metrology and attosecond science},}\ }\href {\doibase
  10.1038/s41566-020-00741-3} {\bibfield  {journal} {\bibinfo  {journal} {Nat.
  Photonics}\ }\textbf {\bibinfo {volume} {15}},\ \bibinfo {pages} {175--186}
  (\bibinfo {year} {2021})}\BibitemShut {NoStop}%
\bibitem [{\citenamefont {Nauta}\ \emph {et~al.}(2021)\citenamefont {Nauta},
  \citenamefont {Oelmann}, \citenamefont {Borodin}, \citenamefont {Ackermann},
  \citenamefont {Knauer}, \citenamefont {Muhammad}, \citenamefont
  {Pappenberger}, \citenamefont {Pfeifer},\ and\ \citenamefont {Crespo
  L{\'o}pez-Urrutia}}]{Nauta2021}%
  \BibitemOpen
  \bibfield  {author} {\bibinfo {author} {\bibfnamefont {J.}~\bibnamefont
  {Nauta}}, \bibinfo {author} {\bibfnamefont {J.-H.}\ \bibnamefont {Oelmann}},
  \bibinfo {author} {\bibfnamefont {A.}~\bibnamefont {Borodin}}, \bibinfo
  {author} {\bibfnamefont {A.}~\bibnamefont {Ackermann}}, \bibinfo {author}
  {\bibfnamefont {P.}~\bibnamefont {Knauer}}, \bibinfo {author} {\bibfnamefont
  {I.~S.}\ \bibnamefont {Muhammad}}, \bibinfo {author} {\bibfnamefont
  {R.}~\bibnamefont {Pappenberger}}, \bibinfo {author} {\bibfnamefont
  {T.}~\bibnamefont {Pfeifer}}, \ and\ \bibinfo {author} {\bibfnamefont
  {J.~R.}\ \bibnamefont {Crespo L{\'o}pez-Urrutia}},\ }\bibfield  {title}
  {\enquote {\bibinfo {title} {{XUV} frequency comb production with an
  astigmatism-compensated enhancement cavity},}\ }\href {\doibase
  10.1364/OE.414987} {\bibfield  {journal} {\bibinfo  {journal} {Opt. Express}\
  }\textbf {\bibinfo {volume} {29}},\ \bibinfo {pages} {2624--2636} (\bibinfo
  {year} {2021})}\BibitemShut {NoStop}%
\bibitem [{\citenamefont {Lyu}\ \emph {et~al.}(2020)\citenamefont {Lyu},
  \citenamefont {Cavaletto}, \citenamefont {Keitel},\ and\ \citenamefont
  {Harman}}]{Lyu2020}%
  \BibitemOpen
  \bibfield  {author} {\bibinfo {author} {\bibfnamefont {C.}~\bibnamefont
  {Lyu}}, \bibinfo {author} {\bibfnamefont {S.~M.}\ \bibnamefont {Cavaletto}},
  \bibinfo {author} {\bibfnamefont {C.~H.}\ \bibnamefont {Keitel}}, \ and\
  \bibinfo {author} {\bibfnamefont {Z.}~\bibnamefont {Harman}},\ }\bibfield
  {title} {\enquote {\bibinfo {title} {Interrogating the temporal coherence of
  {EUV} frequency combs with highly charged ions},}\ }\href {\doibase
  10.1103/PhysRevLett.125.093201} {\bibfield  {journal} {\bibinfo  {journal}
  {Phys. Rev. Lett.}\ }\textbf {\bibinfo {volume} {125}},\ \bibinfo {pages}
  {093201} (\bibinfo {year} {2020})}\BibitemShut {NoStop}%
\bibitem [{\citenamefont {Paul}(1990)}]{Paul1990}%
  \BibitemOpen
  \bibfield  {author} {\bibinfo {author} {\bibfnamefont {W.}~\bibnamefont
  {Paul}},\ }\bibfield  {title} {\enquote {\bibinfo {title} {Electromagnetic
  traps for charged and neutral particles},}\ }\href {\doibase
  10.1103/RevModPhys.62.531} {\bibfield  {journal} {\bibinfo  {journal} {Rev.
  Mod. Phys.}\ }\textbf {\bibinfo {volume} {62}},\ \bibinfo {pages} {531--540}
  (\bibinfo {year} {1990})}\BibitemShut {NoStop}%
\bibitem [{\citenamefont {Dehmelt}(1983)}]{Dehmelt1983}%
  \BibitemOpen
  \bibfield  {author} {\bibinfo {author} {\bibfnamefont {H.}~\bibnamefont
  {Dehmelt}},\ }\enquote {\bibinfo {title} {Stored-ion spectroscopy},}\ in\
  \href {\doibase 10.1007/978-1-4613-3715-7_6} {\emph {\bibinfo {booktitle}
  {Advances in Laser Spectroscopy}}},\ \bibinfo {editor} {edited by\ \bibinfo
  {editor} {\bibfnamefont {F.~T.}\ \bibnamefont {Arecchi}}, \bibinfo {editor}
  {\bibfnamefont {F.}~\bibnamefont {Strumia}}, \ and\ \bibinfo {editor}
  {\bibfnamefont {H.}~\bibnamefont {Walther}}}\ (\bibinfo  {publisher}
  {Springer US},\ \bibinfo {address} {Boston, MA},\ \bibinfo {year} {1983})\
  pp.\ \bibinfo {pages} {153--187}\BibitemShut {NoStop}%
\bibitem [{\citenamefont {Diedrich}\ and\ \citenamefont
  {Walther}(1987)}]{Diedrich1987}%
  \BibitemOpen
  \bibfield  {author} {\bibinfo {author} {\bibfnamefont {F.}~\bibnamefont
  {Diedrich}}\ and\ \bibinfo {author} {\bibfnamefont {H.}~\bibnamefont
  {Walther}},\ }\bibfield  {title} {\enquote {\bibinfo {title} {Nonclassical
  radiation of a single stored ion},}\ }\href {\doibase
  10.1103/PhysRevLett.58.203} {\bibfield  {journal} {\bibinfo  {journal} {Phys.
  Rev. Lett.}\ }\textbf {\bibinfo {volume} {58}},\ \bibinfo {pages} {203--206}
  (\bibinfo {year} {1987})}\BibitemShut {NoStop}%
\bibitem [{\citenamefont {Cirac}\ and\ \citenamefont
  {Zoller}(1995)}]{Cirac1995}%
  \BibitemOpen
  \bibfield  {author} {\bibinfo {author} {\bibfnamefont {J.~I.}\ \bibnamefont
  {Cirac}}\ and\ \bibinfo {author} {\bibfnamefont {P.}~\bibnamefont {Zoller}},\
  }\bibfield  {title} {\enquote {\bibinfo {title} {Quantum computations with
  cold trapped ions},}\ }\href {\doibase 10.1103/PhysRevLett.74.4091}
  {\bibfield  {journal} {\bibinfo  {journal} {Phys. Rev. Lett.}\ }\textbf
  {\bibinfo {volume} {74}},\ \bibinfo {pages} {4091--4094} (\bibinfo {year}
  {1995})}\BibitemShut {NoStop}%
\bibitem [{\citenamefont {Monroe}\ \emph {et~al.}(1995)\citenamefont {Monroe},
  \citenamefont {Meekhof}, \citenamefont {King}, \citenamefont {Itano},\ and\
  \citenamefont {Wineland}}]{Monroe1995}%
  \BibitemOpen
  \bibfield  {author} {\bibinfo {author} {\bibfnamefont {C.}~\bibnamefont
  {Monroe}}, \bibinfo {author} {\bibfnamefont {D.~M.}\ \bibnamefont {Meekhof}},
  \bibinfo {author} {\bibfnamefont {B.~E.}\ \bibnamefont {King}}, \bibinfo
  {author} {\bibfnamefont {W.~M.}\ \bibnamefont {Itano}}, \ and\ \bibinfo
  {author} {\bibfnamefont {D.~J.}\ \bibnamefont {Wineland}},\ }\bibfield
  {title} {\enquote {\bibinfo {title} {Demonstration of a fundamental quantum
  logic gate},}\ }\href {\doibase 10.1103/PhysRevLett.75.4714} {\bibfield
  {journal} {\bibinfo  {journal} {Phys. Rev. Lett.}\ }\textbf {\bibinfo
  {volume} {75}},\ \bibinfo {pages} {4714--4717} (\bibinfo {year}
  {1995})}\BibitemShut {NoStop}%
\bibitem [{\citenamefont {Georgescu}(2020)}]{Georgescu2020}%
  \BibitemOpen
  \bibfield  {author} {\bibinfo {author} {\bibfnamefont {I.}~\bibnamefont
  {Georgescu}},\ }\bibfield  {title} {\enquote {\bibinfo {title} {Trapped ion
  quantum computing turns 25},}\ }\href {\doibase 10.1038/s42254-020-0189-1}
  {\bibfield  {journal} {\bibinfo  {journal} {Nat. Rev. Phys.}\ }\textbf
  {\bibinfo {volume} {2}},\ \bibinfo {pages} {278--278} (\bibinfo {year}
  {2020})}\BibitemShut {NoStop}%
\bibitem [{\citenamefont {Ludlow}\ \emph {et~al.}(2015)\citenamefont {Ludlow},
  \citenamefont {Boyd}, \citenamefont {Ye}, \citenamefont {Peik},\ and\
  \citenamefont {Schmidt}}]{Ludlow2015}%
  \BibitemOpen
  \bibfield  {author} {\bibinfo {author} {\bibfnamefont {A.~D.}\ \bibnamefont
  {Ludlow}}, \bibinfo {author} {\bibfnamefont {M.~M.}\ \bibnamefont {Boyd}},
  \bibinfo {author} {\bibfnamefont {J.}~\bibnamefont {Ye}}, \bibinfo {author}
  {\bibfnamefont {E.}~\bibnamefont {Peik}}, \ and\ \bibinfo {author}
  {\bibfnamefont {P.~O.}\ \bibnamefont {Schmidt}},\ }\bibfield  {title}
  {\enquote {\bibinfo {title} {Optical atomic clocks},}\ }\href {\doibase
  10.1103/RevModPhys.87.637} {\bibfield  {journal} {\bibinfo  {journal} {Rev.
  Mod. Phys.}\ }\textbf {\bibinfo {volume} {87}},\ \bibinfo {pages} {637--701}
  (\bibinfo {year} {2015})}\BibitemShut {NoStop}%
\bibitem [{\citenamefont {Stark}\ \emph {et~al.}(2021)\citenamefont {Stark},
  \citenamefont {Warnecke}, \citenamefont {Bogen}, \citenamefont {Chen},
  \citenamefont {Dijck}, \citenamefont {K{\"u}hn}, \citenamefont {Rosner},
  \citenamefont {Graf}, \citenamefont {Nauta}, \citenamefont {Oelmann},
  \citenamefont {Schm{\"o}ger}, \citenamefont {Schwarz}, \citenamefont
  {Liebert}, \citenamefont {Spie{\ss}}, \citenamefont {King}, \citenamefont
  {Leopold}, \citenamefont {Micke}, \citenamefont {Schmidt}, \citenamefont
  {Pfeifer},\ and\ \citenamefont {Crespo L{\'o}pez-Urrutia}}]{Stark2021}%
  \BibitemOpen
  \bibfield  {author} {\bibinfo {author} {\bibfnamefont {J.}~\bibnamefont
  {Stark}}, \bibinfo {author} {\bibfnamefont {C.}~\bibnamefont {Warnecke}},
  \bibinfo {author} {\bibfnamefont {S.}~\bibnamefont {Bogen}}, \bibinfo
  {author} {\bibfnamefont {S.}~\bibnamefont {Chen}}, \bibinfo {author}
  {\bibfnamefont {E.~A.}\ \bibnamefont {Dijck}}, \bibinfo {author}
  {\bibfnamefont {S.}~\bibnamefont {K{\"u}hn}}, \bibinfo {author}
  {\bibfnamefont {M.~K.}\ \bibnamefont {Rosner}}, \bibinfo {author}
  {\bibfnamefont {A.}~\bibnamefont {Graf}}, \bibinfo {author} {\bibfnamefont
  {J.}~\bibnamefont {Nauta}}, \bibinfo {author} {\bibfnamefont {J.-H.}\
  \bibnamefont {Oelmann}}, \bibinfo {author} {\bibfnamefont {L.}~\bibnamefont
  {Schm{\"o}ger}}, \bibinfo {author} {\bibfnamefont {M.}~\bibnamefont
  {Schwarz}}, \bibinfo {author} {\bibfnamefont {D.}~\bibnamefont {Liebert}},
  \bibinfo {author} {\bibfnamefont {L.~J.}\ \bibnamefont {Spie{\ss}}}, \bibinfo
  {author} {\bibfnamefont {S.~A.}\ \bibnamefont {King}}, \bibinfo {author}
  {\bibfnamefont {T.}~\bibnamefont {Leopold}}, \bibinfo {author} {\bibfnamefont
  {P.}~\bibnamefont {Micke}}, \bibinfo {author} {\bibfnamefont {P.~O.}\
  \bibnamefont {Schmidt}}, \bibinfo {author} {\bibfnamefont {T.}~\bibnamefont
  {Pfeifer}}, \ and\ \bibinfo {author} {\bibfnamefont {J.~R.}\ \bibnamefont
  {Crespo L{\'o}pez-Urrutia}},\ }\bibfield  {title} {\enquote {\bibinfo {title}
  {An ultralow-noise superconducting radio-frequency ion trap for frequency
  metrology with highly charged ions},}\ }\href {\doibase 10.1063/5.0046569}
  {\bibfield  {journal} {\bibinfo  {journal} {Rev. Sci. Instrum.}\ }\textbf
  {\bibinfo {volume} {92}},\ \bibinfo {pages} {083203} (\bibinfo {year}
  {2021})}\BibitemShut {NoStop}%
\bibitem [{\citenamefont {Micke}\ \emph {et~al.}(2018)\citenamefont {Micke},
  \citenamefont {K{\"u}hn}, \citenamefont {Buchauer}, \citenamefont {Harries},
  \citenamefont {B{\"u}cking}, \citenamefont {Blaum}, \citenamefont {Cieluch},
  \citenamefont {Egl}, \citenamefont {Hollain}, \citenamefont {Kraemer},
  \citenamefont {Pfeifer}, \citenamefont {Schmidt}, \citenamefont
  {Sch{\"u}ssler}, \citenamefont {Schweiger}, \citenamefont {St{\"o}hlker},
  \citenamefont {Sturm}, \citenamefont {Wolf}, \citenamefont {Bernitt},\ and\
  \citenamefont {Crespo L{\'o}pez-Urrutia}}]{Micke2018}%
  \BibitemOpen
  \bibfield  {author} {\bibinfo {author} {\bibfnamefont {P.}~\bibnamefont
  {Micke}}, \bibinfo {author} {\bibfnamefont {S.}~\bibnamefont {K{\"u}hn}},
  \bibinfo {author} {\bibfnamefont {L.}~\bibnamefont {Buchauer}}, \bibinfo
  {author} {\bibfnamefont {J.~R.}\ \bibnamefont {Harries}}, \bibinfo {author}
  {\bibfnamefont {T.~M.}\ \bibnamefont {B{\"u}cking}}, \bibinfo {author}
  {\bibfnamefont {K.}~\bibnamefont {Blaum}}, \bibinfo {author} {\bibfnamefont
  {A.}~\bibnamefont {Cieluch}}, \bibinfo {author} {\bibfnamefont
  {A.}~\bibnamefont {Egl}}, \bibinfo {author} {\bibfnamefont {D.}~\bibnamefont
  {Hollain}}, \bibinfo {author} {\bibfnamefont {S.}~\bibnamefont {Kraemer}},
  \bibinfo {author} {\bibfnamefont {T.}~\bibnamefont {Pfeifer}}, \bibinfo
  {author} {\bibfnamefont {P.~O.}\ \bibnamefont {Schmidt}}, \bibinfo {author}
  {\bibfnamefont {R.~X.}\ \bibnamefont {Sch{\"u}ssler}}, \bibinfo {author}
  {\bibfnamefont {{\relax Ch}.}~\bibnamefont {Schweiger}}, \bibinfo {author}
  {\bibfnamefont {T.}~\bibnamefont {St{\"o}hlker}}, \bibinfo {author}
  {\bibfnamefont {S.}~\bibnamefont {Sturm}}, \bibinfo {author} {\bibfnamefont
  {R.~N.}\ \bibnamefont {Wolf}}, \bibinfo {author} {\bibfnamefont
  {S.}~\bibnamefont {Bernitt}}, \ and\ \bibinfo {author} {\bibfnamefont
  {J.~R.}\ \bibnamefont {Crespo L{\'o}pez-Urrutia}},\ }\bibfield  {title}
  {\enquote {\bibinfo {title} {The {H}eidelberg compact electron beam ion
  traps},}\ }\href {\doibase 10.1063/1.5026961} {\bibfield  {journal} {\bibinfo
   {journal} {Rev. Sci. Instrum.}\ }\textbf {\bibinfo {volume} {89}},\ \bibinfo
  {pages} {063109} (\bibinfo {year} {2018})}\BibitemShut {NoStop}%
\bibitem [{\citenamefont {Mandal}, \citenamefont {Sikler},\ and\ \citenamefont
  {Mukherjee}(2011)}]{Mandal2011}%
  \BibitemOpen
  \bibfield  {author} {\bibinfo {author} {\bibfnamefont {P.}~\bibnamefont
  {Mandal}}, \bibinfo {author} {\bibfnamefont {G.}~\bibnamefont {Sikler}}, \
  and\ \bibinfo {author} {\bibfnamefont {M.}~\bibnamefont {Mukherjee}},\
  }\bibfield  {title} {\enquote {\bibinfo {title} {Simulation study and
  analysis of a compact einzel lens-deflector for low energy ion beam},}\
  }\href {\doibase 10.1088/1748-0221/6/02/P02004} {\bibfield  {journal}
  {\bibinfo  {journal} {J. Instrum.}\ }\textbf {\bibinfo {volume} {6}},\
  \bibinfo {pages} {P02004} (\bibinfo {year} {2011})}\BibitemShut {NoStop}%
\bibitem [{\citenamefont {Brownnutt}\ \emph {et~al.}(2015)\citenamefont
  {Brownnutt}, \citenamefont {Kumph}, \citenamefont {Rabl},\ and\ \citenamefont
  {Blatt}}]{Brownnutt2015}%
  \BibitemOpen
  \bibfield  {author} {\bibinfo {author} {\bibfnamefont {M.}~\bibnamefont
  {Brownnutt}}, \bibinfo {author} {\bibfnamefont {M.}~\bibnamefont {Kumph}},
  \bibinfo {author} {\bibfnamefont {P.}~\bibnamefont {Rabl}}, \ and\ \bibinfo
  {author} {\bibfnamefont {R.}~\bibnamefont {Blatt}},\ }\bibfield  {title}
  {\enquote {\bibinfo {title} {Ion-trap measurements of electric-field noise
  near surfaces},}\ }\href {\doibase 10.1103/RevModPhys.87.1419} {\bibfield
  {journal} {\bibinfo  {journal} {Rev. Mod. Phys.}\ }\textbf {\bibinfo {volume}
  {87}},\ \bibinfo {pages} {1419--1482} (\bibinfo {year} {2015})}\BibitemShut
  {NoStop}%
\bibitem [{\citenamefont {Paasche}\ \emph {et~al.}(2003)\citenamefont
  {Paasche}, \citenamefont {Angelescu}, \citenamefont {Ananthamurthy},
  \citenamefont {Biswas}, \citenamefont {Valenzuela},\ and\ \citenamefont
  {Werth}}]{Paasche2003}%
  \BibitemOpen
  \bibfield  {author} {\bibinfo {author} {\bibfnamefont {P.}~\bibnamefont
  {Paasche}}, \bibinfo {author} {\bibfnamefont {C.}~\bibnamefont {Angelescu}},
  \bibinfo {author} {\bibfnamefont {S.}~\bibnamefont {Ananthamurthy}}, \bibinfo
  {author} {\bibfnamefont {D.}~\bibnamefont {Biswas}}, \bibinfo {author}
  {\bibfnamefont {T.}~\bibnamefont {Valenzuela}}, \ and\ \bibinfo {author}
  {\bibfnamefont {G.}~\bibnamefont {Werth}},\ }\bibfield  {title} {\enquote
  {\bibinfo {title} {Instabilities of an electron cloud in a {P}enning trap},}\
  }\href {\doibase 10.1140/epjd/e2002-00239-3} {\bibfield  {journal} {\bibinfo
  {journal} {Eur. Phys. J. D}\ }\textbf {\bibinfo {volume} {22}},\ \bibinfo
  {pages} {183--188} (\bibinfo {year} {2003})}\BibitemShut {NoStop}%
\bibitem [{\citenamefont {Alt}(2002)}]{Alt2002}%
  \BibitemOpen
  \bibfield  {author} {\bibinfo {author} {\bibfnamefont {W.}~\bibnamefont
  {Alt}},\ }\bibfield  {title} {\enquote {\bibinfo {title} {An objective lens
  for efficient fluorescence detection of single atoms},}\ }\href {\doibase
  10.1078/0030-4026-00133} {\bibfield  {journal} {\bibinfo  {journal} {Optik}\
  }\textbf {\bibinfo {volume} {113}},\ \bibinfo {pages} {142--144} (\bibinfo
  {year} {2002})}\BibitemShut {NoStop}%
\bibitem [{\citenamefont {Noek}\ \emph {et~al.}(2013)\citenamefont {Noek},
  \citenamefont {Vrijsen}, \citenamefont {Gaultney}, \citenamefont {Mount},
  \citenamefont {Kim}, \citenamefont {Maunz},\ and\ \citenamefont
  {Kim}}]{Noek2013}%
  \BibitemOpen
  \bibfield  {author} {\bibinfo {author} {\bibfnamefont {R.}~\bibnamefont
  {Noek}}, \bibinfo {author} {\bibfnamefont {G.}~\bibnamefont {Vrijsen}},
  \bibinfo {author} {\bibfnamefont {D.}~\bibnamefont {Gaultney}}, \bibinfo
  {author} {\bibfnamefont {E.}~\bibnamefont {Mount}}, \bibinfo {author}
  {\bibfnamefont {T.}~\bibnamefont {Kim}}, \bibinfo {author} {\bibfnamefont
  {P.}~\bibnamefont {Maunz}}, \ and\ \bibinfo {author} {\bibfnamefont
  {J.}~\bibnamefont {Kim}},\ }\bibfield  {title} {\enquote {\bibinfo {title}
  {High speed, high fidelity detection of an atomic hyperfine qubit},}\ }\href
  {\doibase 10.1364/ol.38.004735} {\bibfield  {journal} {\bibinfo  {journal}
  {Opt. Lett.}\ }\textbf {\bibinfo {volume} {38}},\ \bibinfo {pages} {4735}
  (\bibinfo {year} {2013})}\BibitemShut {NoStop}%
\bibitem [{\citenamefont {Pyka}\ \emph {et~al.}(2013)\citenamefont {Pyka},
  \citenamefont {Herschbach}, \citenamefont {Keller},\ and\ \citenamefont
  {Mehlst{\"a}ubler}}]{Pyka2013}%
  \BibitemOpen
  \bibfield  {author} {\bibinfo {author} {\bibfnamefont {K.}~\bibnamefont
  {Pyka}}, \bibinfo {author} {\bibfnamefont {N.}~\bibnamefont {Herschbach}},
  \bibinfo {author} {\bibfnamefont {J.}~\bibnamefont {Keller}}, \ and\ \bibinfo
  {author} {\bibfnamefont {T.~E.}\ \bibnamefont {Mehlst{\"a}ubler}},\
  }\bibfield  {title} {\enquote {\bibinfo {title} {A high-precision segmented
  {P}aul trap with minimized micromotion for an optical multiple-ion clock},}\
  }\href {\doibase 10.1007/s00340-013-5580-5} {\bibfield  {journal} {\bibinfo
  {journal} {Appl. Phys. B}\ }\textbf {\bibinfo {volume} {114}},\ \bibinfo
  {pages} {231--241} (\bibinfo {year} {2013})}\BibitemShut {NoStop}%
\bibitem [{\citenamefont {Wong-Campos}\ \emph {et~al.}(2016)\citenamefont
  {Wong-Campos}, \citenamefont {Johnson}, \citenamefont {Neyenhuis},
  \citenamefont {Mizrahi},\ and\ \citenamefont {Monroe}}]{WongCampos2016}%
  \BibitemOpen
  \bibfield  {author} {\bibinfo {author} {\bibfnamefont {J.~D.}\ \bibnamefont
  {Wong-Campos}}, \bibinfo {author} {\bibfnamefont {K.~G.}\ \bibnamefont
  {Johnson}}, \bibinfo {author} {\bibfnamefont {B.}~\bibnamefont {Neyenhuis}},
  \bibinfo {author} {\bibfnamefont {J.}~\bibnamefont {Mizrahi}}, \ and\
  \bibinfo {author} {\bibfnamefont {C.}~\bibnamefont {Monroe}},\ }\bibfield
  {title} {\enquote {\bibinfo {title} {High-resolution adaptive imaging of a
  single atom},}\ }\href {\doibase 10.1038/nphoton.2016.136} {\bibfield
  {journal} {\bibinfo  {journal} {Nat. Photonics}\ }\textbf {\bibinfo {volume}
  {10}},\ \bibinfo {pages} {606--610} (\bibinfo {year} {2016})}\BibitemShut
  {NoStop}%
\bibitem [{\citenamefont {Li}\ \emph {et~al.}(2020)\citenamefont {Li},
  \citenamefont {Li}, \citenamefont {Wu}, \citenamefont {Fan}, \citenamefont
  {Tian}, \citenamefont {Yang}, \citenamefont {Zhang},\ and\ \citenamefont
  {Zhang}}]{Li2020}%
  \BibitemOpen
  \bibfield  {author} {\bibinfo {author} {\bibfnamefont {S.}~\bibnamefont
  {Li}}, \bibinfo {author} {\bibfnamefont {G.}~\bibnamefont {Li}}, \bibinfo
  {author} {\bibfnamefont {W.}~\bibnamefont {Wu}}, \bibinfo {author}
  {\bibfnamefont {Q.}~\bibnamefont {Fan}}, \bibinfo {author} {\bibfnamefont
  {Y.}~\bibnamefont {Tian}}, \bibinfo {author} {\bibfnamefont {P.}~\bibnamefont
  {Yang}}, \bibinfo {author} {\bibfnamefont {P.}~\bibnamefont {Zhang}}, \ and\
  \bibinfo {author} {\bibfnamefont {T.}~\bibnamefont {Zhang}},\ }\bibfield
  {title} {\enquote {\bibinfo {title} {High-numerical-aperture and
  long-working-distance objective for single-atom experiments},}\ }\href
  {\doibase 10.1063/5.0001637} {\bibfield  {journal} {\bibinfo  {journal} {Rev.
  Sci. Instrum.}\ }\textbf {\bibinfo {volume} {91}},\ \bibinfo {pages} {043104}
  (\bibinfo {year} {2020})}\BibitemShut {NoStop}%
\bibitem [{\citenamefont {Nordmann}\ \emph {et~al.}(2023)\citenamefont
  {Nordmann}, \citenamefont {Wickenhagen}, \citenamefont {Dole{\v z}al},\ and\
  \citenamefont {Mehlst{\"a}ubler}}]{Nordmann2023}%
  \BibitemOpen
  \bibfield  {author} {\bibinfo {author} {\bibfnamefont {T.}~\bibnamefont
  {Nordmann}}, \bibinfo {author} {\bibfnamefont {S.}~\bibnamefont
  {Wickenhagen}}, \bibinfo {author} {\bibfnamefont {M.}~\bibnamefont {Dole{\v
  z}al}}, \ and\ \bibinfo {author} {\bibfnamefont {T.~E.}\ \bibnamefont
  {Mehlst{\"a}ubler}},\ }\href {\doibase 10.48550/ARXIV.2302.02489} {\enquote
  {\bibinfo {title} {Bichromatic {UV} detection system for atomically-resolved
  imaging of ions},}\ } (\bibinfo {year} {2023}),\ \Eprint
  {http://arxiv.org/abs/2302.02489} {arXiv:2302.02489 [physics.atom-ph]}
  \BibitemShut {NoStop}%
\bibitem [{\citenamefont {Leopold}\ \emph {et~al.}(2019)\citenamefont
  {Leopold}, \citenamefont {King}, \citenamefont {Micke}, \citenamefont
  {Bautista-Salvador}, \citenamefont {Heip}, \citenamefont {Ospelkaus},
  \citenamefont {Crespo L{\'o}pez-Urrutia},\ and\ \citenamefont
  {Schmidt}}]{Leopold2019}%
  \BibitemOpen
  \bibfield  {author} {\bibinfo {author} {\bibfnamefont {T.}~\bibnamefont
  {Leopold}}, \bibinfo {author} {\bibfnamefont {S.~A.}\ \bibnamefont {King}},
  \bibinfo {author} {\bibfnamefont {P.}~\bibnamefont {Micke}}, \bibinfo
  {author} {\bibfnamefont {A.}~\bibnamefont {Bautista-Salvador}}, \bibinfo
  {author} {\bibfnamefont {J.~C.}\ \bibnamefont {Heip}}, \bibinfo {author}
  {\bibfnamefont {C.}~\bibnamefont {Ospelkaus}}, \bibinfo {author}
  {\bibfnamefont {J.~R.}\ \bibnamefont {Crespo L{\'o}pez-Urrutia}}, \ and\
  \bibinfo {author} {\bibfnamefont {P.~O.}\ \bibnamefont {Schmidt}},\
  }\bibfield  {title} {\enquote {\bibinfo {title} {A cryogenic radio-frequency
  ion trap for quantum logic spectroscopy of highly charged ions},}\ }\href
  {\doibase 10.1063/1.5100594} {\bibfield  {journal} {\bibinfo  {journal} {Rev.
  Sci. Instrum.}\ }\textbf {\bibinfo {volume} {90}},\ \bibinfo {pages} {073201}
  (\bibinfo {year} {2019})}\BibitemShut {NoStop}%
\bibitem [{\citenamefont {Dubielzig}\ \emph {et~al.}(2021)\citenamefont
  {Dubielzig}, \citenamefont {Halama}, \citenamefont {Hahn}, \citenamefont
  {Zarantonello}, \citenamefont {Niemann}, \citenamefont {Bautista-Salvador},\
  and\ \citenamefont {Ospelkaus}}]{Dubielzig2021}%
  \BibitemOpen
  \bibfield  {author} {\bibinfo {author} {\bibfnamefont {T.}~\bibnamefont
  {Dubielzig}}, \bibinfo {author} {\bibfnamefont {S.}~\bibnamefont {Halama}},
  \bibinfo {author} {\bibfnamefont {H.}~\bibnamefont {Hahn}}, \bibinfo {author}
  {\bibfnamefont {G.}~\bibnamefont {Zarantonello}}, \bibinfo {author}
  {\bibfnamefont {M.}~\bibnamefont {Niemann}}, \bibinfo {author} {\bibfnamefont
  {A.}~\bibnamefont {Bautista-Salvador}}, \ and\ \bibinfo {author}
  {\bibfnamefont {C.}~\bibnamefont {Ospelkaus}},\ }\bibfield  {title} {\enquote
  {\bibinfo {title} {Ultra-low-vibration closed-cycle cryogenic
  surface-electrode ion trap apparatus},}\ }\href {\doibase 10.1063/5.0024423}
  {\bibfield  {journal} {\bibinfo  {journal} {Rev. Sci. Instrum.}\ }\textbf
  {\bibinfo {volume} {92}},\ \bibinfo {pages} {043201} (\bibinfo {year}
  {2021})}\BibitemShut {NoStop}%
\bibitem [{\citenamefont {Asfour}\ \emph {et~al.}(2017)\citenamefont {Asfour},
  \citenamefont {Weidner}, \citenamefont {Bodendorf}, \citenamefont {Bode},
  \citenamefont {Poleshchuk}, \citenamefont {Nasyrov}, \citenamefont {Grupp},\
  and\ \citenamefont {Bender}}]{Asfour2017}%
  \BibitemOpen
  \bibfield  {author} {\bibinfo {author} {\bibfnamefont {J.-M.}\ \bibnamefont
  {Asfour}}, \bibinfo {author} {\bibfnamefont {F.}~\bibnamefont {Weidner}},
  \bibinfo {author} {\bibfnamefont {C.}~\bibnamefont {Bodendorf}}, \bibinfo
  {author} {\bibfnamefont {A.}~\bibnamefont {Bode}}, \bibinfo {author}
  {\bibfnamefont {A.~G.}\ \bibnamefont {Poleshchuk}}, \bibinfo {author}
  {\bibfnamefont {R.~K.}\ \bibnamefont {Nasyrov}}, \bibinfo {author}
  {\bibfnamefont {F.}~\bibnamefont {Grupp}}, \ and\ \bibinfo {author}
  {\bibfnamefont {R.}~\bibnamefont {Bender}},\ }\bibfield  {title} {\enquote
  {\bibinfo {title} {Diffractive optics for precision alignment of {E}uclid
  space telescope optics (conference presentation)},}\ }in\ \href {\doibase
  10.1117/12.2274349} {\emph {\bibinfo {booktitle} {Astronomical Optics:
  Design, Manufacture, and Test of Space and Ground Systems}}},\ \bibinfo
  {editor} {edited by\ \bibinfo {editor} {\bibfnamefont {P.}~\bibnamefont
  {Hallibert}}, \bibinfo {editor} {\bibfnamefont {T.~B.}\ \bibnamefont {Hull}},
  \ and\ \bibinfo {editor} {\bibfnamefont {D.~W.}\ \bibnamefont {Kim}}}\
  (\bibinfo  {publisher} {{SPIE}},\ \bibinfo {year} {2017})\BibitemShut
  {NoStop}%
\bibitem [{\citenamefont {Gloger}\ \emph {et~al.}(2015)\citenamefont {Gloger},
  \citenamefont {Kaufmann}, \citenamefont {Kaufmann}, \citenamefont {Baig},
  \citenamefont {Collath}, \citenamefont {Johanning},\ and\ \citenamefont
  {Wunderlich}}]{Gloger2015}%
  \BibitemOpen
  \bibfield  {author} {\bibinfo {author} {\bibfnamefont {T.~F.}\ \bibnamefont
  {Gloger}}, \bibinfo {author} {\bibfnamefont {P.}~\bibnamefont {Kaufmann}},
  \bibinfo {author} {\bibfnamefont {D.}~\bibnamefont {Kaufmann}}, \bibinfo
  {author} {\bibfnamefont {M.~T.}\ \bibnamefont {Baig}}, \bibinfo {author}
  {\bibfnamefont {T.}~\bibnamefont {Collath}}, \bibinfo {author} {\bibfnamefont
  {M.}~\bibnamefont {Johanning}}, \ and\ \bibinfo {author} {\bibfnamefont
  {C.}~\bibnamefont {Wunderlich}},\ }\bibfield  {title} {\enquote {\bibinfo
  {title} {Ion-trajectory analysis for micromotion minimization and the
  measurement of small forces},}\ }\href {\doibase 10.1103/PhysRevA.92.043421}
  {\bibfield  {journal} {\bibinfo  {journal} {Phys. Rev. A}\ }\textbf {\bibinfo
  {volume} {92}},\ \bibinfo {pages} {043421} (\bibinfo {year}
  {2015})}\BibitemShut {NoStop}%
\bibitem [{\citenamefont {Berkeland}\ \emph {et~al.}(1998)\citenamefont
  {Berkeland}, \citenamefont {Miller}, \citenamefont {Bergquist}, \citenamefont
  {Itano},\ and\ \citenamefont {Wineland}}]{Berkeland1998}%
  \BibitemOpen
  \bibfield  {author} {\bibinfo {author} {\bibfnamefont {D.~J.}\ \bibnamefont
  {Berkeland}}, \bibinfo {author} {\bibfnamefont {J.~D.}\ \bibnamefont
  {Miller}}, \bibinfo {author} {\bibfnamefont {J.~C.}\ \bibnamefont
  {Bergquist}}, \bibinfo {author} {\bibfnamefont {W.~M.}\ \bibnamefont
  {Itano}}, \ and\ \bibinfo {author} {\bibfnamefont {D.~J.}\ \bibnamefont
  {Wineland}},\ }\bibfield  {title} {\enquote {\bibinfo {title} {Minimization
  of ion micromotion in a {P}aul trap},}\ }\href {\doibase 10.1063/1.367318}
  {\bibfield  {journal} {\bibinfo  {journal} {J. Appl. Phys.}\ }\textbf
  {\bibinfo {volume} {83}},\ \bibinfo {pages} {5025--5033} (\bibinfo {year}
  {1998})}\BibitemShut {NoStop}%
\bibitem [{\citenamefont {Keller}\ \emph {et~al.}(2015)\citenamefont {Keller},
  \citenamefont {Partner}, \citenamefont {Burgermeister},\ and\ \citenamefont
  {Mehlst{\"a}ubler}}]{Keller2015}%
  \BibitemOpen
  \bibfield  {author} {\bibinfo {author} {\bibfnamefont {J.}~\bibnamefont
  {Keller}}, \bibinfo {author} {\bibfnamefont {H.~L.}\ \bibnamefont {Partner}},
  \bibinfo {author} {\bibfnamefont {T.}~\bibnamefont {Burgermeister}}, \ and\
  \bibinfo {author} {\bibfnamefont {T.~E.}\ \bibnamefont {Mehlst{\"a}ubler}},\
  }\bibfield  {title} {\enquote {\bibinfo {title} {Precise determination of
  micromotion for trapped-ion optical clocks},}\ }\href {\doibase
  10.1063/1.4930037} {\bibfield  {journal} {\bibinfo  {journal} {J. Appl.
  Phys.}\ }\textbf {\bibinfo {volume} {118}},\ \bibinfo {pages} {104501}
  (\bibinfo {year} {2015})}\BibitemShut {NoStop}%
\bibitem [{\citenamefont {Thompson}(2015)}]{Thompson2015}%
  \BibitemOpen
  \bibfield  {author} {\bibinfo {author} {\bibfnamefont {R.~C.}\ \bibnamefont
  {Thompson}},\ }\bibfield  {title} {\enquote {\bibinfo {title} {Ion {C}oulomb
  crystals},}\ }\href {\doibase 10.1080/00107514.2014.989715} {\bibfield
  {journal} {\bibinfo  {journal} {Contemp. Phys.}\ }\textbf {\bibinfo {volume}
  {56}},\ \bibinfo {pages} {63--79} (\bibinfo {year} {2015})}\BibitemShut
  {NoStop}%
\bibitem [{\citenamefont {Kn{\"u}nz}\ \emph {et~al.}(2012)\citenamefont
  {Kn{\"u}nz}, \citenamefont {Herrmann}, \citenamefont {Batteiger},
  \citenamefont {Saathoff}, \citenamefont {H{\"a}nsch},\ and\ \citenamefont
  {Udem}}]{Knuenz2012}%
  \BibitemOpen
  \bibfield  {author} {\bibinfo {author} {\bibfnamefont {S.}~\bibnamefont
  {Kn{\"u}nz}}, \bibinfo {author} {\bibfnamefont {M.}~\bibnamefont {Herrmann}},
  \bibinfo {author} {\bibfnamefont {V.}~\bibnamefont {Batteiger}}, \bibinfo
  {author} {\bibfnamefont {G.}~\bibnamefont {Saathoff}}, \bibinfo {author}
  {\bibfnamefont {T.~W.}\ \bibnamefont {H{\"a}nsch}}, \ and\ \bibinfo {author}
  {\bibfnamefont {{\relax Th}.}~\bibnamefont {Udem}},\ }\bibfield  {title}
  {\enquote {\bibinfo {title} {Sub-millikelvin spatial thermometry of a single
  {D}oppler-cooled ion in a {P}aul trap},}\ }\href {\doibase
  10.1103/PhysRevA.85.023427} {\bibfield  {journal} {\bibinfo  {journal} {Phys.
  Rev. A}\ }\textbf {\bibinfo {volume} {85}},\ \bibinfo {pages} {023427}
  (\bibinfo {year} {2012})}\BibitemShut {NoStop}%
\bibitem [{\citenamefont {Rajagopal}\ \emph {et~al.}(2016)\citenamefont
  {Rajagopal}, \citenamefont {Marler}, \citenamefont {Kokish},\ and\
  \citenamefont {Odom}}]{Rajagopal2016}%
  \BibitemOpen
  \bibfield  {author} {\bibinfo {author} {\bibfnamefont {V.}~\bibnamefont
  {Rajagopal}}, \bibinfo {author} {\bibfnamefont {J.~P.}\ \bibnamefont
  {Marler}}, \bibinfo {author} {\bibfnamefont {M.~G.}\ \bibnamefont {Kokish}},
  \ and\ \bibinfo {author} {\bibfnamefont {B.~C.}\ \bibnamefont {Odom}},\
  }\bibfield  {title} {\enquote {\bibinfo {title} {Trapped ion chain
  thermometry and mass spectrometry through imaging},}\ }\href {\doibase
  10.1255/ejms.1408} {\bibfield  {journal} {\bibinfo  {journal} {Eur. J. Mass
  Spectrom.}\ }\textbf {\bibinfo {volume} {22}},\ \bibinfo {pages} {1}
  (\bibinfo {year} {2016})}\BibitemShut {NoStop}%
\bibitem [{\citenamefont {Blatt}\ \emph {et~al.}(1986)\citenamefont {Blatt},
  \citenamefont {Zoller}, \citenamefont {Holzm{\"u}ller},\ and\ \citenamefont
  {Siemers}}]{Blatt1986}%
  \BibitemOpen
  \bibfield  {author} {\bibinfo {author} {\bibfnamefont {R.}~\bibnamefont
  {Blatt}}, \bibinfo {author} {\bibfnamefont {P.}~\bibnamefont {Zoller}},
  \bibinfo {author} {\bibfnamefont {G.}~\bibnamefont {Holzm{\"u}ller}}, \ and\
  \bibinfo {author} {\bibfnamefont {I.}~\bibnamefont {Siemers}},\ }\bibfield
  {title} {\enquote {\bibinfo {title} {{B}rownian motion of a parametric
  oscillator: {A} model for ion confinement in radio frequency traps},}\ }\href
  {\doibase 10.1007/BF01437349} {\bibfield  {journal} {\bibinfo  {journal} {Z.
  Phys. D}\ }\textbf {\bibinfo {volume} {4}},\ \bibinfo {pages} {121--126}
  (\bibinfo {year} {1986})}\BibitemShut {NoStop}%
\bibitem [{\citenamefont {Meissner}\ and\ \citenamefont
  {Ochsenfeld}(1933)}]{MeissnerOchsenfeld1933}%
  \BibitemOpen
  \bibfield  {author} {\bibinfo {author} {\bibfnamefont {W.}~\bibnamefont
  {Meissner}}\ and\ \bibinfo {author} {\bibfnamefont {R.}~\bibnamefont
  {Ochsenfeld}},\ }\bibfield  {title} {\enquote {\bibinfo {title} {Ein neuer
  {E}ffekt bei {E}intritt der {S}upraleitf{\"a}higkeit},}\ }\href {\doibase
  10.1007/BF01504252} {\bibfield  {journal} {\bibinfo  {journal}
  {Naturwissenschaften}\ }\textbf {\bibinfo {volume} {21}},\ \bibinfo {pages}
  {787--788} (\bibinfo {year} {1933})}\BibitemShut {NoStop}%
\bibitem [{\citenamefont {Vallet}\ \emph {et~al.}(1992)\citenamefont {Vallet},
  \citenamefont {Bolor{\'e}}, \citenamefont {Bonin}, \citenamefont {Charrier},
  \citenamefont {Daillant}, \citenamefont {Gratadour}, \citenamefont
  {Koechlin},\ and\ \citenamefont {Safa}}]{Vallet1992}%
  \BibitemOpen
  \bibfield  {author} {\bibinfo {author} {\bibfnamefont {C.}~\bibnamefont
  {Vallet}}, \bibinfo {author} {\bibfnamefont {M.}~\bibnamefont {Bolor{\'e}}},
  \bibinfo {author} {\bibfnamefont {B.}~\bibnamefont {Bonin}}, \bibinfo
  {author} {\bibfnamefont {J.~P.}\ \bibnamefont {Charrier}}, \bibinfo {author}
  {\bibfnamefont {B.}~\bibnamefont {Daillant}}, \bibinfo {author}
  {\bibfnamefont {J.}~\bibnamefont {Gratadour}}, \bibinfo {author}
  {\bibfnamefont {F.}~\bibnamefont {Koechlin}}, \ and\ \bibinfo {author}
  {\bibfnamefont {H.}~\bibnamefont {Safa}},\ }\bibfield  {title} {\enquote
  {\bibinfo {title} {Flux trapping in superconducting cavities},}\ }in\
  \href@noop {} {\emph {\bibinfo {booktitle} {Proc. EPAC '92}}},\ \bibinfo
  {editor} {edited by\ \bibinfo {editor} {\bibfnamefont {H.}~\bibnamefont
  {Heino}}, \bibinfo {editor} {\bibfnamefont {H.}~\bibnamefont {Homeyer}}, \
  and\ \bibinfo {editor} {\bibfnamefont {C.}~\bibnamefont {Petit-Jean-Genaz}}}\
  (\bibinfo {year} {1992})\ pp.\ \bibinfo {pages} {1295--1297}\BibitemShut
  {NoStop}%
\bibitem [{\citenamefont {Padamsee}(2001)}]{Padamsee2001}%
  \BibitemOpen
  \bibfield  {author} {\bibinfo {author} {\bibfnamefont {H.}~\bibnamefont
  {Padamsee}},\ }\bibfield  {title} {\enquote {\bibinfo {title} {The science
  and technology of superconducting cavities for accelerators},}\ }\href
  {\doibase 10.1088/0953-2048/14/4/202} {\bibfield  {journal} {\bibinfo
  {journal} {Supercond. Sci. Technol.}\ }\textbf {\bibinfo {volume} {14}},\
  \bibinfo {pages} {R28} (\bibinfo {year} {2001})}\BibitemShut {NoStop}%
\bibitem [{\citenamefont {Kim}, \citenamefont {Hempstead},\ and\ \citenamefont
  {Strnad}(1965)}]{Kim1965}%
  \BibitemOpen
  \bibfield  {author} {\bibinfo {author} {\bibfnamefont {Y.~B.}\ \bibnamefont
  {Kim}}, \bibinfo {author} {\bibfnamefont {C.~F.}\ \bibnamefont {Hempstead}},
  \ and\ \bibinfo {author} {\bibfnamefont {A.~R.}\ \bibnamefont {Strnad}},\
  }\bibfield  {title} {\enquote {\bibinfo {title} {Flux-flow resistance in
  type-{II} superconductors},}\ }\href {\doibase 10.1103/PhysRev.139.A1163}
  {\bibfield  {journal} {\bibinfo  {journal} {Phys. Rev.}\ }\textbf {\bibinfo
  {volume} {139}},\ \bibinfo {pages} {A1163--A1172} (\bibinfo {year}
  {1965})}\BibitemShut {NoStop}%
\bibitem [{\citenamefont {Bonin}\ and\ \citenamefont
  {R{\"o}th}(1992)}]{Bonin1992}%
  \BibitemOpen
  \bibfield  {author} {\bibinfo {author} {\bibfnamefont {B.}~\bibnamefont
  {Bonin}}\ and\ \bibinfo {author} {\bibfnamefont {R.~W.}\ \bibnamefont
  {R{\"o}th}},\ }\bibfield  {title} {\enquote {\bibinfo {title} {{{$Q$}
  degradation of niobium cavities due to hydrogen contamination}},}\ }\href
  {http://cds.cern.ch/record/1055117} {\bibfield  {journal} {\bibinfo
  {journal} {Part. Accel.}\ }\textbf {\bibinfo {volume} {40}},\ \bibinfo
  {pages} {59--83} (\bibinfo {year} {1992})}\BibitemShut {NoStop}%
\bibitem [{\citenamefont {Gaebler}\ \emph {et~al.}(2016)\citenamefont
  {Gaebler}, \citenamefont {Tan}, \citenamefont {Lin}, \citenamefont {Wan},
  \citenamefont {Bowler}, \citenamefont {Keith}, \citenamefont {Glancy},
  \citenamefont {Coakley}, \citenamefont {Knill}, \citenamefont {Leibfried},\
  and\ \citenamefont {Wineland}}]{Gaebler2016}%
  \BibitemOpen
  \bibfield  {author} {\bibinfo {author} {\bibfnamefont {J.~P.}\ \bibnamefont
  {Gaebler}}, \bibinfo {author} {\bibfnamefont {T.~R.}\ \bibnamefont {Tan}},
  \bibinfo {author} {\bibfnamefont {Y.}~\bibnamefont {Lin}}, \bibinfo {author}
  {\bibfnamefont {Y.}~\bibnamefont {Wan}}, \bibinfo {author} {\bibfnamefont
  {R.}~\bibnamefont {Bowler}}, \bibinfo {author} {\bibfnamefont {A.~C.}\
  \bibnamefont {Keith}}, \bibinfo {author} {\bibfnamefont {S.}~\bibnamefont
  {Glancy}}, \bibinfo {author} {\bibfnamefont {K.}~\bibnamefont {Coakley}},
  \bibinfo {author} {\bibfnamefont {E.}~\bibnamefont {Knill}}, \bibinfo
  {author} {\bibfnamefont {D.}~\bibnamefont {Leibfried}}, \ and\ \bibinfo
  {author} {\bibfnamefont {D.~J.}\ \bibnamefont {Wineland}},\ }\bibfield
  {title} {\enquote {\bibinfo {title} {High-fidelity universal gate set for
  ${^{9}\mathrm{Be}}^{+}$ ion qubits},}\ }\href {\doibase
  10.1103/PhysRevLett.117.060505} {\bibfield  {journal} {\bibinfo  {journal}
  {Phys. Rev. Lett.}\ }\textbf {\bibinfo {volume} {117}},\ \bibinfo {pages}
  {060505} (\bibinfo {year} {2016})}\BibitemShut {NoStop}%
\bibitem [{\citenamefont {Ramsey}(1950)}]{Ramsey1950}%
  \BibitemOpen
  \bibfield  {author} {\bibinfo {author} {\bibfnamefont {N.~F.}\ \bibnamefont
  {Ramsey}},\ }\bibfield  {title} {\enquote {\bibinfo {title} {A molecular beam
  resonance method with separated oscillating fields},}\ }\href {\doibase
  10.1103/PhysRev.78.695} {\bibfield  {journal} {\bibinfo  {journal} {Phys.
  Rev.}\ }\textbf {\bibinfo {volume} {78}},\ \bibinfo {pages} {695--699}
  (\bibinfo {year} {1950})}\BibitemShut {NoStop}%
\bibitem [{\citenamefont {{NIST Ion Storage Group}}(2017)}]{ARTIQ}%
  \BibitemOpen
  \bibfield  {author} {\bibinfo {author} {\bibnamefont {{NIST Ion Storage
  Group}}},\ }\href {\doibase 10.5281/zenodo.591804} {\enquote {\bibinfo
  {title} {{ARTIQ} ({A}dvanced {R}eal-{T}ime {I}nfrastructure for {Q}uantum
  {P}hysics)},}\ } (\bibinfo {year} {2017})\BibitemShut {NoStop}%
\bibitem [{\citenamefont {Chuchelov}\ \emph {et~al.}(2019)\citenamefont
  {Chuchelov}, \citenamefont {Tsygankov}, \citenamefont {Zibrov}, \citenamefont
  {Vaskovskaya}, \citenamefont {Vassiliev}, \citenamefont {Zibrov},
  \citenamefont {Yudin}, \citenamefont {Taichenachev},\ and\ \citenamefont
  {Velichansky}}]{Chuchelov2019}%
  \BibitemOpen
  \bibfield  {author} {\bibinfo {author} {\bibfnamefont {D.~S.}\ \bibnamefont
  {Chuchelov}}, \bibinfo {author} {\bibfnamefont {E.~A.}\ \bibnamefont
  {Tsygankov}}, \bibinfo {author} {\bibfnamefont {S.~A.}\ \bibnamefont
  {Zibrov}}, \bibinfo {author} {\bibfnamefont {M.~I.}\ \bibnamefont
  {Vaskovskaya}}, \bibinfo {author} {\bibfnamefont {V.~V.}\ \bibnamefont
  {Vassiliev}}, \bibinfo {author} {\bibfnamefont {A.~S.}\ \bibnamefont
  {Zibrov}}, \bibinfo {author} {\bibfnamefont {V.~I.}\ \bibnamefont {Yudin}},
  \bibinfo {author} {\bibfnamefont {A.~V.}\ \bibnamefont {Taichenachev}}, \
  and\ \bibinfo {author} {\bibfnamefont {V.~L.}\ \bibnamefont {Velichansky}},\
  }\bibfield  {title} {\enquote {\bibinfo {title} {Central {R}amsey fringe
  identification by means of an auxiliary optical field},}\ }\href {\doibase
  10.1063/1.5111312} {\bibfield  {journal} {\bibinfo  {journal} {J. Appl.
  Phys.}\ }\textbf {\bibinfo {volume} {126}},\ \bibinfo {pages} {054503}
  (\bibinfo {year} {2019})}\BibitemShut {NoStop}%
\bibitem [{\citenamefont {Shiga}, \citenamefont {Itano},\ and\ \citenamefont
  {Bollinger}(2011)}]{Shiga2011}%
  \BibitemOpen
  \bibfield  {author} {\bibinfo {author} {\bibfnamefont {N.}~\bibnamefont
  {Shiga}}, \bibinfo {author} {\bibfnamefont {W.~M.}\ \bibnamefont {Itano}}, \
  and\ \bibinfo {author} {\bibfnamefont {J.~J.}\ \bibnamefont {Bollinger}},\
  }\bibfield  {title} {\enquote {\bibinfo {title} {Diamagnetic correction to
  the $^\text{9}\text{Be}^\text{+}$ ground-state hyperfine constant},}\ }\href
  {\doibase 10.1103/PhysRevA.84.012510} {\bibfield  {journal} {\bibinfo
  {journal} {Phys. Rev. A}\ }\textbf {\bibinfo {volume} {84}},\ \bibinfo
  {pages} {012510} (\bibinfo {year} {2011})}\BibitemShut {NoStop}%
\bibitem [{\citenamefont {Ruster}\ \emph {et~al.}(2016)\citenamefont {Ruster},
  \citenamefont {Schmiegelow}, \citenamefont {Kaufmann}, \citenamefont
  {Warschburger}, \citenamefont {Schmidt-Kaler},\ and\ \citenamefont
  {Poschinger}}]{Ruster2016}%
  \BibitemOpen
  \bibfield  {author} {\bibinfo {author} {\bibfnamefont {T.}~\bibnamefont
  {Ruster}}, \bibinfo {author} {\bibfnamefont {C.~T.}\ \bibnamefont
  {Schmiegelow}}, \bibinfo {author} {\bibfnamefont {H.}~\bibnamefont
  {Kaufmann}}, \bibinfo {author} {\bibfnamefont {C.}~\bibnamefont
  {Warschburger}}, \bibinfo {author} {\bibfnamefont {F.}~\bibnamefont
  {Schmidt-Kaler}}, \ and\ \bibinfo {author} {\bibfnamefont {U.~G.}\
  \bibnamefont {Poschinger}},\ }\bibfield  {title} {\enquote {\bibinfo {title}
  {A long-lived {Z}eeman trapped-ion qubit},}\ }\href {\doibase
  10.1007/s00340-016-6527-4} {\bibfield  {journal} {\bibinfo  {journal} {Appl.
  Phys. B}\ }\textbf {\bibinfo {volume} {122}},\ \bibinfo {pages} {254}
  (\bibinfo {year} {2016})}\BibitemShut {NoStop}%
\bibitem [{\citenamefont {Altarev}\ \emph {et~al.}(2014)\citenamefont
  {Altarev}, \citenamefont {Babcock}, \citenamefont {Beck}, \citenamefont
  {Burghoff}, \citenamefont {Chesnevskaya}, \citenamefont {Chupp},
  \citenamefont {Degenkolb}, \citenamefont {Fan}, \citenamefont {Fierlinger},
  \citenamefont {Frei}, \citenamefont {Gutsmiedl}, \citenamefont
  {Knappe-Gr{\"u}neberg}, \citenamefont {Kuchler}, \citenamefont {Lauer},
  \citenamefont {Link}, \citenamefont {Lins}, \citenamefont {Marino},
  \citenamefont {McAndrew}, \citenamefont {Niessen}, \citenamefont {Paul},
  \citenamefont {Petzoldt}, \citenamefont {Schl{\"a}pfer}, \citenamefont
  {Schnabel}, \citenamefont {Sharma}, \citenamefont {Singh}, \citenamefont
  {Stoepler}, \citenamefont {Stuiber}, \citenamefont {Sturm}, \citenamefont
  {Taubenheim}, \citenamefont {Trahms}, \citenamefont {Voigt},\ and\
  \citenamefont {Zechlau}}]{Altarev2014}%
  \BibitemOpen
  \bibfield  {author} {\bibinfo {author} {\bibfnamefont {I.}~\bibnamefont
  {Altarev}}, \bibinfo {author} {\bibfnamefont {E.}~\bibnamefont {Babcock}},
  \bibinfo {author} {\bibfnamefont {D.}~\bibnamefont {Beck}}, \bibinfo {author}
  {\bibfnamefont {M.}~\bibnamefont {Burghoff}}, \bibinfo {author}
  {\bibfnamefont {S.}~\bibnamefont {Chesnevskaya}}, \bibinfo {author}
  {\bibfnamefont {T.}~\bibnamefont {Chupp}}, \bibinfo {author} {\bibfnamefont
  {S.}~\bibnamefont {Degenkolb}}, \bibinfo {author} {\bibfnamefont
  {I.}~\bibnamefont {Fan}}, \bibinfo {author} {\bibfnamefont {P.}~\bibnamefont
  {Fierlinger}}, \bibinfo {author} {\bibfnamefont {A.}~\bibnamefont {Frei}},
  \bibinfo {author} {\bibfnamefont {E.}~\bibnamefont {Gutsmiedl}}, \bibinfo
  {author} {\bibfnamefont {S.}~\bibnamefont {Knappe-Gr{\"u}neberg}}, \bibinfo
  {author} {\bibfnamefont {F.}~\bibnamefont {Kuchler}}, \bibinfo {author}
  {\bibfnamefont {T.}~\bibnamefont {Lauer}}, \bibinfo {author} {\bibfnamefont
  {P.}~\bibnamefont {Link}}, \bibinfo {author} {\bibfnamefont {T.}~\bibnamefont
  {Lins}}, \bibinfo {author} {\bibfnamefont {M.}~\bibnamefont {Marino}},
  \bibinfo {author} {\bibfnamefont {J.}~\bibnamefont {McAndrew}}, \bibinfo
  {author} {\bibfnamefont {B.}~\bibnamefont {Niessen}}, \bibinfo {author}
  {\bibfnamefont {S.}~\bibnamefont {Paul}}, \bibinfo {author} {\bibfnamefont
  {G.}~\bibnamefont {Petzoldt}}, \bibinfo {author} {\bibfnamefont
  {U.}~\bibnamefont {Schl{\"a}pfer}}, \bibinfo {author} {\bibfnamefont
  {A.}~\bibnamefont {Schnabel}}, \bibinfo {author} {\bibfnamefont
  {S.}~\bibnamefont {Sharma}}, \bibinfo {author} {\bibfnamefont
  {J.}~\bibnamefont {Singh}}, \bibinfo {author} {\bibfnamefont
  {R.}~\bibnamefont {Stoepler}}, \bibinfo {author} {\bibfnamefont
  {S.}~\bibnamefont {Stuiber}}, \bibinfo {author} {\bibfnamefont
  {M.}~\bibnamefont {Sturm}}, \bibinfo {author} {\bibfnamefont
  {B.}~\bibnamefont {Taubenheim}}, \bibinfo {author} {\bibfnamefont
  {L.}~\bibnamefont {Trahms}}, \bibinfo {author} {\bibfnamefont
  {J.}~\bibnamefont {Voigt}}, \ and\ \bibinfo {author} {\bibfnamefont
  {T.}~\bibnamefont {Zechlau}},\ }\bibfield  {title} {\enquote {\bibinfo
  {title} {A magnetically shielded room with ultra low residual field and
  gradient},}\ }\href {\doibase 10.1063/1.4886146} {\bibfield  {journal}
  {\bibinfo  {journal} {Rev. Sci. Instrum.}\ }\textbf {\bibinfo {volume}
  {85}},\ \bibinfo {pages} {075106} (\bibinfo {year} {2014})}\BibitemShut
  {NoStop}%
\bibitem [{\citenamefont {Ladd}\ \emph {et~al.}(2010)\citenamefont {Ladd},
  \citenamefont {Jelezko}, \citenamefont {Laflamme}, \citenamefont {Nakamura},
  \citenamefont {Monroe},\ and\ \citenamefont {O'Brien}}]{Ladd2010}%
  \BibitemOpen
  \bibfield  {author} {\bibinfo {author} {\bibfnamefont {T.~D.}\ \bibnamefont
  {Ladd}}, \bibinfo {author} {\bibfnamefont {F.}~\bibnamefont {Jelezko}},
  \bibinfo {author} {\bibfnamefont {R.}~\bibnamefont {Laflamme}}, \bibinfo
  {author} {\bibfnamefont {Y.}~\bibnamefont {Nakamura}}, \bibinfo {author}
  {\bibfnamefont {C.}~\bibnamefont {Monroe}}, \ and\ \bibinfo {author}
  {\bibfnamefont {J.~L.}\ \bibnamefont {O'Brien}},\ }\bibfield  {title}
  {\enquote {\bibinfo {title} {Quantum computers},}\ }\href {\doibase
  10.1038/nature08812} {\bibfield  {journal} {\bibinfo  {journal} {Nature}\
  }\textbf {\bibinfo {volume} {464}},\ \bibinfo {pages} {45--53} (\bibinfo
  {year} {2010})}\BibitemShut {NoStop}%
\bibitem [{\citenamefont {Monz}\ \emph {et~al.}(2011)\citenamefont {Monz},
  \citenamefont {Schindler}, \citenamefont {Barreiro}, \citenamefont {Chwalla},
  \citenamefont {Nigg}, \citenamefont {Coish}, \citenamefont {Harlander},
  \citenamefont {H{\"a}nsel}, \citenamefont {Hennrich},\ and\ \citenamefont
  {Blatt}}]{Monz2011}%
  \BibitemOpen
  \bibfield  {author} {\bibinfo {author} {\bibfnamefont {T.}~\bibnamefont
  {Monz}}, \bibinfo {author} {\bibfnamefont {P.}~\bibnamefont {Schindler}},
  \bibinfo {author} {\bibfnamefont {J.~T.}\ \bibnamefont {Barreiro}}, \bibinfo
  {author} {\bibfnamefont {M.}~\bibnamefont {Chwalla}}, \bibinfo {author}
  {\bibfnamefont {D.}~\bibnamefont {Nigg}}, \bibinfo {author} {\bibfnamefont
  {W.~A.}\ \bibnamefont {Coish}}, \bibinfo {author} {\bibfnamefont
  {M.}~\bibnamefont {Harlander}}, \bibinfo {author} {\bibfnamefont
  {W.}~\bibnamefont {H{\"a}nsel}}, \bibinfo {author} {\bibfnamefont
  {M.}~\bibnamefont {Hennrich}}, \ and\ \bibinfo {author} {\bibfnamefont
  {R.}~\bibnamefont {Blatt}},\ }\bibfield  {title} {\enquote {\bibinfo {title}
  {14-qubit entanglement: {C}reation and coherence},}\ }\href {\doibase
  10.1103/PhysRevLett.106.130506} {\bibfield  {journal} {\bibinfo  {journal}
  {Phys. Rev. Lett.}\ }\textbf {\bibinfo {volume} {106}},\ \bibinfo {pages}
  {130506} (\bibinfo {year} {2011})}\BibitemShut {NoStop}%
\bibitem [{\citenamefont {Vandersypen}\ and\ \citenamefont
  {Chuang}(2005)}]{Vandersypen2005}%
  \BibitemOpen
  \bibfield  {author} {\bibinfo {author} {\bibfnamefont {L.~M.~K.}\
  \bibnamefont {Vandersypen}}\ and\ \bibinfo {author} {\bibfnamefont {I.~L.}\
  \bibnamefont {Chuang}},\ }\bibfield  {title} {\enquote {\bibinfo {title}
  {{NMR} techniques for quantum control and computation},}\ }\href {\doibase
  10.1103/RevModPhys.76.1037} {\bibfield  {journal} {\bibinfo  {journal} {Rev.
  Mod. Phys.}\ }\textbf {\bibinfo {volume} {76}},\ \bibinfo {pages}
  {1037--1069} (\bibinfo {year} {2005})}\BibitemShut {NoStop}%
\bibitem [{\citenamefont {Leibfried}\ \emph {et~al.}(2003)\citenamefont
  {Leibfried}, \citenamefont {Blatt}, \citenamefont {Monroe},\ and\
  \citenamefont {Wineland}}]{Leibfried2003}%
  \BibitemOpen
  \bibfield  {author} {\bibinfo {author} {\bibfnamefont {D.}~\bibnamefont
  {Leibfried}}, \bibinfo {author} {\bibfnamefont {R.}~\bibnamefont {Blatt}},
  \bibinfo {author} {\bibfnamefont {C.}~\bibnamefont {Monroe}}, \ and\ \bibinfo
  {author} {\bibfnamefont {D.}~\bibnamefont {Wineland}},\ }\bibfield  {title}
  {\enquote {\bibinfo {title} {Quantum dynamics of single trapped ions},}\
  }\href {\doibase 10.1103/RevModPhys.75.281} {\bibfield  {journal} {\bibinfo
  {journal} {Rev. Mod. Phys.}\ }\textbf {\bibinfo {volume} {75}},\ \bibinfo
  {pages} {281--324} (\bibinfo {year} {2003})}\BibitemShut {NoStop}%
\bibitem [{\citenamefont {Kotler}\ \emph {et~al.}(2011)\citenamefont {Kotler},
  \citenamefont {Akerman}, \citenamefont {Glickman}, \citenamefont {Keselman},\
  and\ \citenamefont {Ozeri}}]{Kotler2011}%
  \BibitemOpen
  \bibfield  {author} {\bibinfo {author} {\bibfnamefont {S.}~\bibnamefont
  {Kotler}}, \bibinfo {author} {\bibfnamefont {N.}~\bibnamefont {Akerman}},
  \bibinfo {author} {\bibfnamefont {Y.}~\bibnamefont {Glickman}}, \bibinfo
  {author} {\bibfnamefont {A.}~\bibnamefont {Keselman}}, \ and\ \bibinfo
  {author} {\bibfnamefont {R.}~\bibnamefont {Ozeri}},\ }\bibfield  {title}
  {\enquote {\bibinfo {title} {Single-ion quantum lock-in amplifier},}\ }\href
  {\doibase 10.1038/nature10010} {\bibfield  {journal} {\bibinfo  {journal}
  {Nature}\ }\textbf {\bibinfo {volume} {473}},\ \bibinfo {pages} {61--65}
  (\bibinfo {year} {2011})}\BibitemShut {NoStop}%
\bibitem [{\citenamefont {Harty}\ \emph {et~al.}(2014)\citenamefont {Harty},
  \citenamefont {Allcock}, \citenamefont {Ballance}, \citenamefont {Guidoni},
  \citenamefont {Janacek}, \citenamefont {Linke}, \citenamefont {Stacey},\ and\
  \citenamefont {Lucas}}]{Harty2014}%
  \BibitemOpen
  \bibfield  {author} {\bibinfo {author} {\bibfnamefont {T.~P.}\ \bibnamefont
  {Harty}}, \bibinfo {author} {\bibfnamefont {D.~T.~C.}\ \bibnamefont
  {Allcock}}, \bibinfo {author} {\bibfnamefont {C.~J.}\ \bibnamefont
  {Ballance}}, \bibinfo {author} {\bibfnamefont {L.}~\bibnamefont {Guidoni}},
  \bibinfo {author} {\bibfnamefont {H.~A.}\ \bibnamefont {Janacek}}, \bibinfo
  {author} {\bibfnamefont {N.~M.}\ \bibnamefont {Linke}}, \bibinfo {author}
  {\bibfnamefont {D.~N.}\ \bibnamefont {Stacey}}, \ and\ \bibinfo {author}
  {\bibfnamefont {D.~M.}\ \bibnamefont {Lucas}},\ }\bibfield  {title} {\enquote
  {\bibinfo {title} {High-fidelity preparation, gates, memory, and readout of a
  trapped-ion quantum bit},}\ }\href {\doibase 10.1103/PhysRevLett.113.220501}
  {\bibfield  {journal} {\bibinfo  {journal} {Phys. Rev. Lett.}\ }\textbf
  {\bibinfo {volume} {113}},\ \bibinfo {pages} {220501} (\bibinfo {year}
  {2014})}\BibitemShut {NoStop}%
\bibitem [{\citenamefont {Kotler}\ \emph {et~al.}(2014)\citenamefont {Kotler},
  \citenamefont {Akerman}, \citenamefont {Navon}, \citenamefont {Glickman},\
  and\ \citenamefont {Ozeri}}]{Kotler2014}%
  \BibitemOpen
  \bibfield  {author} {\bibinfo {author} {\bibfnamefont {S.}~\bibnamefont
  {Kotler}}, \bibinfo {author} {\bibfnamefont {N.}~\bibnamefont {Akerman}},
  \bibinfo {author} {\bibfnamefont {N.}~\bibnamefont {Navon}}, \bibinfo
  {author} {\bibfnamefont {Y.}~\bibnamefont {Glickman}}, \ and\ \bibinfo
  {author} {\bibfnamefont {R.}~\bibnamefont {Ozeri}},\ }\bibfield  {title}
  {\enquote {\bibinfo {title} {Measurement of the magnetic interaction between
  two bound electrons of two separate ions},}\ }\href {\doibase
  10.1038/nature13403} {\bibfield  {journal} {\bibinfo  {journal} {Nature}\
  }\textbf {\bibinfo {volume} {510}},\ \bibinfo {pages} {376--380} (\bibinfo
  {year} {2014})}\BibitemShut {NoStop}%
\bibitem [{\citenamefont {Wang}\ \emph {et~al.}(2021)\citenamefont {Wang},
  \citenamefont {Luan}, \citenamefont {Qiao}, \citenamefont {Um}, \citenamefont
  {Zhang}, \citenamefont {Wang}, \citenamefont {Yuan}, \citenamefont {Gu},
  \citenamefont {Zhang},\ and\ \citenamefont {Kim}}]{Wang2021}%
  \BibitemOpen
  \bibfield  {author} {\bibinfo {author} {\bibfnamefont {P.}~\bibnamefont
  {Wang}}, \bibinfo {author} {\bibfnamefont {C.-Y.}\ \bibnamefont {Luan}},
  \bibinfo {author} {\bibfnamefont {M.}~\bibnamefont {Qiao}}, \bibinfo {author}
  {\bibfnamefont {M.}~\bibnamefont {Um}}, \bibinfo {author} {\bibfnamefont
  {J.}~\bibnamefont {Zhang}}, \bibinfo {author} {\bibfnamefont
  {Y.}~\bibnamefont {Wang}}, \bibinfo {author} {\bibfnamefont {X.}~\bibnamefont
  {Yuan}}, \bibinfo {author} {\bibfnamefont {M.}~\bibnamefont {Gu}}, \bibinfo
  {author} {\bibfnamefont {J.}~\bibnamefont {Zhang}}, \ and\ \bibinfo {author}
  {\bibfnamefont {K.}~\bibnamefont {Kim}},\ }\bibfield  {title} {\enquote
  {\bibinfo {title} {Single ion qubit with estimated coherence time exceeding
  one hour},}\ }\href {\doibase 10.1038/s41467-020-20330-w} {\bibfield
  {journal} {\bibinfo  {journal} {Nat. Commun.}\ }\textbf {\bibinfo {volume}
  {12}},\ \bibinfo {pages} {233} (\bibinfo {year} {2021})}\BibitemShut
  {NoStop}%
\bibitem [{\citenamefont {H{\"a}ffner}, \citenamefont {Roos},\ and\
  \citenamefont {Blatt}(2008)}]{Haffner2008}%
  \BibitemOpen
  \bibfield  {author} {\bibinfo {author} {\bibfnamefont {H.}~\bibnamefont
  {H{\"a}ffner}}, \bibinfo {author} {\bibfnamefont {C.~F.}\ \bibnamefont
  {Roos}}, \ and\ \bibinfo {author} {\bibfnamefont {R.}~\bibnamefont {Blatt}},\
  }\bibfield  {title} {\enquote {\bibinfo {title} {Quantum computing with
  trapped ions},}\ }\href {\doibase 10.1016/j.physrep.2008.09.003} {\bibfield
  {journal} {\bibinfo  {journal} {Phys. Rep.}\ }\textbf {\bibinfo {volume}
  {469}},\ \bibinfo {pages} {155--203} (\bibinfo {year} {2008})}\BibitemShut
  {NoStop}%
\bibitem [{\citenamefont {Blatt}\ and\ \citenamefont {Roos}(2012)}]{Blatt2012}%
  \BibitemOpen
  \bibfield  {author} {\bibinfo {author} {\bibfnamefont {R.}~\bibnamefont
  {Blatt}}\ and\ \bibinfo {author} {\bibfnamefont {C.~F.}\ \bibnamefont
  {Roos}},\ }\bibfield  {title} {\enquote {\bibinfo {title} {Quantum
  simulations with trapped ions},}\ }\href {\doibase 10.1038/nphys2252}
  {\bibfield  {journal} {\bibinfo  {journal} {Nat. Phys.}\ }\textbf {\bibinfo
  {volume} {8}},\ \bibinfo {pages} {277--284} (\bibinfo {year}
  {2012})}\BibitemShut {NoStop}%
\bibitem [{\citenamefont {Lidar}, \citenamefont {Chuang},\ and\ \citenamefont
  {Whaley}(1998)}]{Lidar1998}%
  \BibitemOpen
  \bibfield  {author} {\bibinfo {author} {\bibfnamefont {D.~A.}\ \bibnamefont
  {Lidar}}, \bibinfo {author} {\bibfnamefont {I.~L.}\ \bibnamefont {Chuang}}, \
  and\ \bibinfo {author} {\bibfnamefont {K.~B.}\ \bibnamefont {Whaley}},\
  }\bibfield  {title} {\enquote {\bibinfo {title} {Decoherence-free subspaces
  for quantum computation},}\ }\href {\doibase 10.1103/PhysRevLett.81.2594}
  {\bibfield  {journal} {\bibinfo  {journal} {Phys. Rev. Lett.}\ }\textbf
  {\bibinfo {volume} {81}},\ \bibinfo {pages} {2594--2597} (\bibinfo {year}
  {1998})}\BibitemShut {NoStop}%
\bibitem [{\citenamefont {Manovitz}\ \emph {et~al.}(2019)\citenamefont
  {Manovitz}, \citenamefont {Shaniv}, \citenamefont {Shapira}, \citenamefont
  {Ozeri},\ and\ \citenamefont {Akerman}}]{Manovitz2019}%
  \BibitemOpen
  \bibfield  {author} {\bibinfo {author} {\bibfnamefont {T.}~\bibnamefont
  {Manovitz}}, \bibinfo {author} {\bibfnamefont {R.}~\bibnamefont {Shaniv}},
  \bibinfo {author} {\bibfnamefont {Y.}~\bibnamefont {Shapira}}, \bibinfo
  {author} {\bibfnamefont {R.}~\bibnamefont {Ozeri}}, \ and\ \bibinfo {author}
  {\bibfnamefont {N.}~\bibnamefont {Akerman}},\ }\bibfield  {title} {\enquote
  {\bibinfo {title} {Precision measurement of atomic isotope shifts using a
  two-isotope entangled state},}\ }\href {\doibase
  10.1103/PhysRevLett.123.203001} {\bibfield  {journal} {\bibinfo  {journal}
  {Phys. Rev. Lett.}\ }\textbf {\bibinfo {volume} {123}},\ \bibinfo {pages}
  {203001} (\bibinfo {year} {2019})}\BibitemShut {NoStop}%
\bibitem [{\citenamefont {Pruttivarasin}\ \emph {et~al.}(2015)\citenamefont
  {Pruttivarasin}, \citenamefont {Ramm}, \citenamefont {Porsev}, \citenamefont
  {Tupitsyn}, \citenamefont {Safronova}, \citenamefont {Hohensee},\ and\
  \citenamefont {H{\"a}ffner}}]{Pruttivarasin2015}%
  \BibitemOpen
  \bibfield  {author} {\bibinfo {author} {\bibfnamefont {T.}~\bibnamefont
  {Pruttivarasin}}, \bibinfo {author} {\bibfnamefont {M.}~\bibnamefont {Ramm}},
  \bibinfo {author} {\bibfnamefont {S.~G.}\ \bibnamefont {Porsev}}, \bibinfo
  {author} {\bibfnamefont {I.~I.}\ \bibnamefont {Tupitsyn}}, \bibinfo {author}
  {\bibfnamefont {M.~S.}\ \bibnamefont {Safronova}}, \bibinfo {author}
  {\bibfnamefont {M.~A.}\ \bibnamefont {Hohensee}}, \ and\ \bibinfo {author}
  {\bibfnamefont {H.}~\bibnamefont {H{\"a}ffner}},\ }\bibfield  {title}
  {\enquote {\bibinfo {title} {{M}ichelson--{M}orley analogue for electrons
  using trapped ions to test {L}orentz symmetry},}\ }\href {\doibase
  10.1038/nature14091} {\bibfield  {journal} {\bibinfo  {journal} {Nature}\
  }\textbf {\bibinfo {volume} {517}},\ \bibinfo {pages} {592--595} (\bibinfo
  {year} {2015})}\BibitemShut {NoStop}%
\bibitem [{\citenamefont {Dreissen}\ \emph {et~al.}(2022)\citenamefont
  {Dreissen}, \citenamefont {Yeh}, \citenamefont {F{\"u}rst}, \citenamefont
  {Grensemann},\ and\ \citenamefont {Mehlst{\"a}ubler}}]{Dreissen2022}%
  \BibitemOpen
  \bibfield  {author} {\bibinfo {author} {\bibfnamefont {L.~S.}\ \bibnamefont
  {Dreissen}}, \bibinfo {author} {\bibfnamefont {C.-H.}\ \bibnamefont {Yeh}},
  \bibinfo {author} {\bibfnamefont {H.~A.}\ \bibnamefont {F{\"u}rst}}, \bibinfo
  {author} {\bibfnamefont {K.~C.}\ \bibnamefont {Grensemann}}, \ and\ \bibinfo
  {author} {\bibfnamefont {T.~E.}\ \bibnamefont {Mehlst{\"a}ubler}},\
  }\bibfield  {title} {\enquote {\bibinfo {title} {Improved bounds on {L}orentz
  violation from composite pulse {R}amsey spectroscopy in a trapped ion},}\
  }\href {\doibase 10.1038/s41467-022-34818-0} {\bibfield  {journal} {\bibinfo
  {journal} {Nat. Commun.}\ }\textbf {\bibinfo {volume} {13}},\ \bibinfo
  {pages} {7314} (\bibinfo {year} {2022})}\BibitemShut {NoStop}%
\end{thebibliography}%

\end{document}